\documentclass[preprint,authoryear,12pt]{elsarticle}

\usepackage{times}
\usepackage{bm}
\usepackage{natbib}
\usepackage{amsmath,amsfonts}
\usepackage{dsfont}
\usepackage{graphicx}
\usepackage{amssymb}
\usepackage[latin1]{inputenc}
\usepackage{setspace}                   % espaçamento flexível
\usepackage{multirow}
\usepackage{color}

\graphicspath{{./figuras/}}             % caminho das figuras (recomendável)
\newcommand{\vg}[1]{\mbox{\boldmath $#1$}}    % Negrito (gregas)
\newcommand{\vm}[1]{\ensuremath{\mathbf{#1}}} % Negrito (romanas)
\journal{Computational Statistics $\&$ Data Analysis}

\begin{document}

\begin{frontmatter}

%\title{Beta regression models with measurement errors}
\title{Errors-in-variables beta regression models}
\author{Jalmar M. F. Carrasco}
\address{Departamento de Estatística, Universidade Federal da Bahia, Brazil}
\author{Silvia L. P. Ferrari
\footnote
{Corresponding author: Departamento de Estatística, Universidade de São Paulo,
Rua do Matão, 1010, 05508-090,  São Paulo, SP, Brazil.
e-mail:silviaferrari.usp@gmail.com}
}
\address{Departamento de Estatística, Universidade de São Paulo, Brazil}
\author{Reinaldo B. Arellano-Valle}
\address{Departamento de Estadística, Pontifícia Universidad Católica de Chile, Chile}
\date{}

%\begin{document}

%\maketitle

\begin{abstract}
Beta regression models provide an adequate approach for modeling continuous outcomes limited to the interval $(0,1)$.
This paper deals with an extension of beta regression models that allow for explanatory variables to be measured with error.
The structural approach, in which the covariates measured with error are assumed to be random variables, is employed.
Three estimation methods are presented, namely maximum likelihood, maximum pseudo-likelihood and regression calibration.
Monte Carlo simulations are used to evaluate the performance of the proposed  estimators and the na\"{i}ve estimator. Also,
a residual analysis for beta regression models with measurement errors is proposed. The results are
illustrated in a real data set.\footnote{Supplementary material presents numerical tables used to produce Figures 1-8.}

\vspace{5mm} \noindent {\em Keywords}: Beta regression model; \ Errors-in-variables model; \ Gauss-Hermite quadrature;
Maximum likelihood; \ Maximum pseudo-likelihood; \ Regression calibration.
\end{abstract}

\end{frontmatter}

\section{Introduction}
\label{sec:intro}
Errors-in-variables models, also called measurement error models, are widely applicable in many research areas since
they allow for the presence of explicative variables that are measured with errors or that cannot be observed directly
(latent variables). Many examples and applications of these models are considered in the books by \citet{Fuller},
\citet{Ca+Ru+Ste} and \citet{Cheng+VanNess}.  It is well known that measurement errors cause biased and inconsistent
parameter estimates and lead to erroneous conclusions in inferential analysis.
Errors-in-variables models are specified in such a way that the
distribution of the response variable, $y$, is assumed to depend on covariates, $x$, which are imprecisely measured,
and observable variables, $w$, are seen as surrogates for the unobservable true covariates.
The classical linear errors-in-variables model has been extensively discussed in the literature, particularly under the
normality assumption for the distribution of the unobservable variables. For a systematic review of such models see
\citet{Fuller} and \citet{Cheng+VanNess}; see also \citet{Arellano+Bolfarine} and \citet{Castro+Galea+Bolfarine3}.

The beta regression models provide an adequate approach for modeling continuous outcomes limited to the interval $(0,1)$,
or more generally, limited to any open interval $(a,b)$ as long as the limits are known \citep{Ferra+Crib}. Although
the literature on beta regression has grown fast in the last few years, errors-in-variables models with beta distributed
outcomes is an unexplored area.

A beta regression model assumes that the response variable, $y$, has a beta distribution with probability density function
\begin{eqnarray}
\label{ferrari}
f(y;\mu,\phi)=\frac{\Gamma(\phi)}{\Gamma(\mu \phi)\Gamma[(1-\mu)\phi]}y^{\mu \phi-1}(1-y)^{(1-\mu)\phi-1}, \ \ 0<y<1,
\end{eqnarray}
where $\Gamma(\cdot)$ is the gamma function, $0<\mu<1$ and $\phi>0$, and we write $y\sim{\rm Beta}(\mu,\phi)$.
Here, $\mu=\textrm{E}(y)$ and $\phi$ is regarded as a precision parameter since $\textrm{Var}(y)=\mu(1-\mu)/(1+\phi)$.
For independent observations $y_{1},y_{2},\ldots,y_{n}$, where each $y_{t}$ follows a beta density (\ref{ferrari})
with mean $\mu_{t}$ and unknown precision parameter $\phi$, the beta regression model defined by \cite{Ferra+Crib}
assumes that
\begin{eqnarray}
\label{f1}
g(\mu_{t})=\vm{z}^{\top}_{t}\vg{\alpha},
\end{eqnarray}
with $\vg{\alpha} \in \mathbb{R}^{p_\alpha}$ being  a column vector of unknown parameters, and with
$\vm{z}^{\top}_{t}=(z_{t1},\ldots,z_{t{p_\alpha}})$ being a vector of $p_\alpha$ fixed covariates $(p_\alpha<n)$.
The link function  $g(\cdot):(0,1)\rightarrow \mathbb{R}$ is assumed  to be a continuous, strictly monotone and twice
differentiable function. There are many possible choices for $g(\cdot)$, for instance, the logit link,
$g(\mu_{t})=\log[\mu_{t}/(1-\mu_{t})]$, the probit link, $g(\mu_{t})=\Phi^{-1}(\mu_{t})$, where $\Phi(\cdot)$ is the
cumulative distribution function of the standard normal distribution, and the complementary log-log link,
$g(\mu_{t})=\log[-\log(1-\mu_{t})]$.

Extensions for the beta regression model proposed by  \cite{Ferra+Crib} that allow the precision parameter to vary across
observations, or that involve non-linear structures for the regression specification of the mean and the precision parameter,
are presented by \cite{Smithson+Verkuilen}, \cite{Sim+Barr+Roch}, among others. The beta regression model with linear
specification for the transformed mean and precision parameter is given by (\ref{ferrari}), (\ref{f1}) and
\begin{eqnarray}
\label{f2}
h(\phi_{t})&=&\vm{v}_{t}^{\top}\vg{\gamma},
\end{eqnarray}
where
$\vg{\gamma}\in \mathbb{R}^{p_\gamma}$  ($p_\alpha+p_\gamma<n$)
is a column vector of unknown parameters, $\vm{v}_{t}=(v_{t1}, \cdots, v_{tp_\gamma})^{\top}$  is a vector of fixed
covariates, $h(\cdot):(0,\infty)\longrightarrow \mathbb{R}$ is a strictly monotone, twice differentiable link function.
A possible choice for  $h(\cdot)$ is $h(\phi_{t})=\log(\phi_{t})$.

The purpose of this paper is to extend the beta regression model (\ref{ferrari})-(\ref{f2}) to the situation where some
covariates are not directly measured or are measured with error. A practical application of  errors-in-variables
beta regression models will be illustrated in a study of the risk of coronary heart disease as a function of low-density
lipoprotein ($LDL$) cholesterol level (``bad cholesterol'') and body mass index ($BMI$). The dataset consists of observations
of systolic blood pressure $(SBP)$, diastolic blood pressure $(DBP)$, $BMI$ and total cholesterol level $(TC)$ in a group of
182 smoking women aged 50 to 87 years. The total cholesterol may be considered as a surrogate of $LDL$, which is a covariate
of interest, and whose direct measure is more expensive and time consuming. The difference between $SBP$ and $DBP$ results
in what is known as the pulse pressure, $PP=SBP-DBP$, and the relative pulse pressure is $RPP=(SBP-DBP)/SBP=PP/SBP$. Small
values of $RPP$, $RPP<25\%$ say, is indicative of risk of heart disease \citep[p. 58]{ATLS}. Notice that the response variable,
$RPP$, is continuous and limited to the unit interval, and that one of the covariates, namely $LDL$, is not measured directly.

This paper is organized as follows. In Section \ref{model}, we present an errors-in-variables beta regression model
under the structural approach, and the corresponding likelihood function. In Section \ref{estimation}, we present three
different estimation methods, namely maximum likelihood, maximum pseudo-likelihood, and regression calibration. In Section
\ref{simula}, we perform a simulation study to evaluate and compare the performance of the three estimation approaches. In
Section \ref{sec:resid}, we propose a residual analysis. Section \ref{apli:corazon} presents an application of the proposed
model. Concluding remarks are presented in Section \ref{conclu}.

\section{Model and likelihood}
\label{model}
Let $y_{1},\ldots,y_{n}$ be independent observable random variables arising from a sample of size $n$, such that $y_{t}$
has a beta distribution with probability density function (\ref{ferrari}) with parameters $\mu=\mu_t$ and $\phi=\phi_t$.
In the following, we assume that $\mu_t$ and $\phi_t$ may depend on covariates and unknown parameters. In practice,
some covariates may not be precisely observed, but, instead, may be obtained with error. The model considered in this
paper assumes a linear structure for the specification of the mean and the precision parameters, and also assumes that
both specifications may involve covariates measured with error. Specifically, we replace the mean submodel (\ref{f1}) and
the precision submodel (\ref{f2}) by
\begin{eqnarray}
\label{g}
g(\mu_{t})&=&\vm{z}^{\top}_{t}\vg{\alpha}+\vm{x}_{t}^{\top}\vg{\beta}, \\
\label{h}
h(\phi_{t})&=&\vm{v}^{\top}_{t}\vg{\gamma}+\vm{m}_{t}^{\top}\vg{\lambda},
\end{eqnarray}
respectively, where $\vg{\beta} \in \mathbb{R}^{p_\beta}$, $\vg{\lambda} \in \mathbb{R}^{p_\lambda}$ are column vectors of
unknown parameters, $\vm{x}_{t}=(x_{t1}, \cdots,x_{t{p_\beta}})^{\top}$ and
$\vm{m}_{t}=(m_{t1},\cdots,m_{t{p_\lambda}})^{\top}$ ($p_\alpha+p_\beta+ p_\gamma+p_\lambda<n$) are unobservable
(latent) covariates, in the sense that they are observed with error. The vectors of covariates measured without error,
$\vm{z}_{t}$ and $\vm{v}_{t}$, may contain variables in common, and likewise, $\vm{x}_{t}$ and $\vm{m}_{t}$. Let
$\vm{s}_{t}$ be the vector containing all the unobservable covariates. For $t=1,\ldots,n$, the random vector $\vm{w}_{t}$
is observed in place of $\vm{s}_{t}$, and it is assumed that
\begin{eqnarray}
\label{modeloEM}
\vm{w}_{t}=\vg{\tau}_{0}+\vg{\tau}_{1}\circ\vm{s}_{t}+\vm{e}_{t},
\end{eqnarray}
where $\vm{e}_{t}$ is a vector of random errors, $\vg{\tau}_{0}$ and $\vg{\tau}_{1}$ are (possibly unknown) parameter vectors
and $\circ$ represents the Hadamard (elementwise) product. The parameter vectors $\vg{\tau}_{0}$ and $\vg{\tau}_{1}$ can be
interpreted as the additive and multiplicative biases of the measurement error mechanism, respectively. If $\vg{\tau}_{0}$ is a
vector of zeros and $\vg{\tau}_{1}$ is a vector of ones, we have the classical additive model $\vm{w}_{t}=\vm{s}_{t}+\vm{e}_{t}$.
Here, we follow the structural approach, in which the unobservable covariates are regarded as random variables, i.e.
we assume that $\vm{s}_{1},\dots, \vm{s}_{n}$ are independent and identically distributed random vectors. In this case, it is also
usual to assume that they are independent of the measurement errors $\vm{e}_{1}, \ldots,\vm{e}_{n}$. Moreover, the normality
assumption for the joint distribution of $\vm{s}_{t}$ and $\vm{e}_{t}$ is assumed.
The parameters of the joint distribution of $\vm{w}_{t}$ and $\vm{s}_{t}$ is denoted by $\vg{\delta}$.

Let $(y_{1},\vm{w}_{1}), \ldots, (y_{n},\vm{w}_{n})$ be the observable variables.
We omit the observable vectors $\vm{z}_{t}$ and $\vm{v}_{t}$ in the notation as they are non-random and known.
The joint density function of $(y_{t},\vm{w}_{t})$, which is the observation for the $t$-th individual, is obtained by integrating
the joint density of the complete data $(y_{t},\vm{w}_{t},\vm{s}_{t})$,
$$
f(y_{t},\vm{w}_{t},\vm{s}_{t}; \vg{\theta}, \vg{\delta})=f(y_t|\vm{w_t},\vm{s}_t;\vg{\theta})f(\vm{s}_t,\vm{w}_t;\vg{\delta}),
$$
with respect to $\vm{s}_{t}$. Here,
$\vg{\theta}=(\vg{\alpha}^{\top},\vg{\beta}^{\top},\vg{\gamma}^{\top},\vg{\lambda}^{\top})^{\top}$
represents the parameter of interest, and $\vg{\delta}$ is the nuisance parameter. The joint density
$f(\vm{w}_t,\vm{s}_t;\vg{\delta})$, which is associated to the measurement error model, can be written as $f(\vm{w}_t,\vm{s}_t;\vg{\delta})=f(\vm{w}_t|\vm{s}_t;\vg{\delta})f(\vm{s}_t|\vg{\delta})$ as well as  $f(\vm{w}_t,\vm{s}_t;\vg{\delta})=f(\vm{s}_t|\vm{w}_t;\vg{\delta})f(\vm{w}_t|\vg{\delta})$.
In this work  we assume  that, given the true  (unobservable) covariates $\vm{s}_t$, the response variable $y_t$ does not depend
on the surrogate covariates $\vm{w}_t$; i.e. $f(y_t|\vm{w}_t,\vm{s}_t;\vg{\theta})=f(y_t|\vm{s}_t;\vg{\theta})$. In other words,
conditionally on $\vm{s}_{t}$, $y_t$ and $\vm{w}_{t}$ are assumed to be independent \citep{Hele+Rei}. Therefore, the density
function of $(y_{t},\vm{w}_{t})$ is given by
\begin{eqnarray*}
\label{f_yw}
f(y_{t},\vm{w}_{t};\vg{\theta}, \vg{\delta})\hspace{-0.3cm}&=&\hspace{-0.3cm}\int_{-\infty}^{\infty} \ldots \int_{-\infty}^{\infty}f(y_{t},\vm{w}_{t},\vm{s}_{t};\vg{\theta}, \vg{\delta})d\vm{s}_{t},\nonumber\\
\hspace{-0.3cm}&=&\hspace{-0.3cm}\int_{-\infty}^{\infty} \ldots \int_{-\infty}^{\infty}f(y_{t}|\vm{s}_{t};\vg{\theta})
f(\vm{w}_{t},\vm{s}_{t};\vg{\delta})
d\vm{s}_{t}.
\end{eqnarray*}

The log-likelihood function for a sample of $n$ observations is given by
\small
\begin{eqnarray}
\label{vero12}
\ell(\vg{\theta}, \vg{\delta})\hspace{-0.3cm}&=&\hspace{-0.3cm}\sum_{t=1}^{n}\log \int_{-\infty}^{\infty} \cdots \int_{-\infty}^{\infty}f(y_{t}|\vm{s}_{t};\vg{\theta})f(\vm{s}_{t}|\vm{w}_{t};\vg{\delta})f(\vm{w}_{t};\vg{\delta})d\vm{s}_{t}, \nonumber \\
             \hspace{-0.3cm}&=&\hspace{-0.3cm}\sum_{t=1}^{n}\log f(\vm{w}_{t};\vg{\delta})+\sum_{i=1}^{n}\log \int_{-\infty}^{\infty}  \cdots   \int_{-\infty}^{\infty}f(y_{t}|\vm{s}_{t};\vg{\theta})f(\vm{s}_{t}|\vm{w}_{t};\vg{\delta})d\vm{s}_{t}.
\end{eqnarray}
\normalsize
In general, the likelihood function involves analytically intractable integrals and, hence, approximate inference methods
need to be considered. In the next section, we present three different approaches to estimate the parameters.

In order to facilitate the description of the estimation methods, we assume that a single covariate, $x_{t}$, is measured
with error, and that it is used for the specifications of both the mean and precision submodels. We then have
$p_\beta=p_\lambda=1$ and $\vm{s}_t=x_{t}=m_{t}$. We also assume independence and normality of random errors. The methodologies
presented in this paper can be extended to the situation where $\vm{x}_{t}$ and $\vm{m}_{t}$ are distinct, or when covariates
measured with error appear only in the specification of the mean or the precision parameter. 

To be specific, from now on, the model under consideration is summarized as follows. For $i=1,\ldots,n$,
\begin{eqnarray}
\label{ydadoxw}
&y_t | x_t, w_t \sim {\rm Beta}(\mu_t,\phi_t),\\
\label{links}
&g(\mu_{t})=\vm{z}^{\top}_{t}\vg{\alpha}+x_{t}\beta, \ \ \ h(\phi_{t})=\vm{v}^{\top}_{t}\vg{\gamma}+x_{t}\lambda,\\
\label{covariate}
&w_{t} = \tau_0+\tau_1 x_{t} + e_{t}, \ \
x_{t} \stackrel{{\rm ind}} \sim N(\mu_{x},\sigma_{x}^{2}),\ \
e_{t} \stackrel{{\rm ind}}\sim N(0,\sigma_{e}^{2}),
\end{eqnarray}
with $x_{t}$ and $e_{t'}$,  for $t,t'=1,\ldots,n$, being independent. The unknown parameter vectors $\vg{\alpha}$ and
$\vg{\gamma}$ were defined above, and $\beta\in\mathbb{R}$, $\lambda\in\mathbb{R}$, $\mu_x\in\mathbb{R}$ and $\sigma^2_x>0$
are unknown parameters. Note that it is assumed that the conditional distribution of $y_t$ given $(x_t, w_t)$ does not depend on
$w_t$. Also, if $\tau_{0}=0$ and $\tau_{1}=1$, (\ref{covariate}) corresponds to the classical additive error model
$w_{t} = x_{t} + e_{t}$. From (\ref{covariate}) we have
\begin{eqnarray}
\label{distrib}
w_t \stackrel{{\rm ind}} \sim {\rm N}(\tau_{0}+\tau_{1}\mu_x,\tau_{1}^{2}\sigma^2_x+\sigma^2_e), \ \
x_t | w_t \stackrel{{\rm ind}} \sim {\rm N}(\mu_{x_{t}|w_{t}},\sigma_{x_{t}|w_{t}}^{2}),\end{eqnarray}
where
\begin{eqnarray}
\label{mu-sigma}
\mu_{x_{t}|w_{t}}=\mu_{x}+k_x[w_{t}-(\tau_{0}+\tau_{1}\mu_{x})], \ \ \sigma_{x_{t}|w_{t}}^{2}=\sigma_{e}^{2}k_x/\tau_{1},
\end{eqnarray}
with $k_x=\tau_{1}\sigma_{x}^{2}/(\tau_{1}^{2}\sigma_{x}^{2}+\sigma_{e}^{2})$ being known as
the reliability ratio. To avoid non-identifiability of parameters we assume that $(\tau_0, \tau_1, \sigma^2_e)$ or
$(\tau_0, \tau_1, k_x)$ is either a known parameter vector or is estimated from supplementary information,
typically replicate measurements or partial observation of the error-free covariate. In any case, either of these
vectors is regarded as a known quantity in the inferential procedure. Hence, the nuisance
parameter vector is $\vg{\delta}=(\mu_x,\sigma^2_x)^\top$.

The log-likelihood function in (\ref{vero12}) for $n$ observations taken from the model described in (\ref{ydadoxw}),
(\ref{links}) and (\ref{covariate}) is given by
\begin{eqnarray}
\label{vero13}
\ell(\vg{\theta},\vg{\delta})=\sum_{t=1}^{n}\ell_{1t}(\vg{\delta})+\sum_{t=1}^{n}\ell_{2t}(\vg{\theta},\vg{\delta}),
\end{eqnarray}
where
\begin{eqnarray}
\label{loglik}
\ell_{1t}(\vg{\delta})
\hspace{-0.3cm}&=&\hspace{-0.3cm}-\frac{1}{2}\log[2\pi(\tau_{1}^{2}\sigma_{x}^{2}+\sigma_{e}^{2})]
-\frac{[w_{t}-(\tau_{0}+\tau_{1}\mu_{x})]^{2}}{2(\tau_{1}^2\sigma_{x}^{2}+\sigma_{e}^{2})},\\
\label{loglikt2}
\ell_{2t}(\vg{\theta},\vg{\delta})
\hspace{-0.3cm}&=&\hspace{-0.3cm} \log \int_{-\infty}^{\infty}f(y_{t}|x_{t};\vg{\theta},\vg{\delta})\frac{1}{\sqrt{2\pi\sigma_{x_{t}|w_{t}}^{2}}}
\exp\left[-\frac{(x_{t}-\mu_{x_{t}|w_{t}})^2}{2\sigma_{x_{t}|w_{t}}^{2}}\right]dx_{t}.
\end{eqnarray}

\section{Estimation}
\label{estimation}

\subsection{Maximum likelihood estimation}
\label{vero}

The second term of the log-likelihood function $\ell(\vg{\theta},\vg{\delta})$ in (\ref{vero13}), which depends on a
non-analytical integral (as can be seen in (\ref{loglikt2})), can be
approximated using the Gauss-Hermite quadrature, which consists of the approximation
\begin{eqnarray}
\label{QGH}
\int_{-\infty}^{\infty}e^{-x^{2}}f(x)dx \approx \sum_{q=1}^{Q}\nu_{q}f(\eta_{q}),
\end{eqnarray}
where $\eta_{q}$ and $\nu_{q}$ represent the  $q$-th zero and weight, respectively, of the orthogonal Hermite polynomial
of order $Q$ (number of quadrature points); see, for instance, \citet[Chapter 22]{Abramowitz+Stegun}.
Using the transformation $u_{t}=(x_{t}-\mu_{x_{t}|w_{t}})/\sqrt{2 \sigma_{x_{t}|w_{t}}^{2}}$ in (\ref{loglikt2}), we have that
$x_{t}=\mu_{x_{t}|w_{t}}+\sqrt{2 \sigma_{x_{t}|w_{t}}^{2}}u_{t}$ and $dx_{t}=\sqrt{2\sigma_{x_{t}|w_{t}}^{2}}du_{t}$.
Hence, by applying (\ref{QGH}) in (\ref{loglikt2}), the log-likelihood function (\ref{vero13}) can be approximated by
\begin{eqnarray}
\label{aproxvero}
\ell_{a}(\vg{\theta},\vg{\delta})=\sum_{t=1}^{n}\ell_{1t}(\vg{\delta})+\sum_{t=1}^{n}\log \left(\sum_{q=1}^{Q}\frac{\nu_{q}}{\sqrt{\pi}}\exp[l(\mu_{tq},\phi_{tq})]\right),
\end{eqnarray}
where
\begin{eqnarray}
\label{loglik-beta}
l(\mu,\phi)&=&\log \Gamma(\phi)-\log \Gamma(\mu \phi)- \log \Gamma[(1-\mu)\phi]\\
&+&(\mu\phi-1)\log(y_{t})
+[(1-\mu)\phi-1]\log(1-y_{t}),\nonumber\\
g(\mu_{tq})&=&\vm{z}^{\top}_{t}\vg{\alpha}+x_{t}^{*}\beta, \ \
h(\phi_{tq})=\vm{v}^{\top}_{t}\vg{\gamma}+x_{t}^{*}\lambda, \ \
x_{t}^{*}=\mu_{x_{t}|w_{t}}+\sqrt{2\sigma_{x_{t}|w_{t}}^{2}}\eta_{q},\nonumber
\end{eqnarray}
where $\mu_{x_{t}|w_{t}}$ and $\sigma^{2}_{x_{t}|w_{t}}$ are given in (\ref{mu-sigma}).

The approximate maximum likelihood estimator of $(\vg{\theta}^\top,\vg{\delta}^\top)^\top$, $(\widehat{\vg{\theta}}^\top,\widehat{\vg{\delta}}^\top)^\top$ say, is obtained by solving the system of equations $\partial{\ell_{a}(\vg{\theta},\vg{\delta})}/\partial\vg{\theta}=0,$ $\partial{\ell_{a}(\vg{\theta},\vg{\delta})}/\partial\vg{\delta}=0$.
For computational implementation, the derivatives of $\ell_{a}(\vg{\theta},\vg{\delta})$ with respect to the
parameters can be analytically obtained or numerical derivatives can be used. Our numerical results were obtained using numerical derivatives.

\subsection{Maximum pseudo-likelihood estimation}
\label{pVero}

The central idea of the maximum pseudo-likelihood estimation method is to replace the nuisance parameters by consistent
estimates in the log-likelihood function (\ref{vero13}). The resulting function can be regarded as a pseudo-log-likelihood
function that depends on the parameters of interest only (\citet{Guolo}, \citet{Gong+Samaniego}, \citet{Skrondal+Kuha}) .

The log-likelihood function (\ref{vero13}) is maximized in two steps. First, we estimate the nuisance parameter vector
$\vg{\delta}$ by maximizing the reduced log-likelihood function
\begin{eqnarray}
\label{lvero_red}
\ell_{r}(\vg{\delta})=\sum_{t=1}^{n}\ell_{1t}(\vg{\delta}),
\end{eqnarray}
where $\ell_{1t}(\vg{\delta})$ is given in (\ref{loglik}).
Second, the estimate $\widehat{\vg{\delta}}$ obtained from the maximization of (\ref{lvero_red}) is inserted in the original
log-likelihood function (\ref{vero13}), which results in the pseudo-log-likelihood function
\begin{eqnarray}
\label{vero2}
\ell_{p}(\vg{\theta};\widehat{\vg{\delta}})=\sum_{t=1}^{n}\ell_{1t}(\widehat{\vg{\delta}})+
\sum_{t=1}^{n}\ell_{2t}(\vg{\theta},\widehat{\vg{\delta}}).
\end{eqnarray}

As in $\ell(\vg{\theta},\vg{\delta})$, the second term in $\ell_{p}(\vg{\theta};\widehat{\vg{\delta}})$ cannot be expressed in closed form
and requires numerical integration. However, unlike
the  integral in $\ell(\vg{\theta},\vg{\delta})$,  the integral in $\ell_{p}(\vg{\theta};\widehat{\vg{\delta}})$ depends on the
parameter of interest only. From (\ref{QGH}), it is possible to approximate $\ell_{2t}(\vg{\theta},\widehat{\vg{\delta}})$ by a
summation. After some algebra, we have that an approximate pseudo-log-likelihood for the beta regression model with one covariate
measured with error  is given by (\ref{vero2}) with $\ell_{2t}(\vg{\theta},\widehat{\vg{\delta}})$ replaced by
\begin{eqnarray*}
\label{lp2}
\ell_{2t}^\dag(\vg{\theta},\widehat{\vg{\delta}})&=& \log \left(\sum_{q=1}^{Q}\frac{\nu_{q}}{\sqrt{\pi}}\exp\left[l(\widehat\mu_{tq},\widehat\phi_{tq})\right]\right),
\end{eqnarray*}
where $l(\mu,\phi)$ is given in (\ref{loglik-beta}),
\begin{eqnarray*}
g(\widehat\mu_{tq})\hspace{-0.3cm}&=&\hspace{-0.3cm}\vm{z}^{\top}_{t}\vg{\alpha}+\widehat{x}_{t}\beta, \ \
h(\widehat\phi_{tq})= \vm{v}^{\top}_{t}\vg{\gamma}+\widehat{x}_{t}\lambda, \ \
\widehat{x}_{t}=\widehat{\mu}_{x_{t}|w_{t}}+\sqrt{2\widehat{\sigma}_{x_{t}|w_{t}}^{2}}\eta_{q},
\end{eqnarray*}
with $\widehat{\mu}_{x_{t}|w_{t}}$ and $\widehat{\sigma}^{2}_{x_{t}|w_{t}}$ being estimates of
${\mu}_{x_{t}|w_{t}}$ and ${\sigma}^{2}_{x_{t}|w_{t}}$, respectively, obtained from the maximization of (\ref{lvero_red}).

The approximate pseudo-likelihood estimator of $\vg{\theta}$ is obtained by maximizing the approximate pseudo-log-likelihood function given above. 
Such an estimator has been proposed in a recent paper by \citet{Skrondal+Kuha} in the context of generalized linear models, and named improved 
regression calibration estimator. 

It can be shown that, under regularity conditions (\citet{Gong+Samaniego} and \citet{Parke}), the approximate pseudo-log-likelihood estimator is 
consistent and the asymptotic distribution of
$\sqrt{n}(\widehat{\vg{\theta}}-\vg{\theta})$ is normal with mean zero and covariance matrix
\begin{eqnarray}\label{Sigma}
\vm{\Sigma}=\textrm{I}_{\vg{\theta} \vg{\theta}}^{-1} + \textrm{I}_{\vg{\theta} \vg{\theta}}^{-1}\textrm{I}_{\vg{\theta} \vg{\delta}}\textrm{I}_{\vg{\delta} \vg{\delta}}^{-1}\vm{\Sigma}_{\vg{\delta} \vg{\delta}}\textrm{I}_{\vg{\delta} \vg{\delta}}^{-1}\textrm{I}_{\vg{\theta} \vg{\delta}}^{\top}  \textrm{I}_{\vg{\theta} \vg{\theta}}^{-1},
\end{eqnarray}
where
\begin{eqnarray*}
\textrm{I}_{\vg{\theta} \vg{\theta}}&=&-\sum_{t=1}^{n}\textrm{E}\left(\frac{\partial^{2}\ell_{pt}(\vg{\theta},\vg{\delta})}
{\partial \vg{\theta} \partial \vg{\theta}^{\top}}\right), \ \
\textrm{I}_{\vg{\theta} \vg{\delta}}=
-\sum_{t=1}^{n}\textrm{E}\left(\frac{\partial^{2}\ell_{pt}(\vg{\theta},\vg{\delta})}
{\partial \vg{\theta} \partial \vg{\delta}^{\top}}\right),\\
\vm{\Sigma}_{\vg{\delta} \vg{\delta}}&=&
\sum_{t=1}^{n}\frac{\partial \ell_{rt}(\vg{\delta})}{\partial \vg{\delta}}
              \left(\frac{\partial \ell_{rt}(\vg{\delta})}{\partial \vg{\delta}}\right)^{\top}, \ \
\textrm{I}_{\vg{\delta} \vg{\delta}}=-\sum_{t=1}^{n}\textrm{E}\left(\frac{\partial^{2}\ell_{pt}(\vg{\theta},\vg{\delta})}
    {\partial \vg{\delta} \partial \vg{\delta}^{\top}}\right),							
\end{eqnarray*}
with $\vg{\delta}$ replaced by $\widehat{\vg{\delta}}$, and with $\ell_{rt}(\vg{\delta})$
and $\ell_{pt}(\vg{\theta},\vg{\delta})$ being the $t$-th element of the
log-likelihood functions $\ell_{r}(\vg{\delta})$ and $\ell_{p}(\vg{\theta},\vg{\delta})$ in (\ref{lvero_red})
and (\ref{vero2}), respectively. Details on the conditions and proof for the consistency and asymptotic normality 
of pseudo-likelihood estimators can be found in \citet[Section 5]{Skrondal+Kuha} and \citet[Sections 24.2.2 and 24.2.4]{Gourieroux+Monfort}.
%\citet[Section A.6.6]{Ca+Ru+Ste}.
For the errors-in-variables beta regression model considered here,  $\textrm{I}_{\vg{\theta} \vg{\theta}}$,
$\textrm{I}_{\vg{\delta} \vg{\delta}}$ and $\textrm{I}_{\vg{\theta} \vg{\delta}}$ do not have closed form.
We suggest to replace the expected information matrix by the observed information matrix.
For computational implementation, the needed derivatives can be analytically or numerically obtained. We
used numerical derivatives in our simulations and applications.

\subsection{Regression calibration estimation}
\label{RC}

The regression calibration method has been widely used in errors-in-variables modeling due to its simplicity; see
\citet[Chap. 4]{Ca+Ru+Ste}, \citet{Laurence+Douglas+Carroll+Kipnis},
\citet{Thurston+Williams+Hauser+Hu+Hernandez-Avila+Spiegelman} and \citet{Guolo}.
The central idea is to replace the unobservable variable, $x_{t}$, by an estimate of the conditional
expected value of $x_{t}$ given $w_{t}$, $\textnormal{E}(x_{t}|w_{t})$, in the likelihood function.
Let $r(w_{t},\vg{\delta})=\textnormal{E}(x_{t}|w_{t})$ be the calibration function.
%, which depends of a paramater vector $\vg{\delta}$.
The replacement of the unobservable covariate $x_{t}$ by $r(w_{t},\widehat{\vg{\delta}})$
establishes a modified model for the data. Here, $\widehat{\vg{\delta}}$ is an estimate of $\vg{\delta}$.

For our errors-in-variables beta regression model, the calibration function is
$r(w_{t},\vg{\delta})=\mu_{x_{t}|w_{t}}$ as defined in (\ref{mu-sigma}) . From (\ref{distrib}) and (\ref{mu-sigma}), we have that
$\overline w=\sum_{t=1}^n w_{t}/n$ and $s^2_w=\sum_{t=1}^n (w_{t}-\overline w)^2/(n-1)$ are optimal estimates of
$\tau_0+\tau_1\mu_x$ and $\tau_1^2\sigma^2_x+\sigma^2_{e}$, respectively. These estimates can be used to estimate
the calibration function.

By inserting the calibration function in the conditional density function of $y_{t}$ given $x_{t}$, we obtain the
modified log-likelihood function
\begin{eqnarray}
\label{lrc}
\ell_{rc}(\vg{\theta})=\sum_{t=1}^{n} l(\widetilde\mu_{t},\widetilde\phi_{t}),
\end{eqnarray}
where $l(\mu,\phi)$ is given in (\ref{loglik-beta}),
\begin{eqnarray*}
g(\widetilde\mu_{t})\hspace{-0.3cm}&=&\hspace{-0.3cm}\vm{z}_{t}^{\top}\vg{\alpha}+\widetilde x_{t}\beta, \ \
h(\widetilde\phi_{t})=\vm{v}_{t}^{\top}\vg{\gamma}+\widetilde x_{t}\lambda,
%\widetilde x_{t}=\overline w+\widehat{k}_{x}(w_{t}-\overline w),
\end{eqnarray*}
with $\widetilde x_{t}$ being the estimated calibration function. Note that the modified log-likelihood function in
(\ref{lrc}) only depends on the parameter of interest, $\vg{\theta}$. The regression calibration estimate of $\vg{\theta}$
is obtained from the system of equations $\partial{\ell_{rc}(\vg{\theta})}/\partial\vg{\theta}=0$, which requires a numerical
algorithm. Since  $\ell_{rc}(\vg{\theta})$ coincides with the log-likelihood function for the usual beta regression model,
$\widetilde x_{t}$ acting as an observable covariate, these equations can be numerically solved from available computational
packages, for instance the  $\tt betareg$ package \citep{Cri+Zei} implemented in the $\tt R$ platform. 
Standard errors for regression calibration estimates can be obtained through bootstrap resampling.

It is well known that regression calibration estimators are, in general, inconsistent. \citet{Skrondal+Kuha} point out that
``the inconsistency is typically small when the true effects of the covariates measured with error are moderate and/or the 
measurement error variance are small, but more pronounced when these conditions do not hold.'' Numerical properties of the 
three estimators described above are investigated in the next section.

\section{Monte Carlo simulation results}
\label{simula}

We now present Monte Carlo simulation results on the performance of the different estimation methods described in
Section \ref{estimation}. All simulation results are based on 5,000 Monte Carlo replications.
We consider errors-in-variables beta regression models with $\log(\mu_{t}/(1-\mu_{t}))=\alpha+\beta x_{t}$,
and $\log(\phi_{t})=\gamma$ (constant precision model) and $\log(\phi_{t})=\gamma+\lambda x_{t}$ (varying precision model),
with $w_{t}$ and $x_{t}$ being generated from (\ref{covariate}). We set $\alpha=$2.0, $\beta=-0.6$, $\lambda=0.5$, $\mu_{x}=2.5$,
$\sigma_{x}^{2}=2.7$, and $\gamma=2.5$ for the constant precision model and $\gamma=4$ for the varying precision model. The
parameters of the measurement error mechanism are assumed to be known, and we set $\tau_{0}=0$, $\tau_{1}=1$, and the following
values for the reliability ratio: $k_x=0.95$ (low measurement error), $k_x=0.75$ (moderate measurement error), and $k_x=0.50$ 
(high measurement error), which correspond to $\sigma^2_e =   \sigma^2_x/10$, $\sigma^2_e = \sigma^2_x/3$, and $\sigma^2_e = \sigma^2_x$,
respectively. The sample sizes are $n=25,$ $50,$ $100$, $200$, and $300$. For each simulated sample the parameters
were estimated under two different settings. First, we ignored the
measurement error in $x_{t}$, i.e. we used what is called the  na\"{i}ve method $~(\ell_{naive})$; second, we recognized that $x_{t}$
is measured with error and estimated the parameters using the three methods proposed in this paper: approximate maximum likelihood
$(\ell_{a})$, (approximate) maximum pseudo-likelihood $(\ell_{p})$, and regression calibration $(\ell_{rc})$. Whenever Gaussian
quadrature was required, we set the number of quadrature points at $Q=50$. The maximization of the relevant
(approximate/pseudo/modified) likelihoods was performed using the quasi-Newton BFGS nonlinear optimization algorithm with numerical
derivatives
%developed by Broyden, Fletcher, Goldfarf \& Shanno (see, for instance, \cite{PRESS}) and
implemented in the function MAXBFGS in the matrix language programming \verb"Ox" \citep{Doornik}. 
The detailed simulation results (not shown here to save space) are collected in the Supplementary Material. 

Figures \ref{Bias095model1}-\ref{Bias050model2} present plots of the bias and the root-mean-square error
of the estimators against sample size, for $k_x=0.95, 0.75,$ and $0.50$ under the constant precision model and the varying
precision model. As expected, the na\"{i}ve estimator is biased and its bias and mean-square error do not converge to zero
as $n$ grows even when the reliability ratio is large (i.e. the variance of the measurement error is small compared to
the variance of $x$). In other words, the plots suggest that the na\"{i}ve estimator is not consistent. For all the cases,
the approximate maximum likelihood and maximum pseudo-likelihood estimators perform similarly. In general, their performance
is clearly better than that of the regression calibration and na\"{i}ve estimators. Under constant precision 
(Figures \ref{Bias095model1}-\ref{Bias050model1}), the regression calibration estimator is as biased as the na\"{i}ve estimator for
estimating the precision parameter. However, for estimating the coefficients associated to the mean submodel, the regression calibration
estimator performs much better than the na\"{i}ve estimator in most of the cases.
%{\color{red}However, for estimating $\beta$, the coefficient associated to the covariate measured
%with error, it performs well if the measurement error variance is small. In fact, the regression calibration, approximate
%maximum likelihood ratio and maximum pseudo-likelihood ratio estimators are virtually unbiased for estimating $\beta$ when
%$k_x=0.95$; also, their mean-square errors converge to zero as $n$ grows.} 
Under the varying precision model (Figures \ref{Bias095model2}-\ref{Bias050model2}), similar conclusions are reached.

We now turn to the evaluation of confidence intervals constructed from the different estimators. 
The standard errors for the approximate maximum likelihood and maximum pseudo-likelihood estimators were calculated from 
%(\ref{Ja}) 
the Hessian matrix of the approximate log-likelihood function ({\ref{aproxvero}}) and from
(\ref{Sigma}), respectively. For the regression calibration estimator, standard errors were obtained through parametric bootstrap 
resampling.
%coupled with their respective asymptotic distributions. 
Figures \ref{Coveragemodel1}-\ref{Coveragemodel2} present plots of the estimated true coverages
of confidence intervals constructed with $95\%$ nominal confidence level, for $k_x=0.95, 0.75,$ and $0.50$ under the constant precision
and varying precision models for $n$ ranging from 25 to 300. For all the cases, the estimated true coverages of the confidence
intervals based on the na\"{i}ve estimator decrease as $n$ grows, and hence they cannot be recommended.
When the measurement error is not very large (eg. $k_x=0.95$ and $k_x=0.75$), the confidence intervals constructed from the approximate maximum likelihood 
and maximum pseudo-likelihood estimators present true coverage close to $95\%$, except for estimating the precision parameter with small samples.
%, more so if the sample size is large. 
For the constant precision model (Figure \ref{Coveragemodel1}), the regression calibration estimator produces reliable confidence intervals for parameters
of the mean submodel when the measurement error variance is small (eg. $k_x=0.95$). However, for estimating the precision parameter,
the regression calibration estimator produces confidence intervals with true coverage much smaller than $95\%$ when $n$ is large.
Under the varying precision model (Figure \ref{Coveragemodel2}), we arrive at similar conclusions, but it is noteworthy that the 
maximum pseudo-likelihood method yields confidence intervals with higher coverage than the approximate maximum likelihood estimation when the sample is not
large.

Overall, we conclude that ignoring the measurement error produces misleading inference. Also, if the measurement error variance
is small, the regression calibration approach is reliable for estimating the parameters of the mean submodel.
Moreover, inference based on the approximate likelihood and the pseudo-likelihood methods present good performance
for the estimation of all the parameters. Since the pseudo-likelihood approach is computationally less demanding than
the approximate maximum likelihood approach, we recommend the maximum pseudo-likelihood estimation for practical applications.

\begin{figure}
\centering
\includegraphics[scale=0.3]{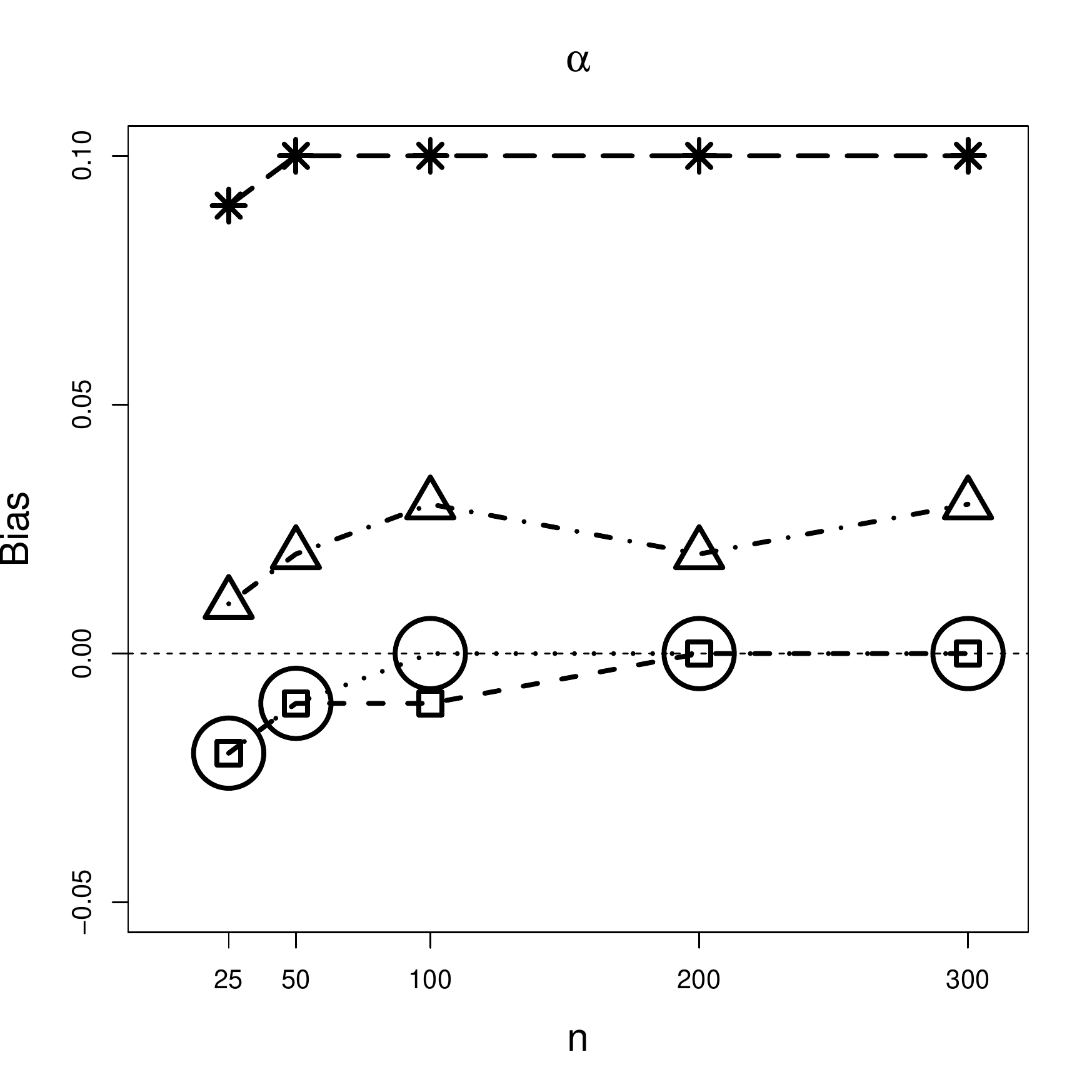} \qquad \includegraphics[scale=0.3]{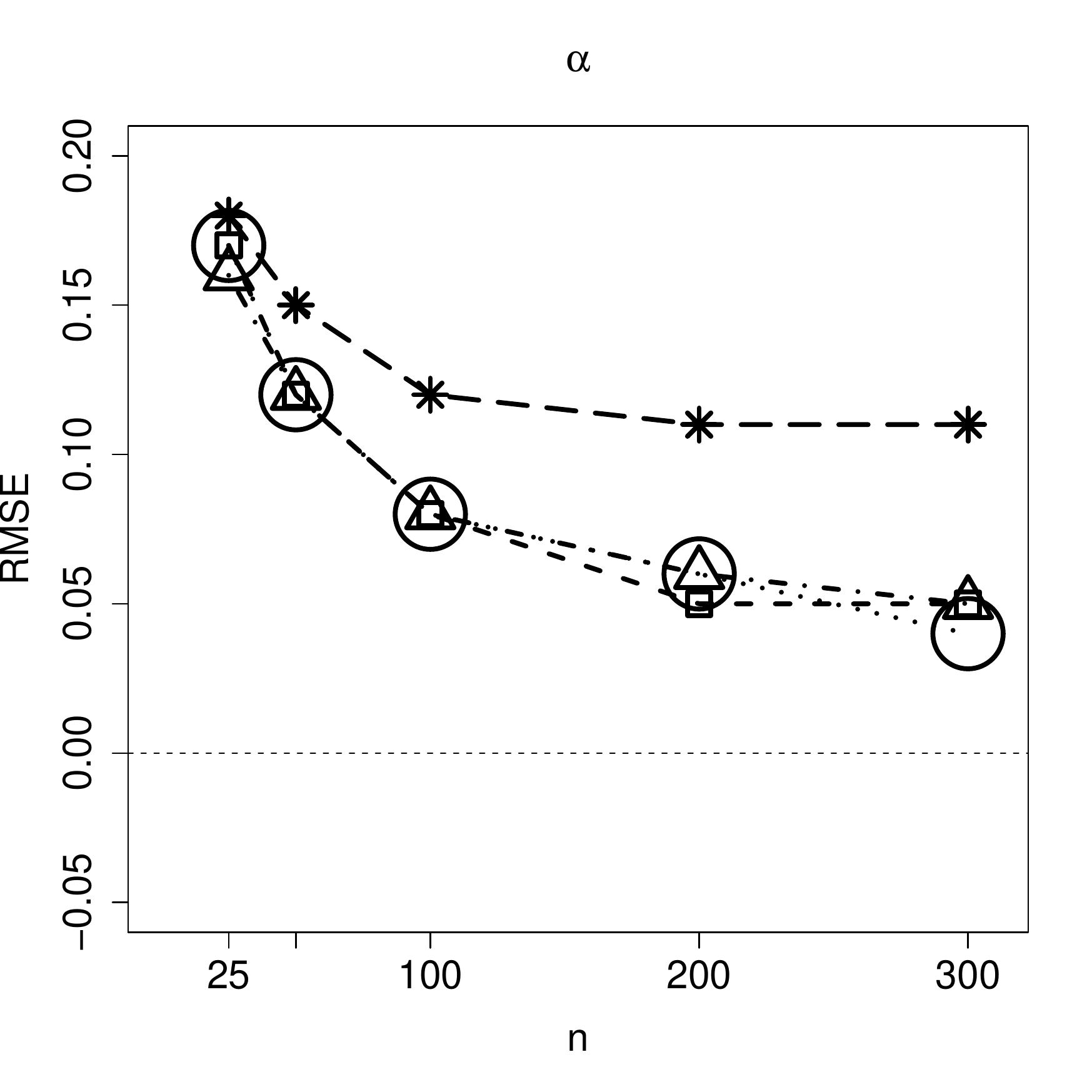}\\
\includegraphics[scale=0.3]{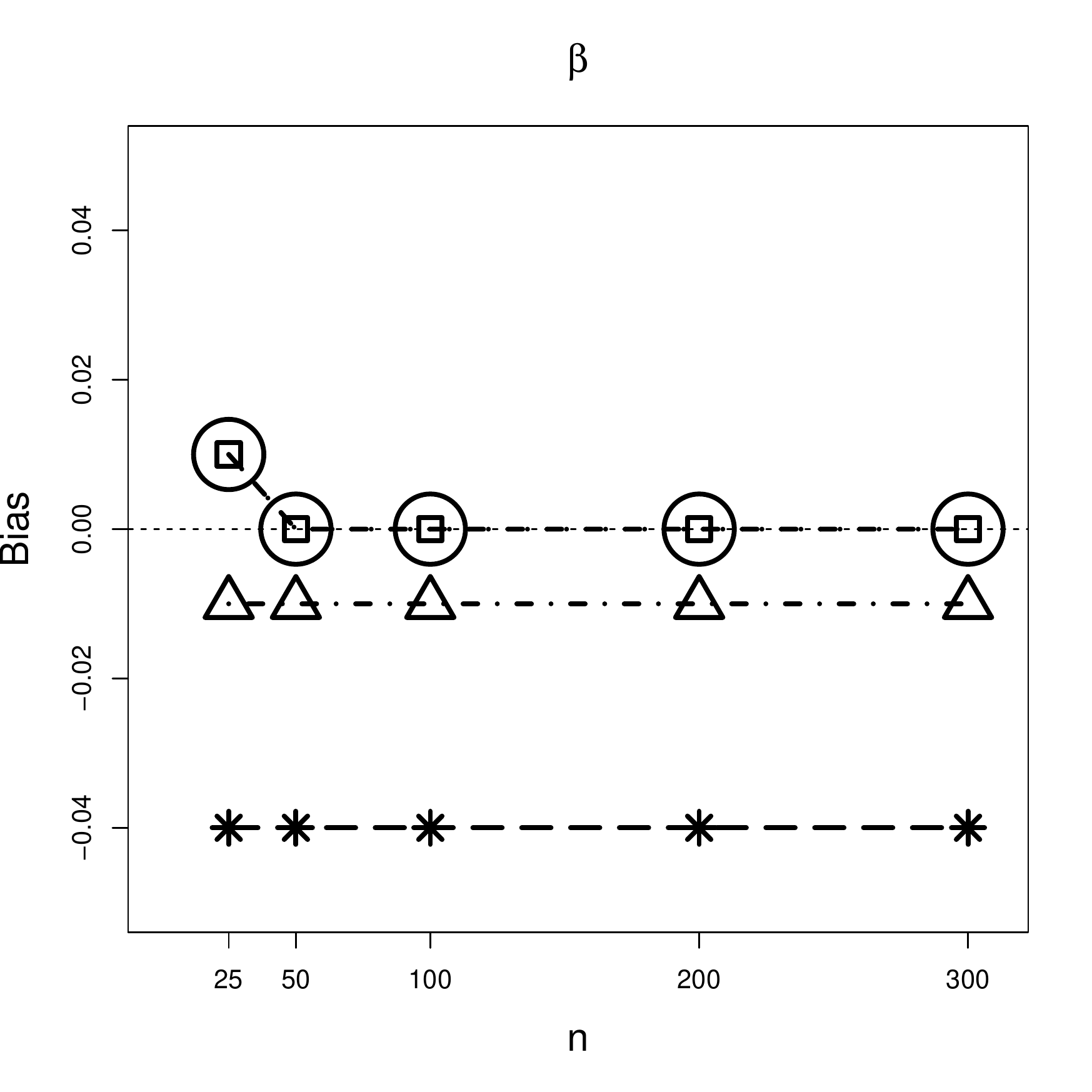}  \qquad \includegraphics[scale=0.3]{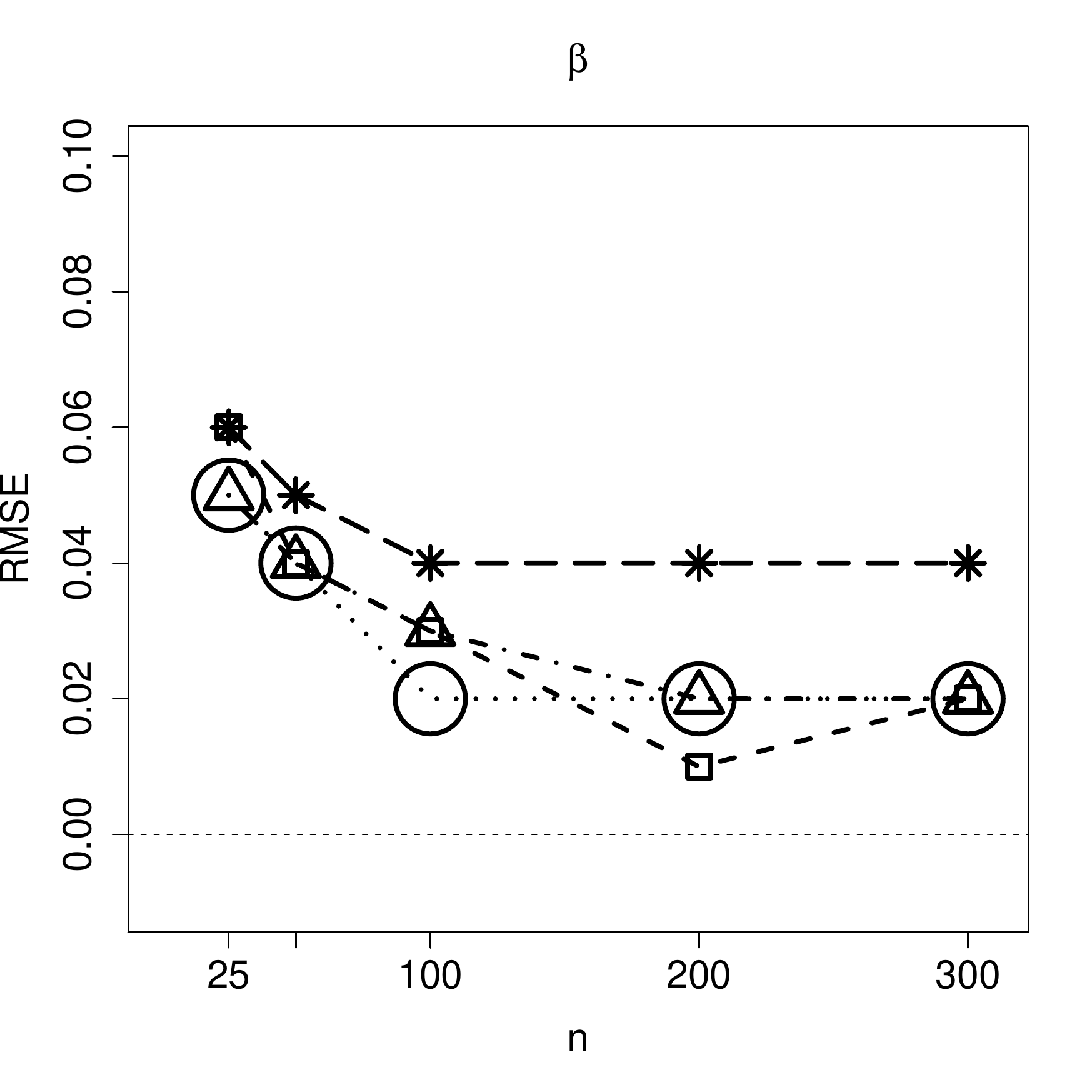}\\
\includegraphics[scale=0.3]{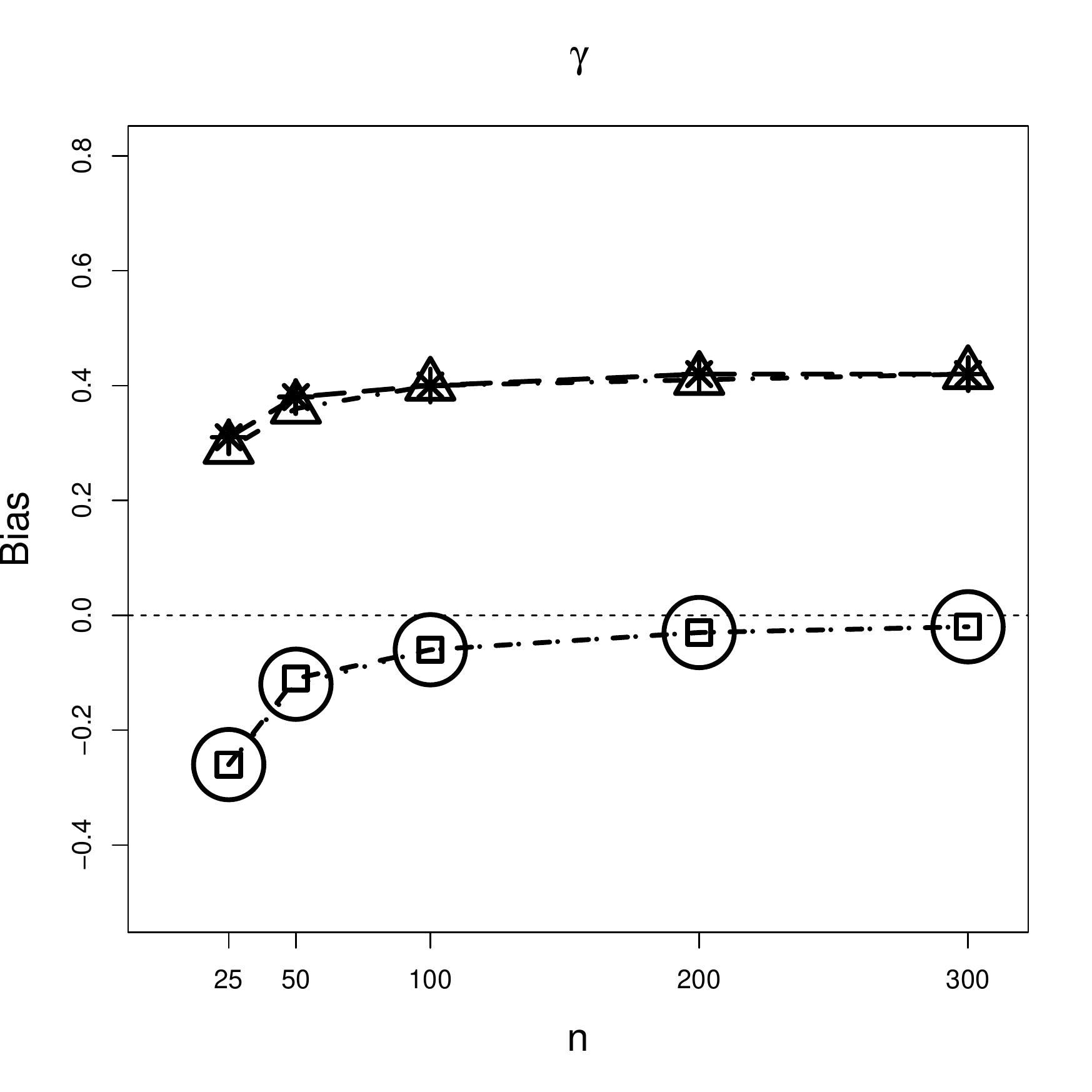}  \qquad \includegraphics[scale=0.3]{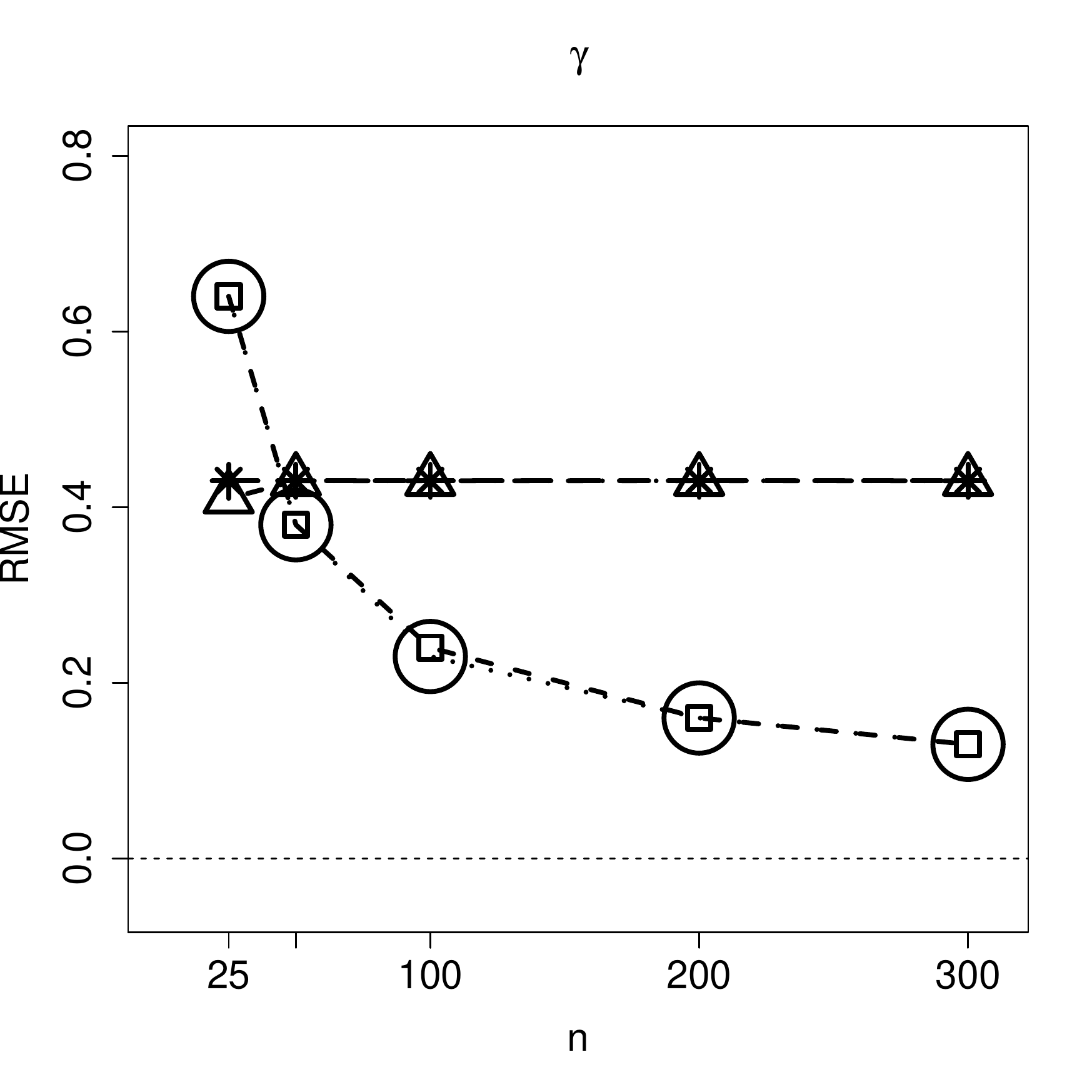}\\
\caption{Bias and RMSE for the estimators of $\alpha$, $\beta$ and $\gamma$ for $k_{x}=0.95$, constant precision model; $\ell_{a}$ (square), $\ell_{p}$ (circle), $\ell_{rc}$ (triangle) and $\ell_{naive}$(star).}
\label{Bias095model1}
\end{figure}

\begin{figure}
\centering
\includegraphics[scale=0.3]{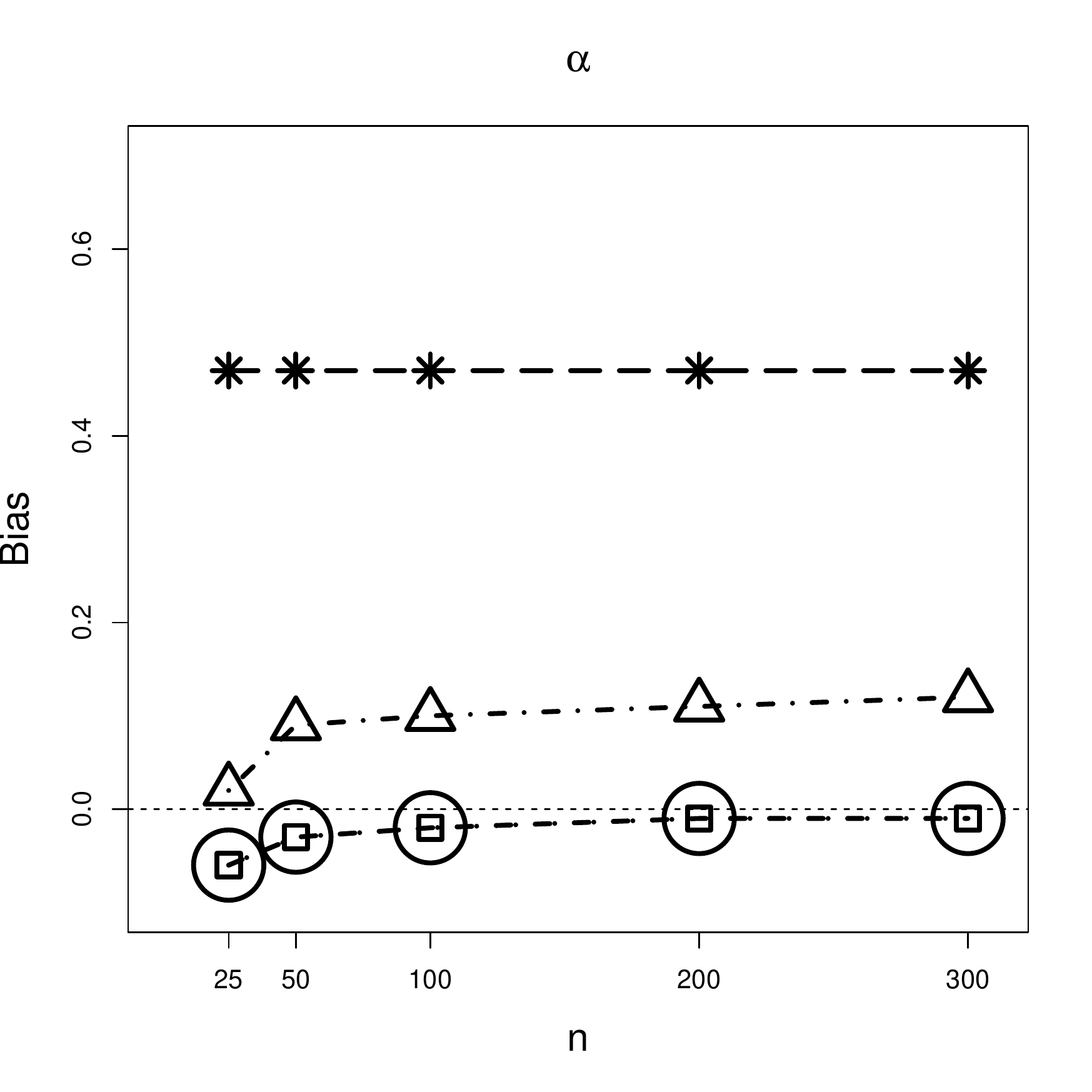} \qquad \includegraphics[scale=0.3]{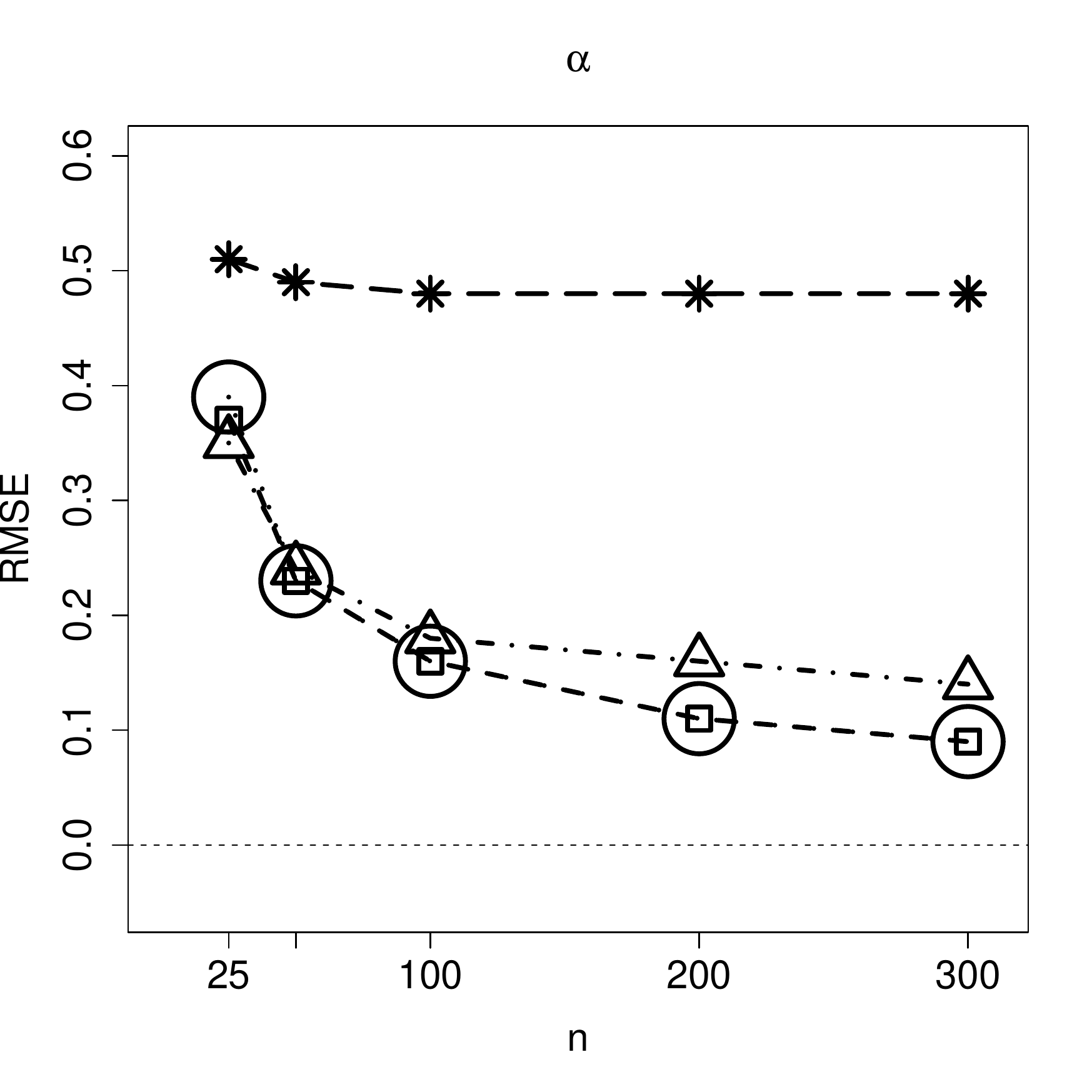}\\
\includegraphics[scale=0.3]{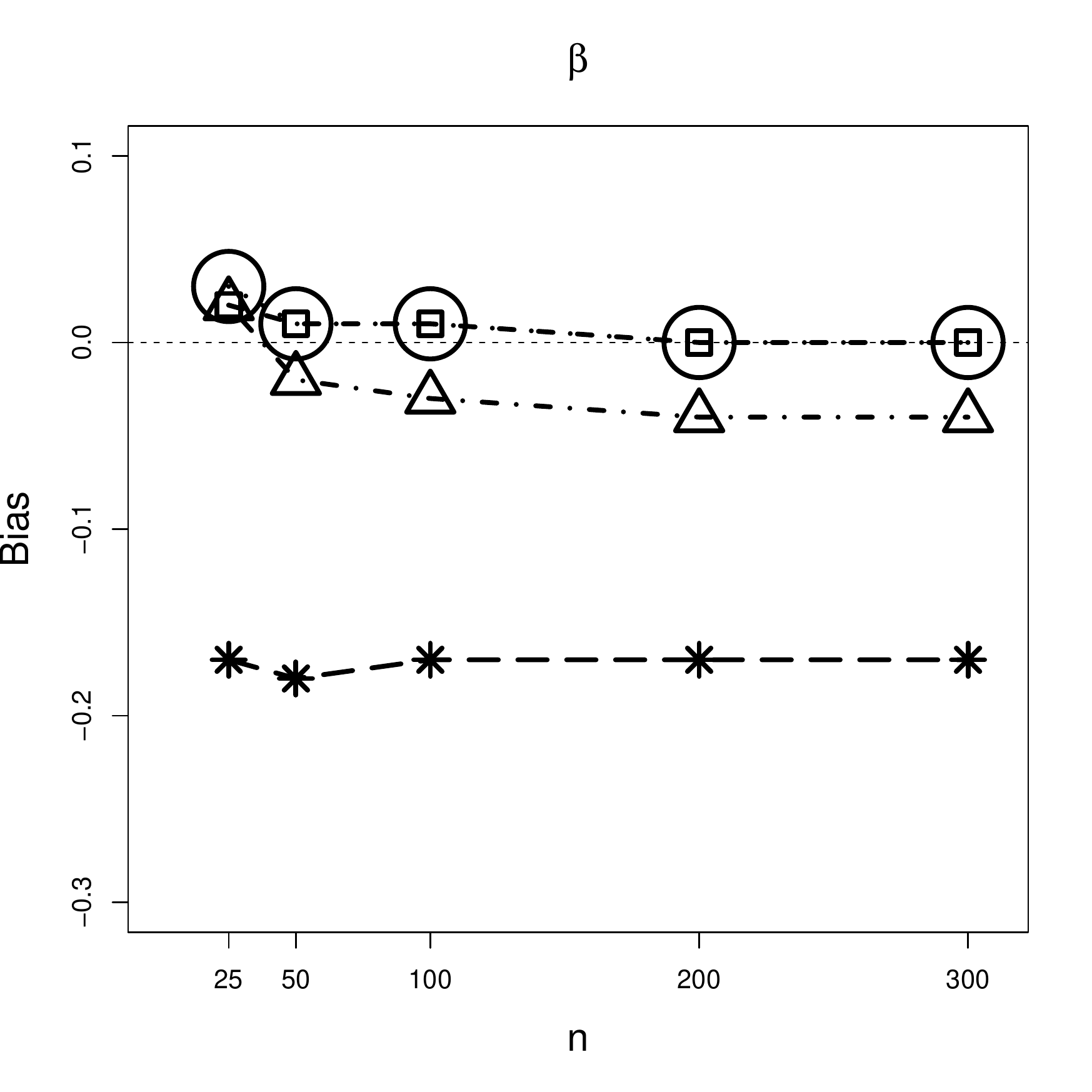}  \qquad \includegraphics[scale=0.3]{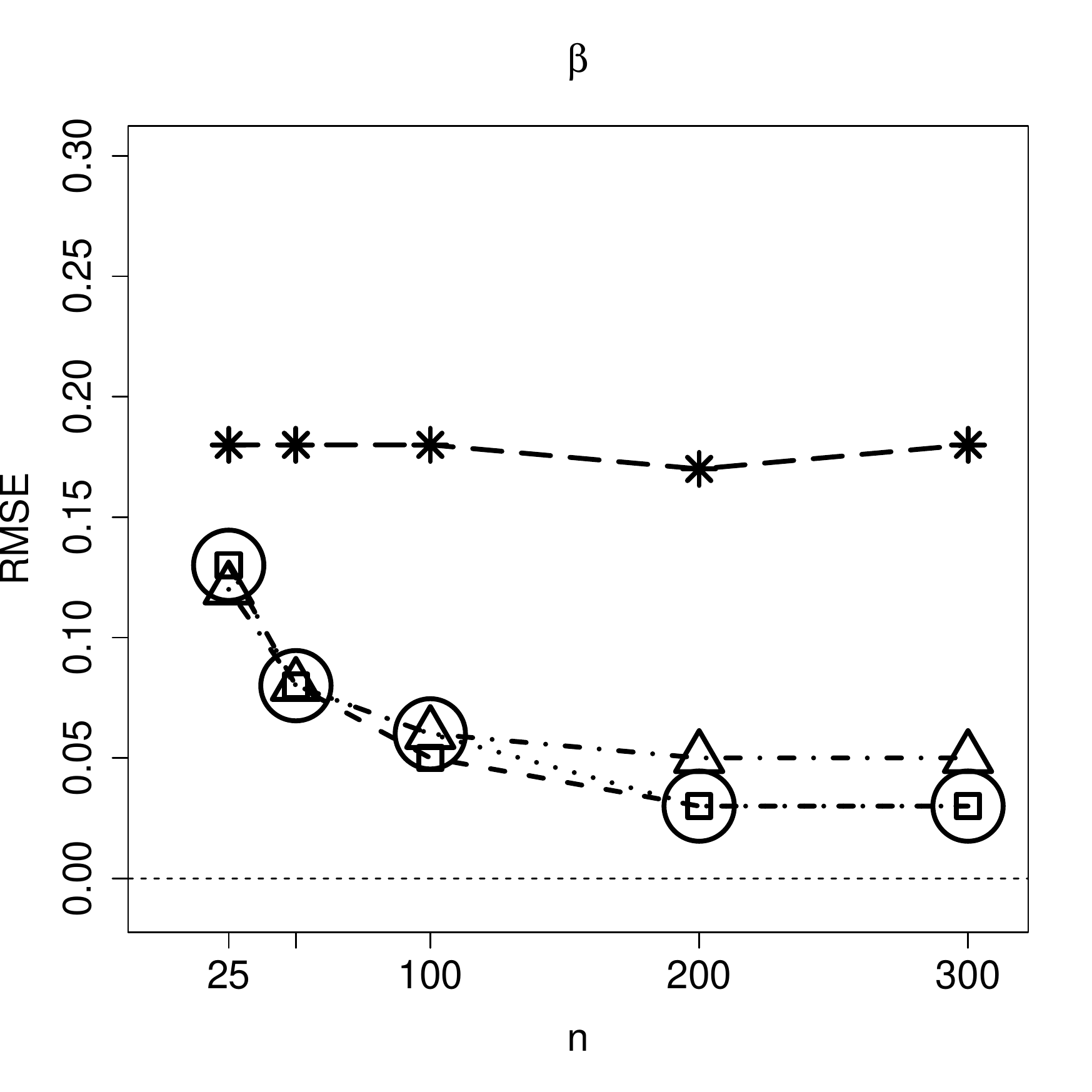}\\
\includegraphics[scale=0.3]{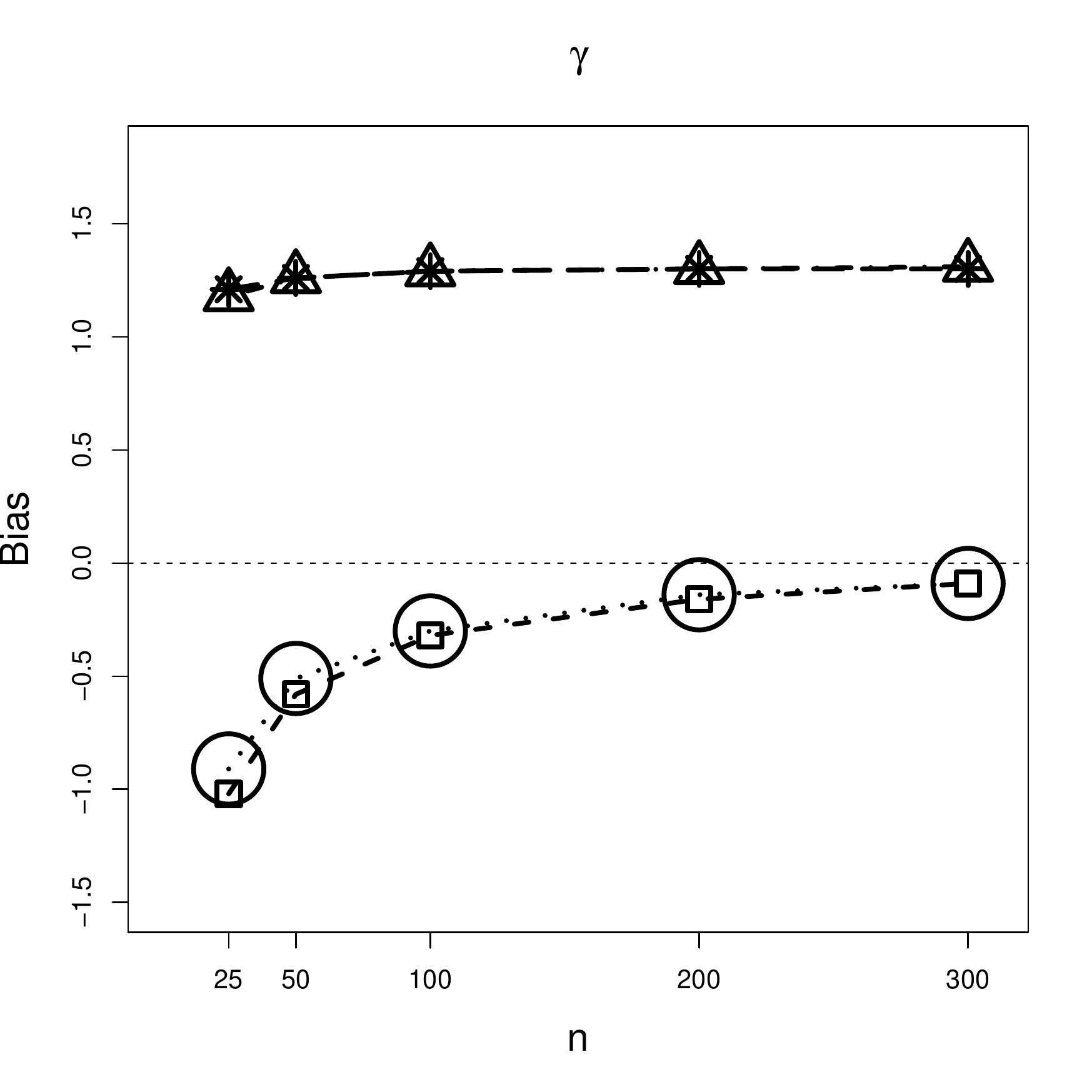}  \qquad \includegraphics[scale=0.3]{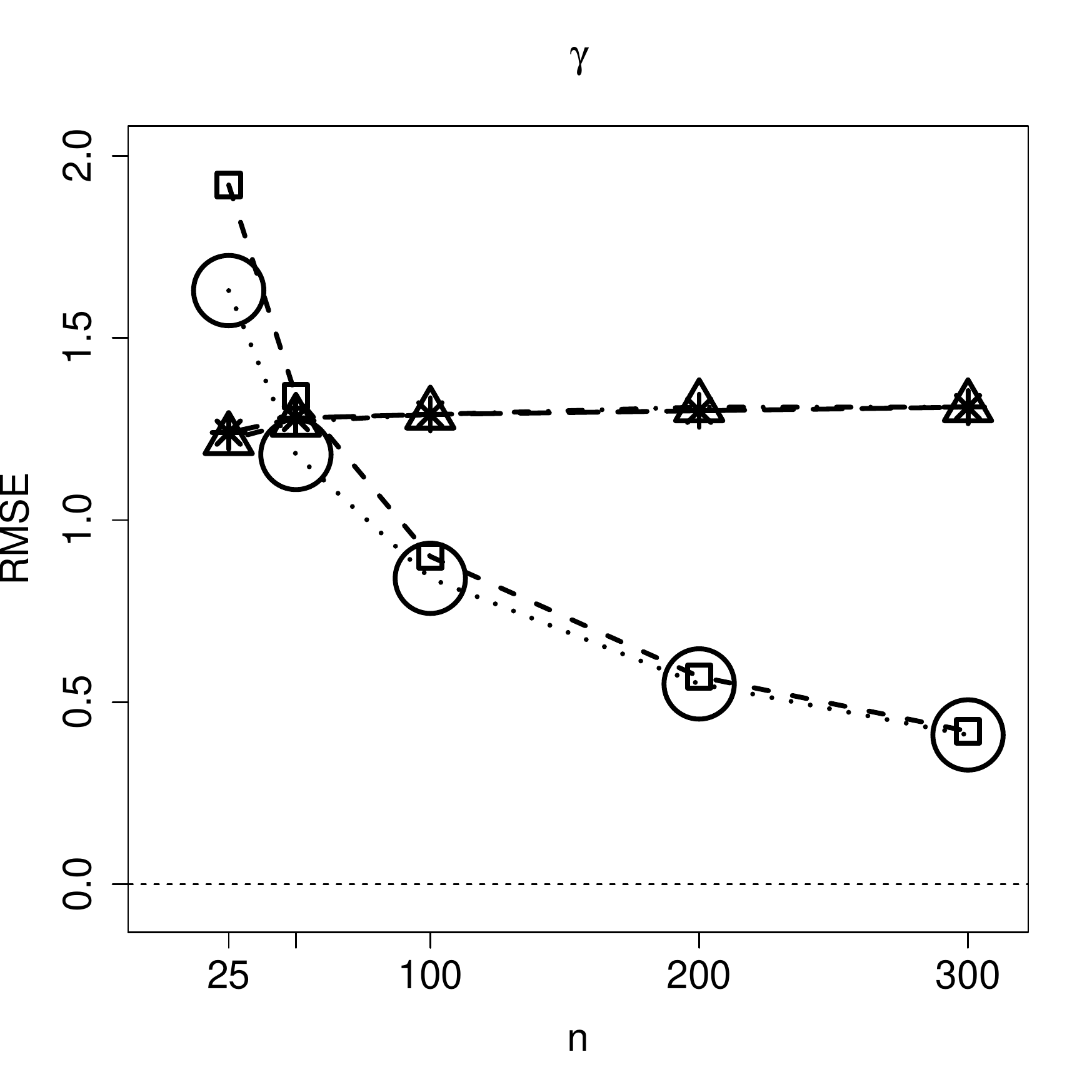}\\
\caption{Bias and RMSE for the estimators of $\alpha$, $\beta$ and $\gamma$ for $k_{x}=0.75$, constant precision model; $\ell_{a}$ (square), $\ell_{p}$ (circle), $\ell_{rc}$ (triangle) and $\ell_{naive}$(star).}
\label{Bias075model1}
\end{figure}

\begin{figure}
\centering
\includegraphics[scale=0.3]{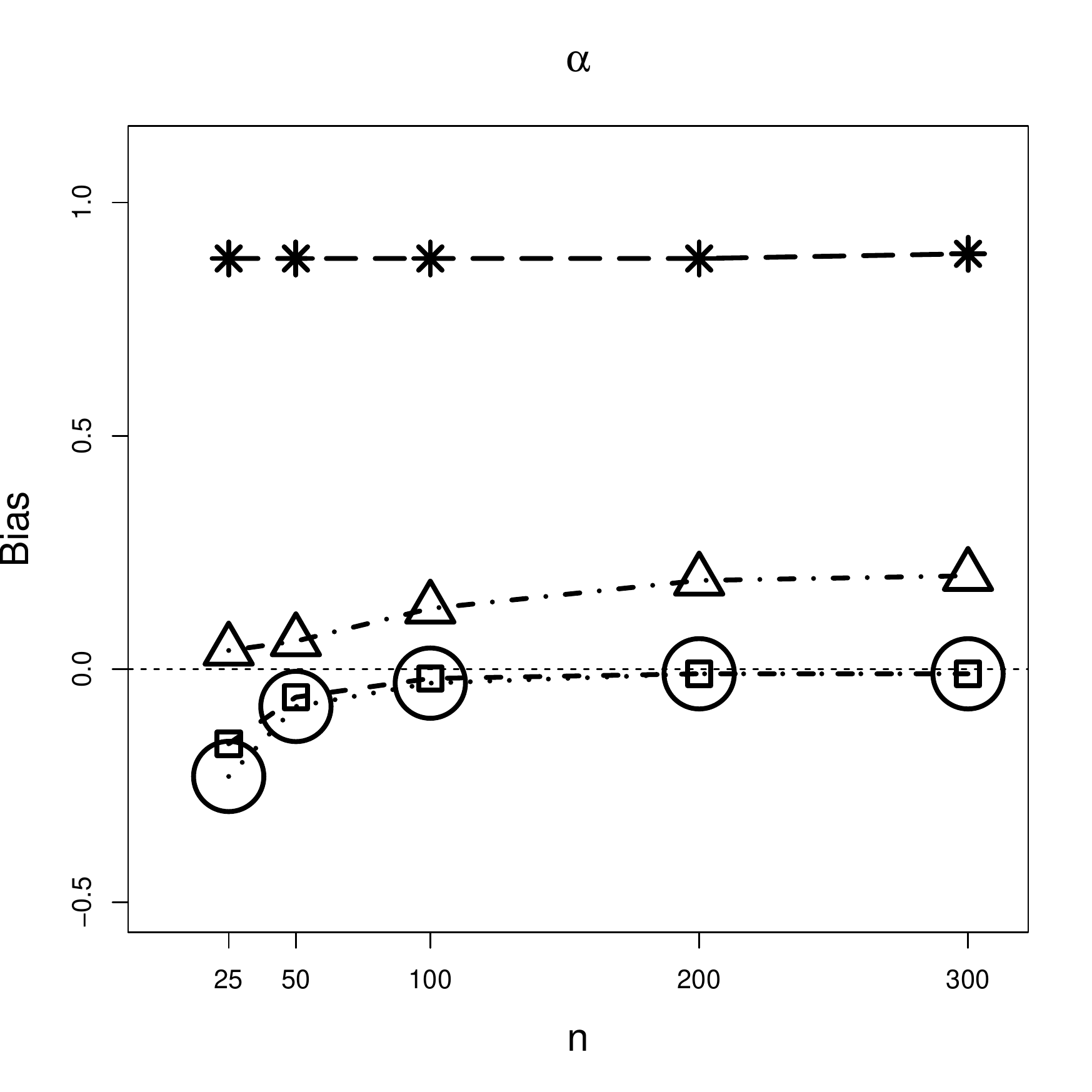} \qquad \includegraphics[scale=0.3]{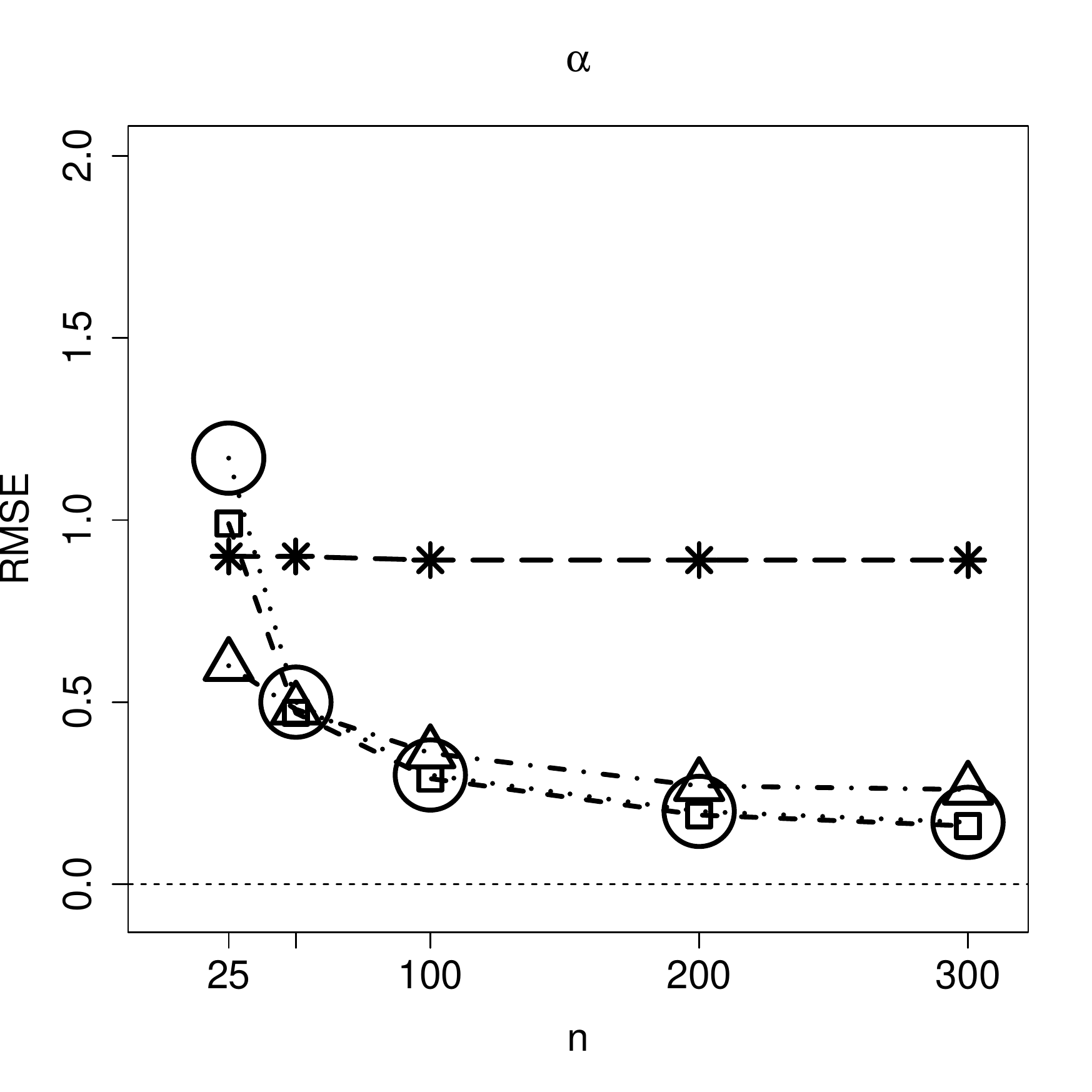}\\
\includegraphics[scale=0.3]{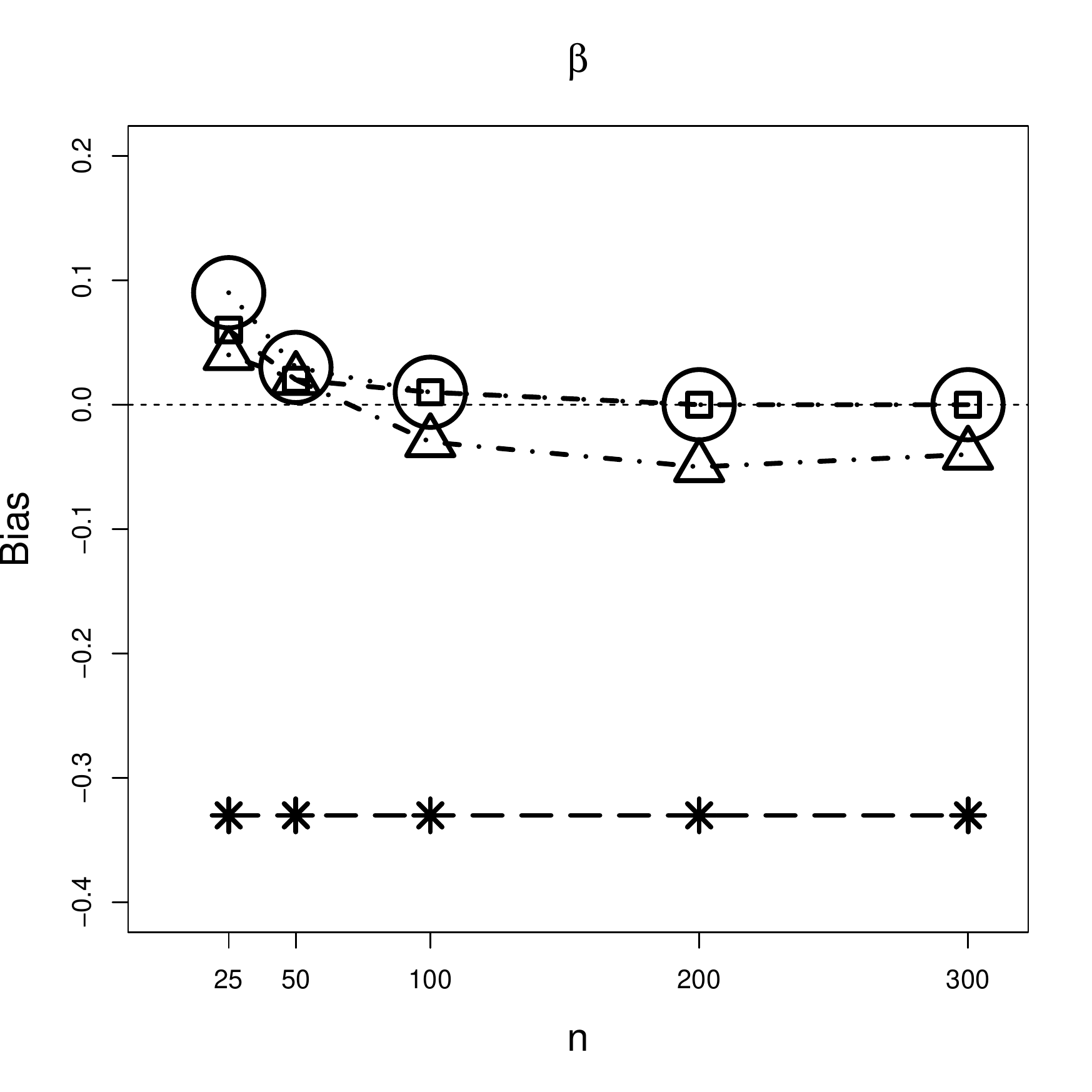}  \qquad \includegraphics[scale=0.3]{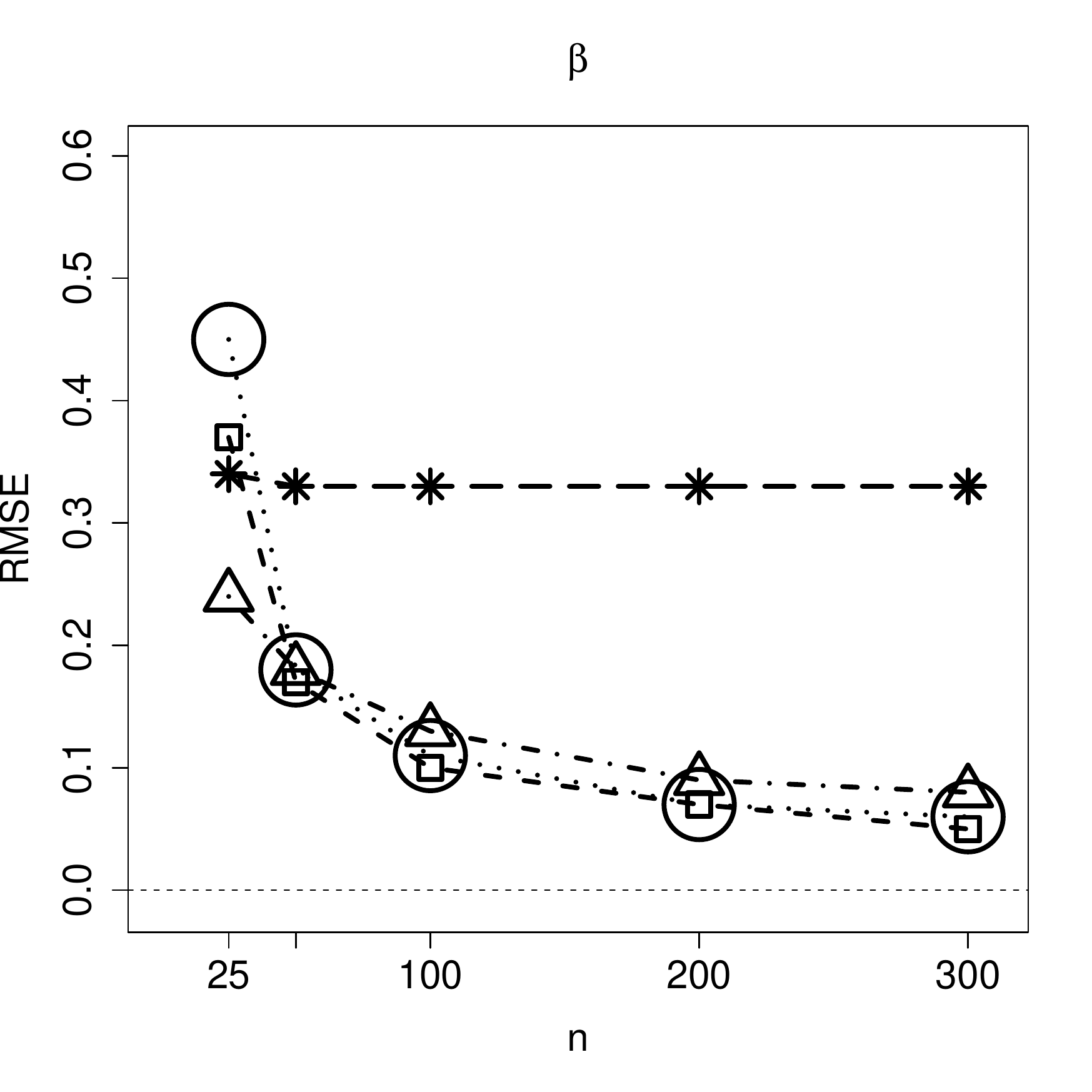}\\
\includegraphics[scale=0.3]{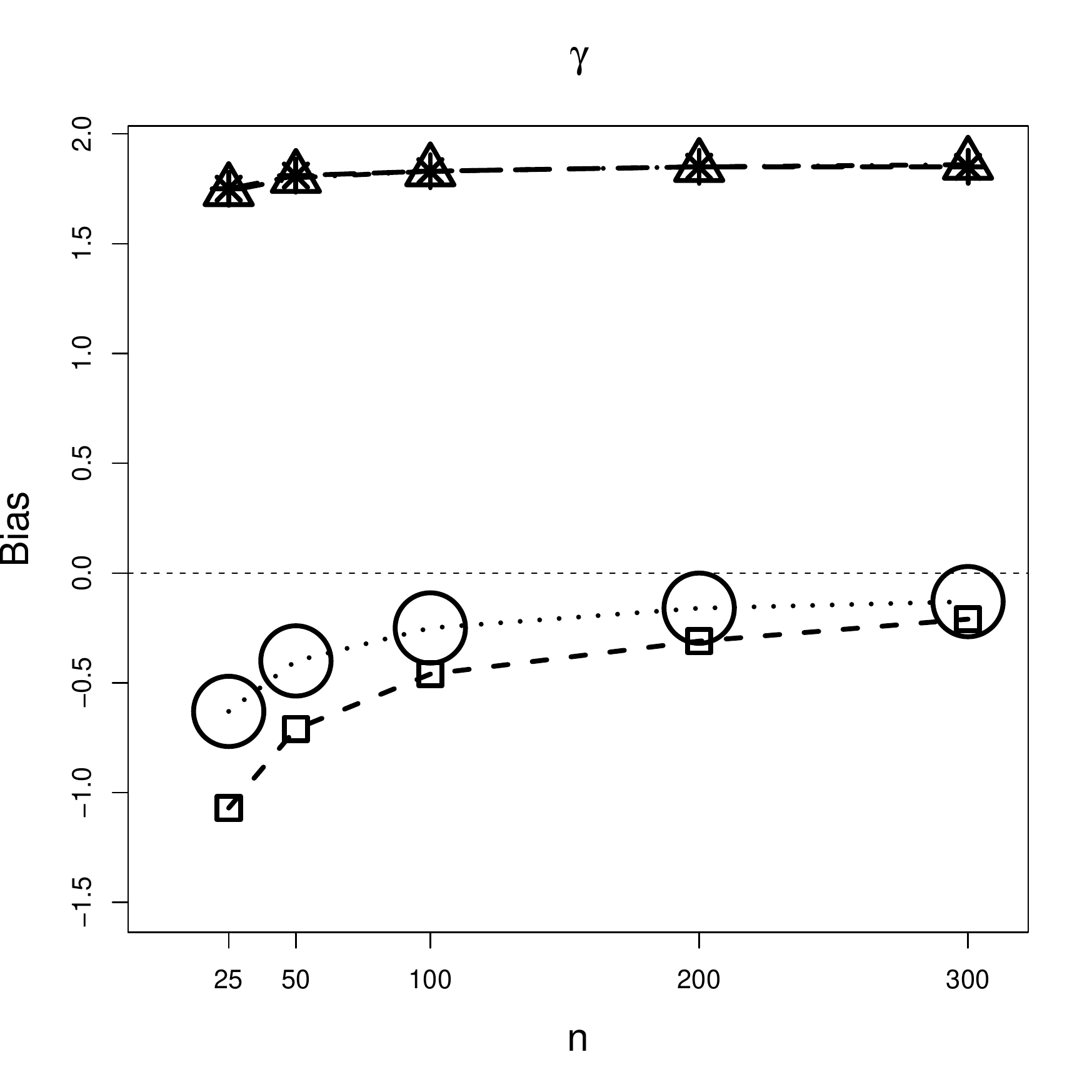}  \qquad \includegraphics[scale=0.3]{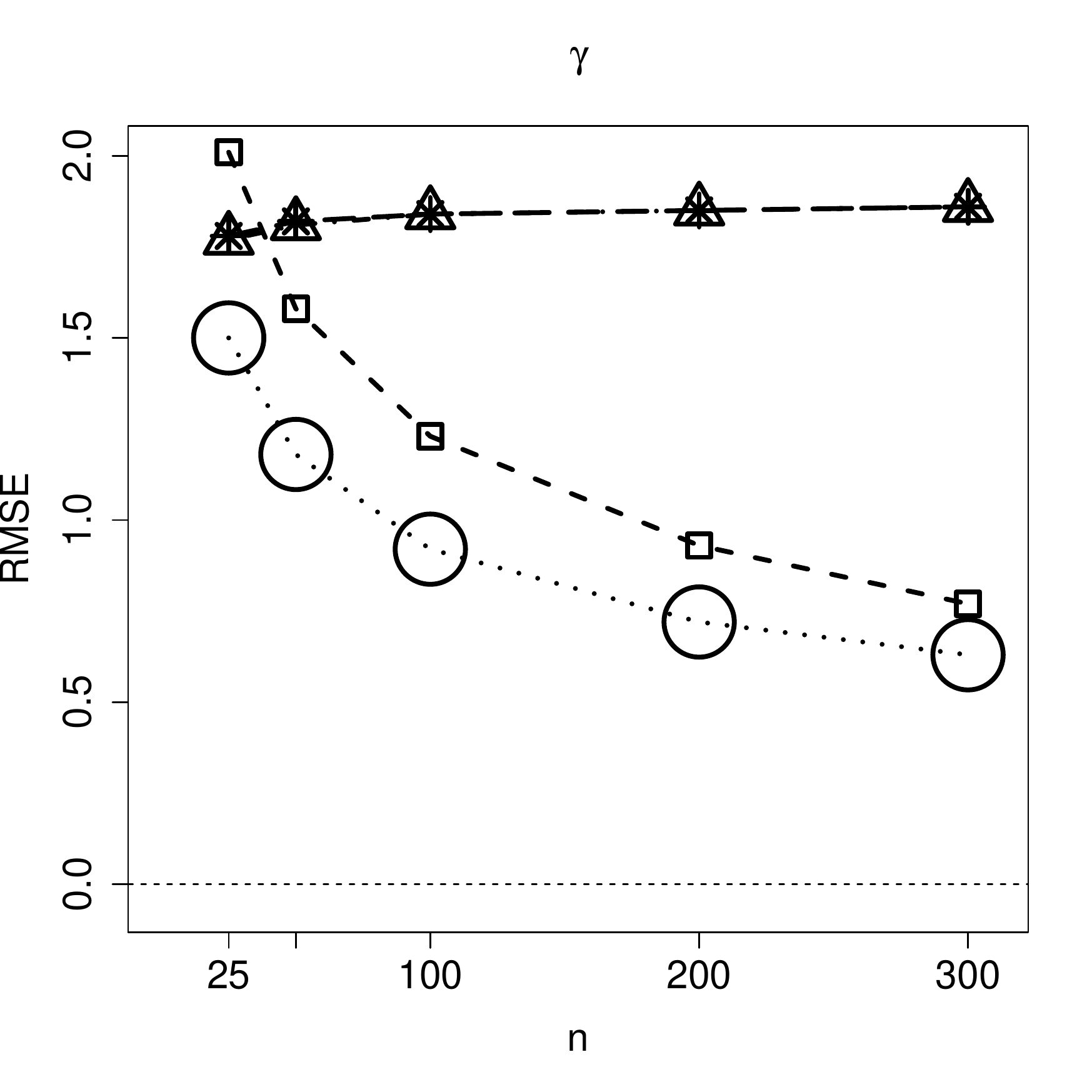}\\
\caption{Bias and RMSE for the estimators of $\alpha$, $\beta$ and $\gamma$ for $k_{x}=0.50$, constant precision model; $\ell_{a}$ (square), $\ell_{p}$ (circle), $\ell_{rc}$ (triangle) and $\ell_{naive}$(star).}
\label{Bias050model1}
\end{figure}

\begin{figure}
\centering
\includegraphics[scale=0.25]{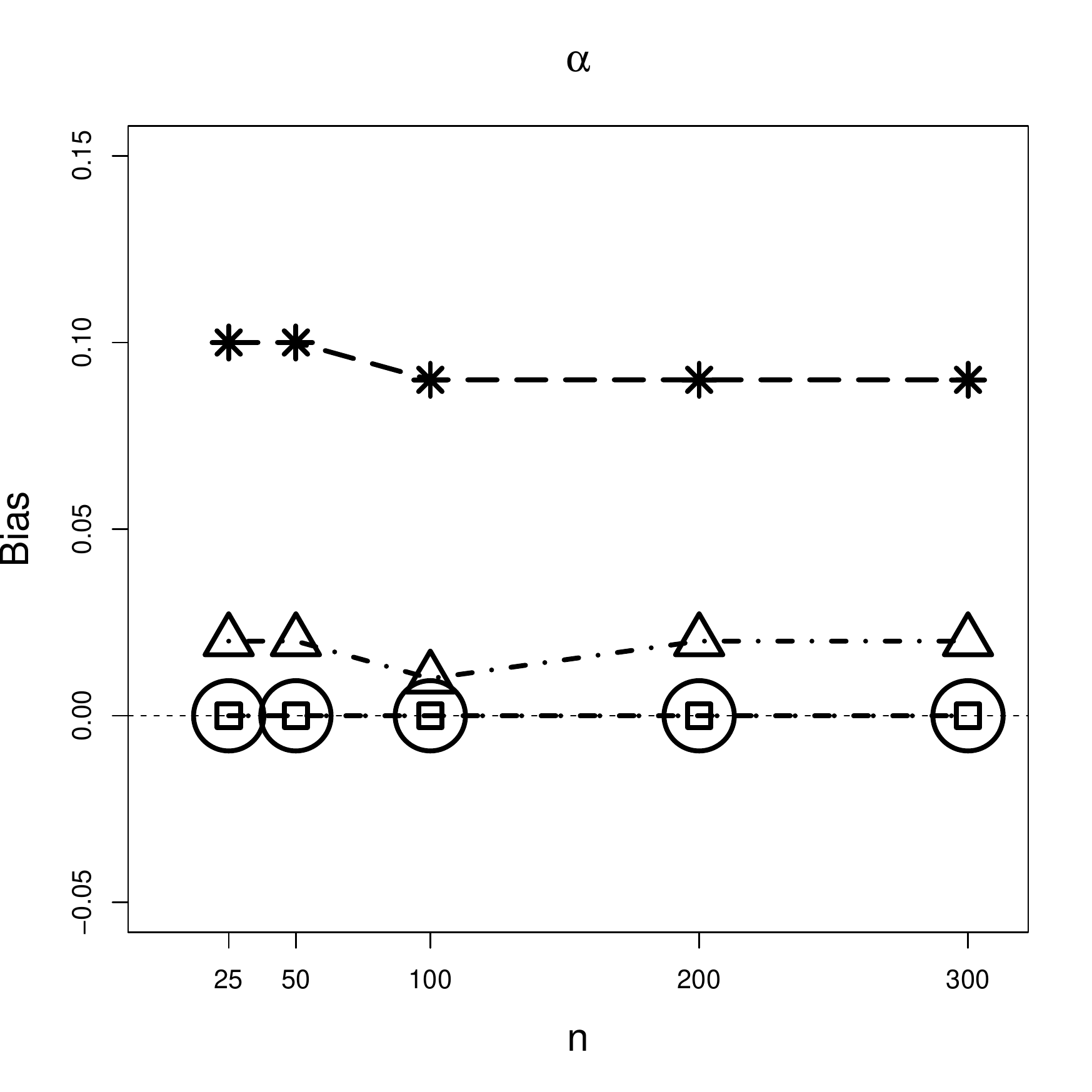} \qquad \includegraphics[scale=0.25]{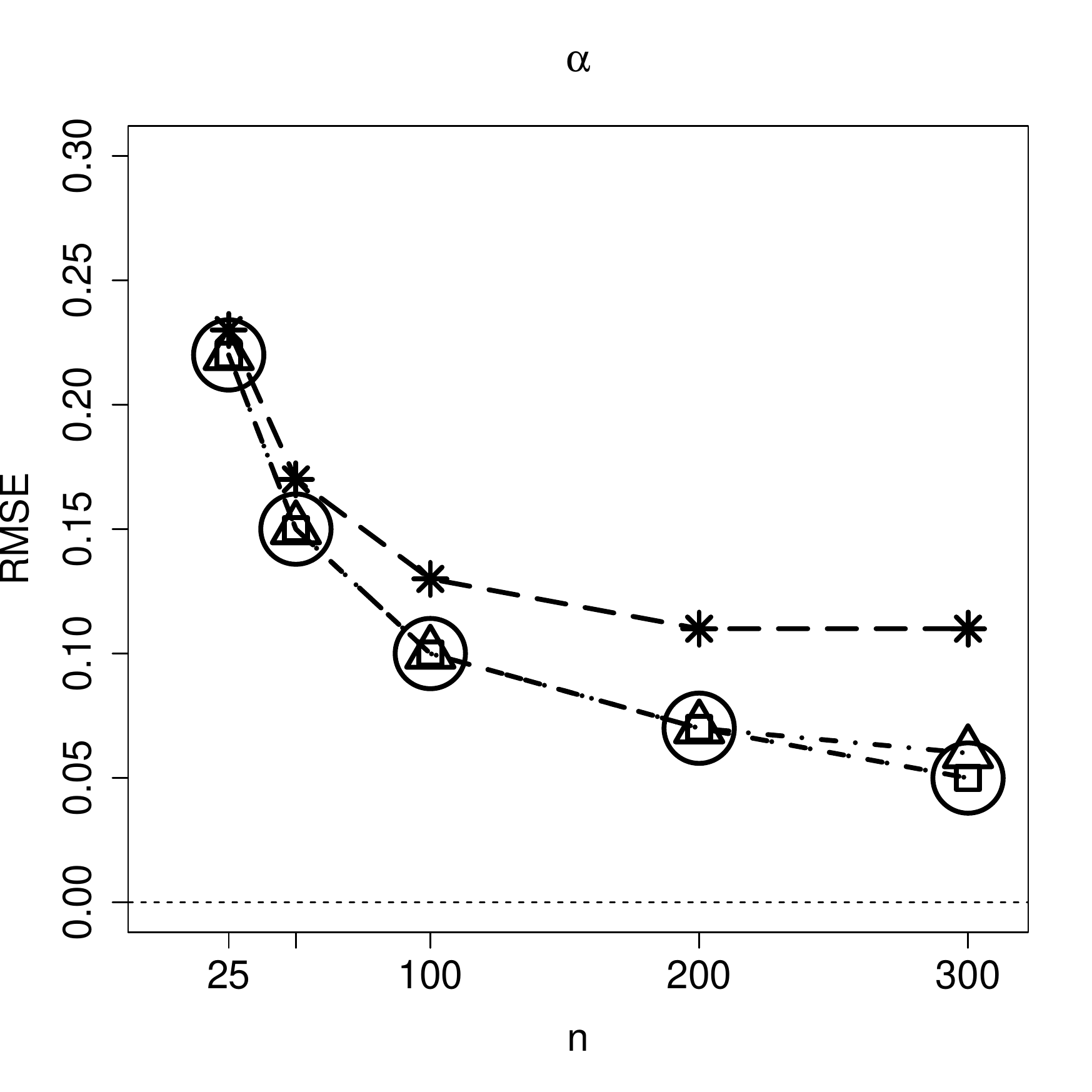}\\
\includegraphics[scale=0.25]{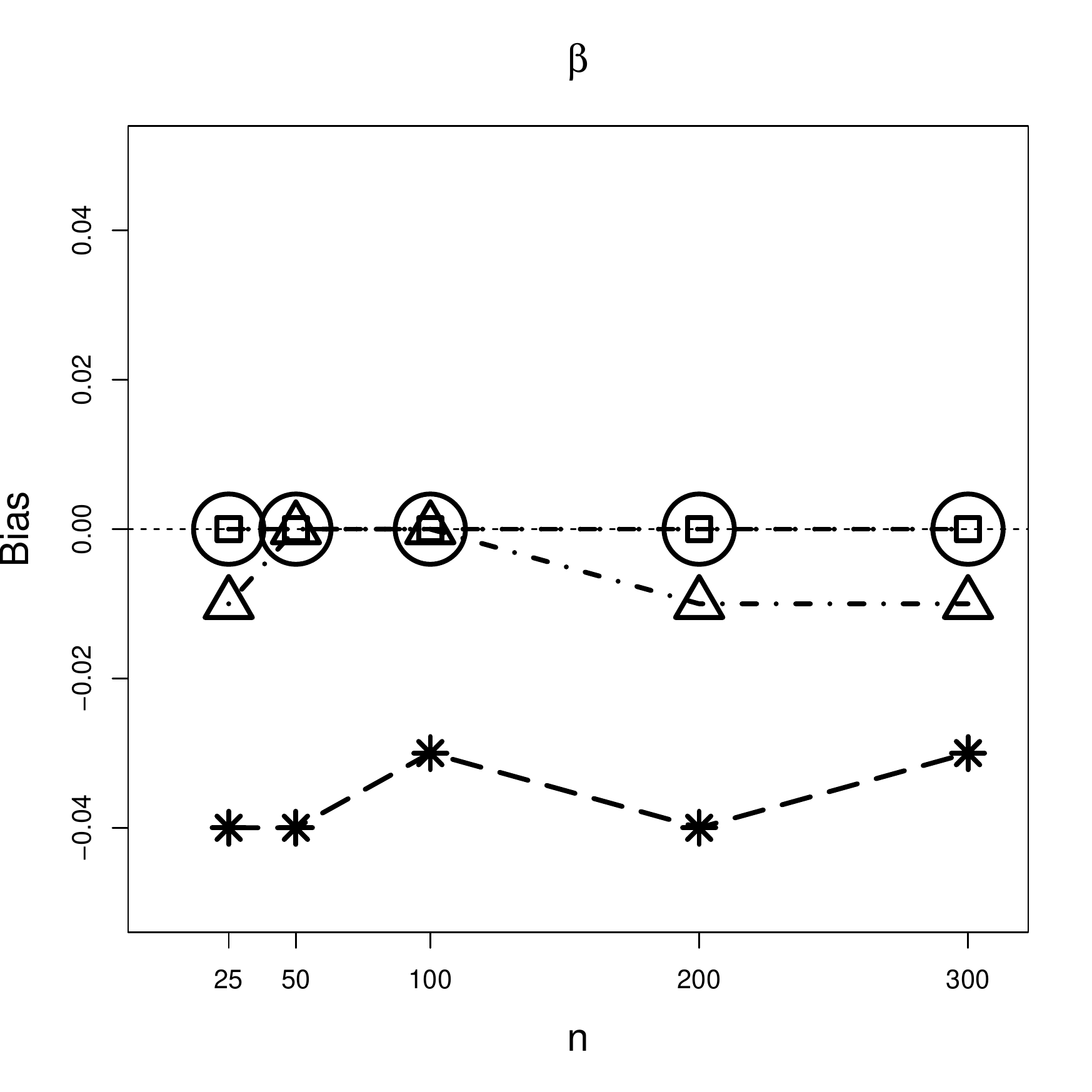}  \qquad \includegraphics[scale=0.25]{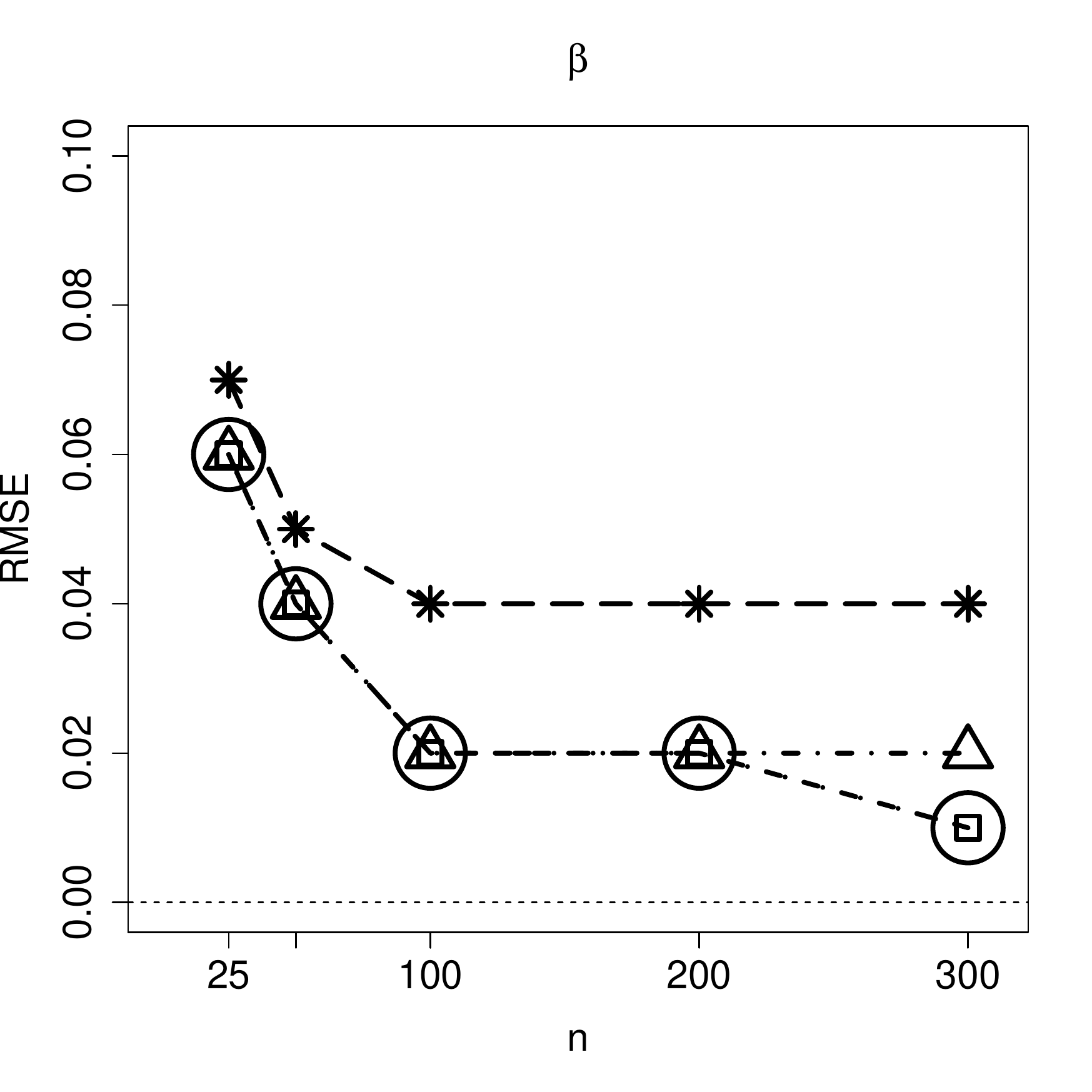}\\
\includegraphics[scale=0.25]{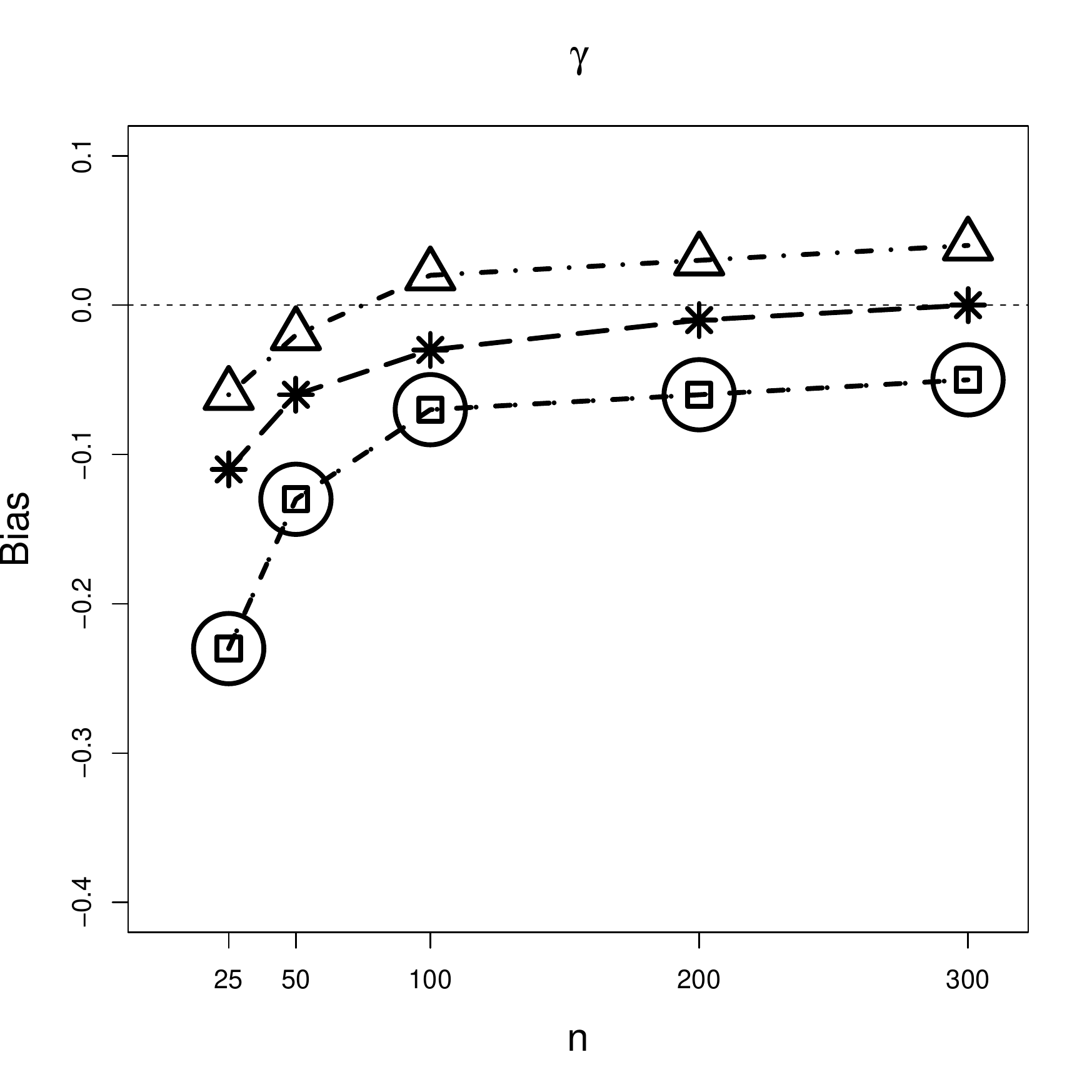}  \qquad \includegraphics[scale=0.25]{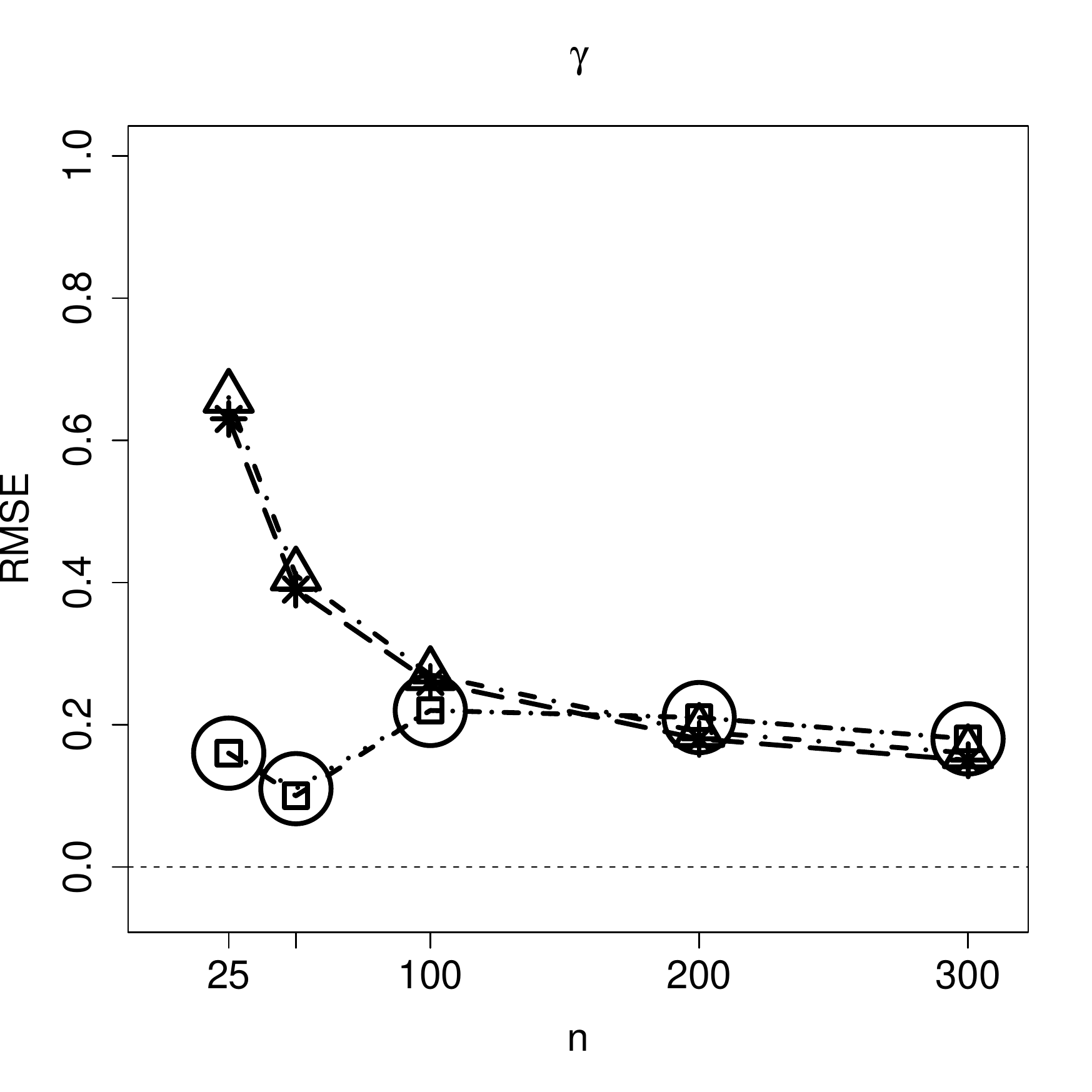}\\
\includegraphics[scale=0.25]{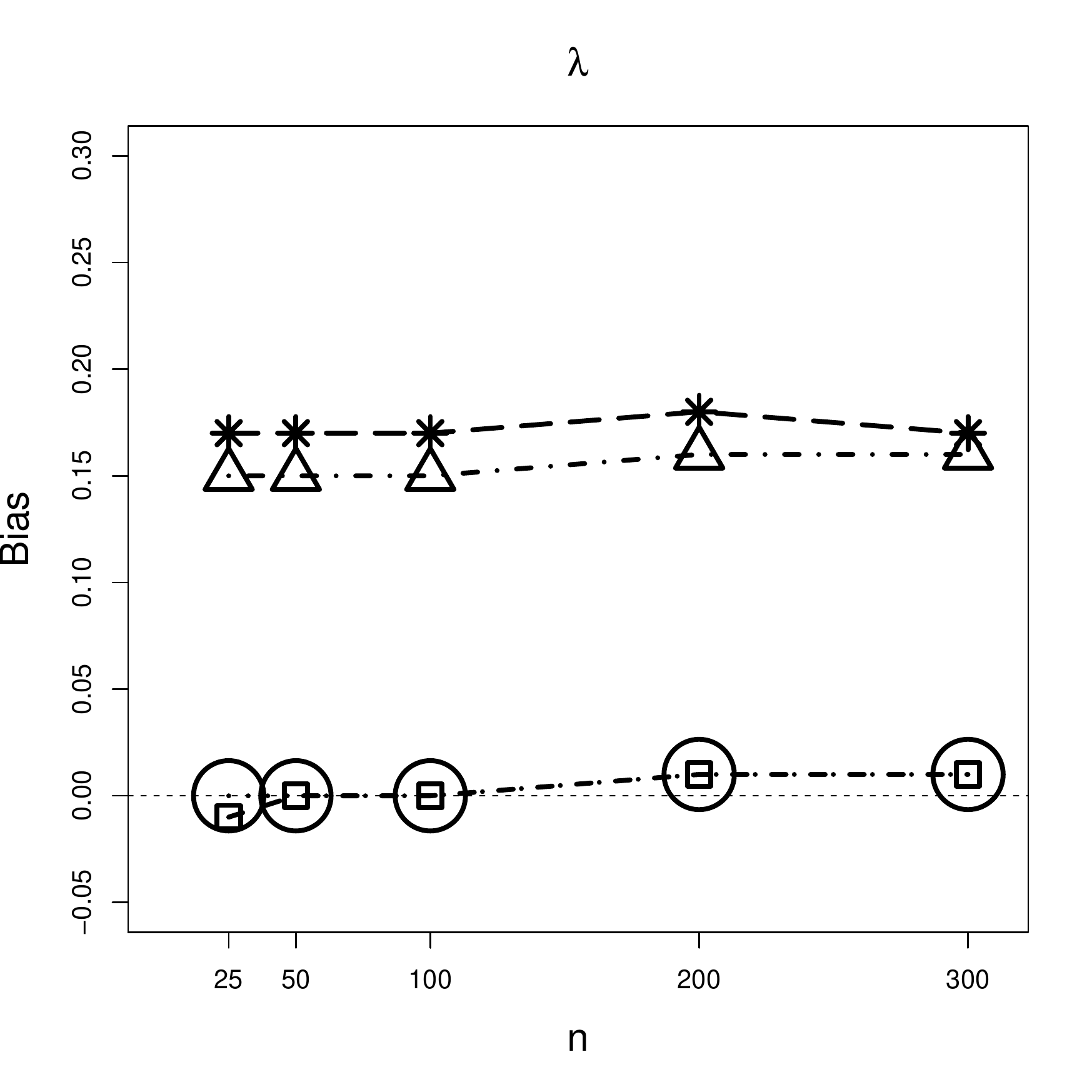} \qquad \includegraphics[scale=0.25]{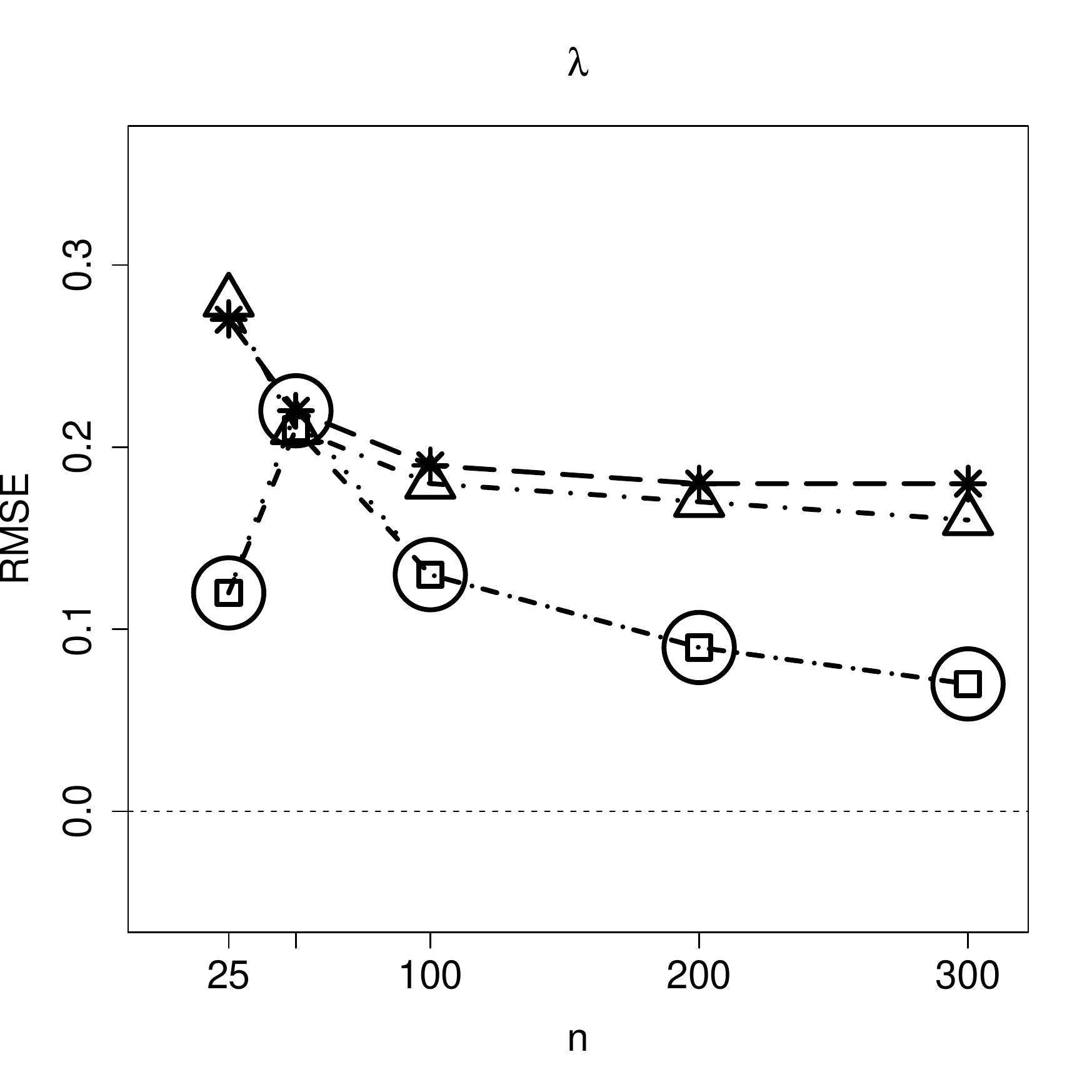}
\caption{Bias and RMSE for the estimators of $\alpha$, $\beta$, $\gamma$ and $\lambda$ for $k_{x}=0.95$, varying precision model; $\ell_{a}$ (square), $\ell_{p}$ (circle), $\ell_{rc}$ (triangle) and $\ell_{naive}$(star).}
\label{Bias095model2}
\end{figure}

\begin{figure}
\centering
\includegraphics[scale=0.25]{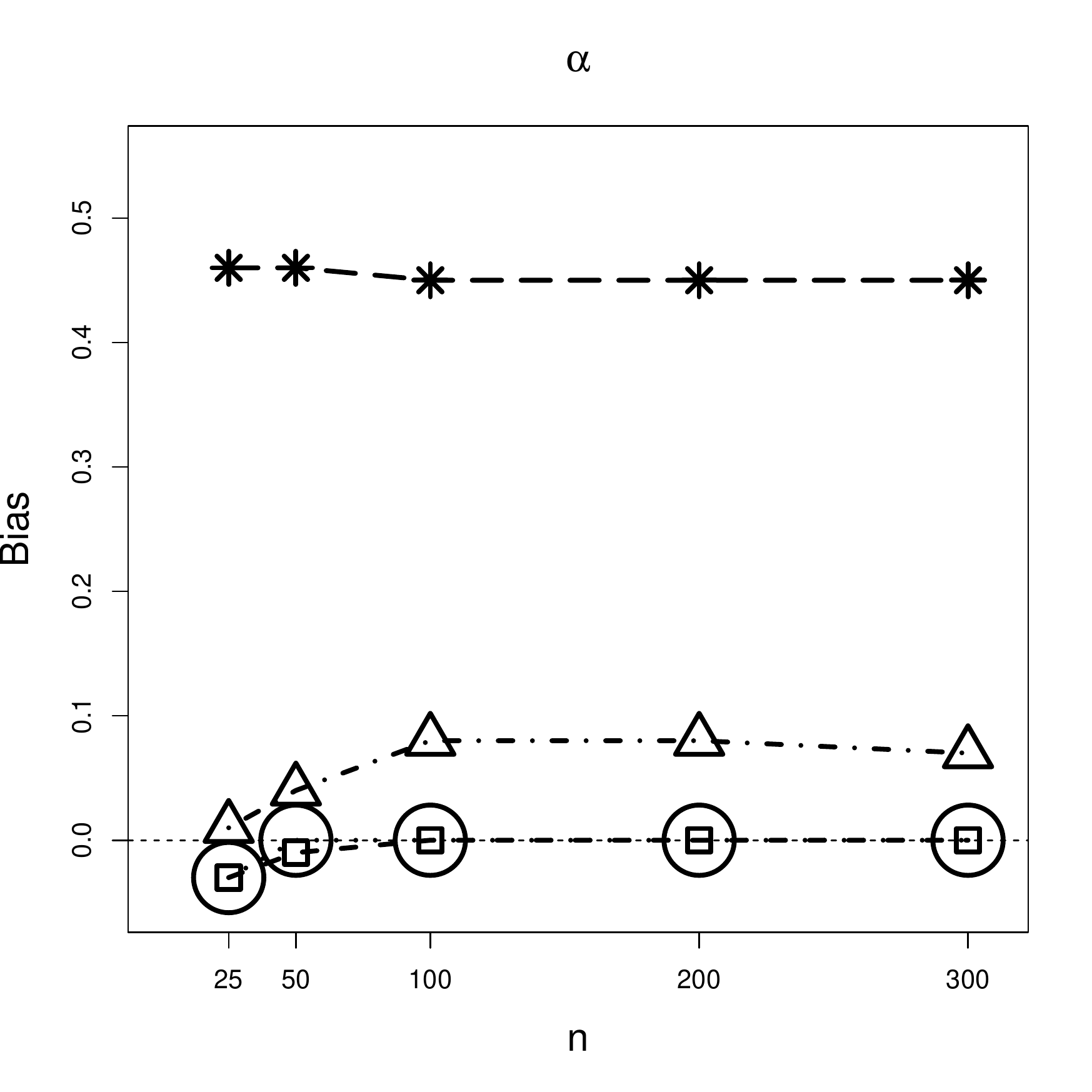} \qquad \includegraphics[scale=0.25]{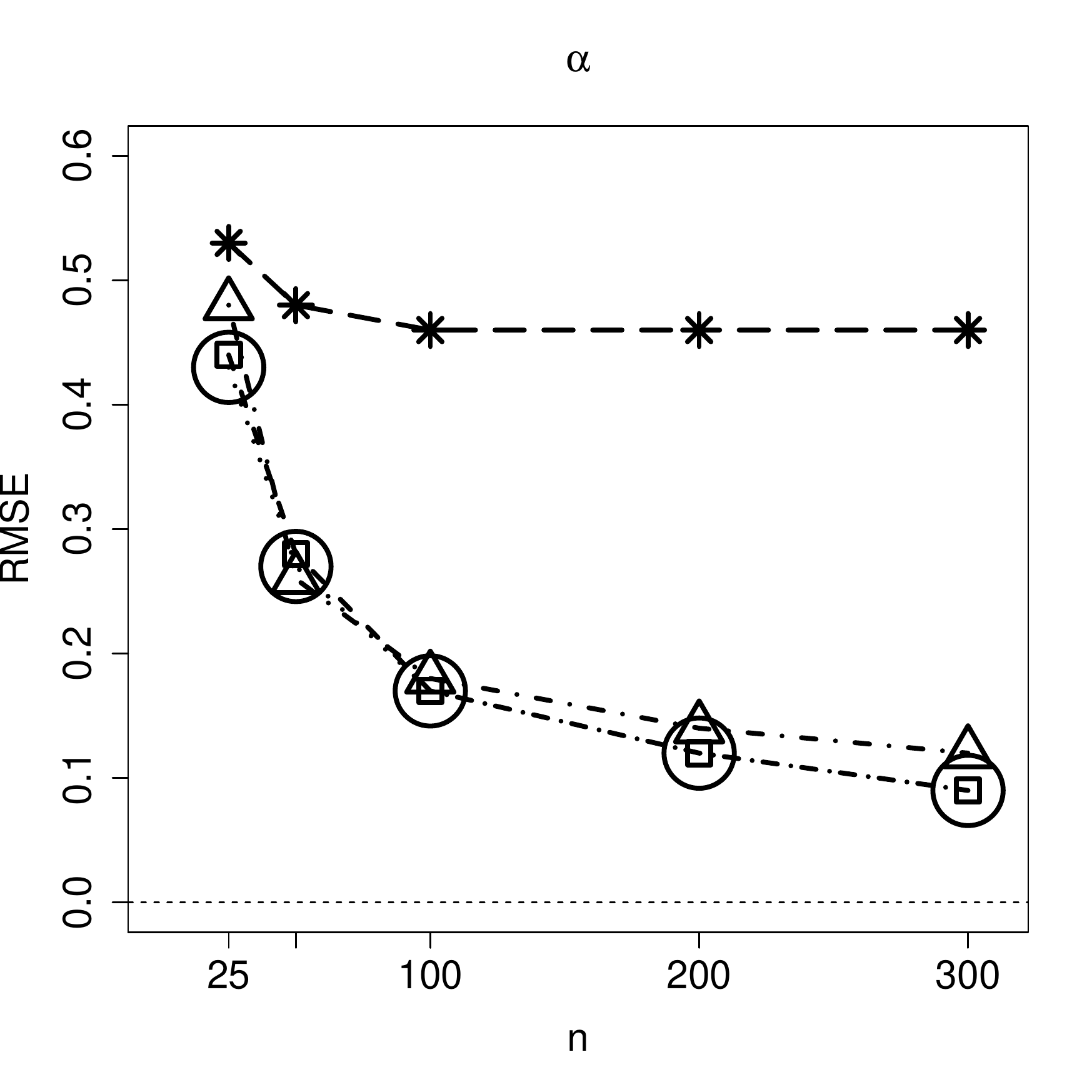}\\
\includegraphics[scale=0.25]{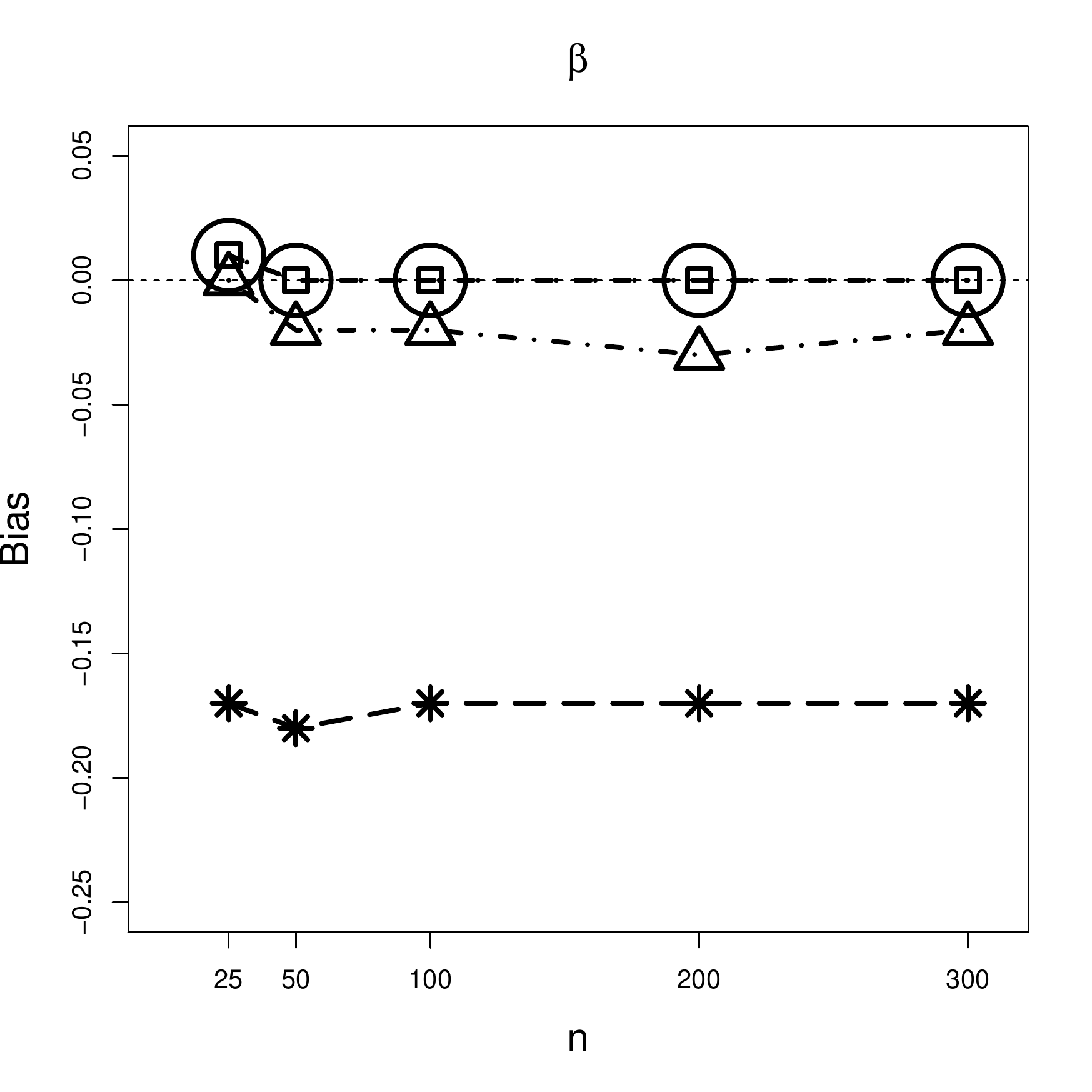}  \qquad \includegraphics[scale=0.25]{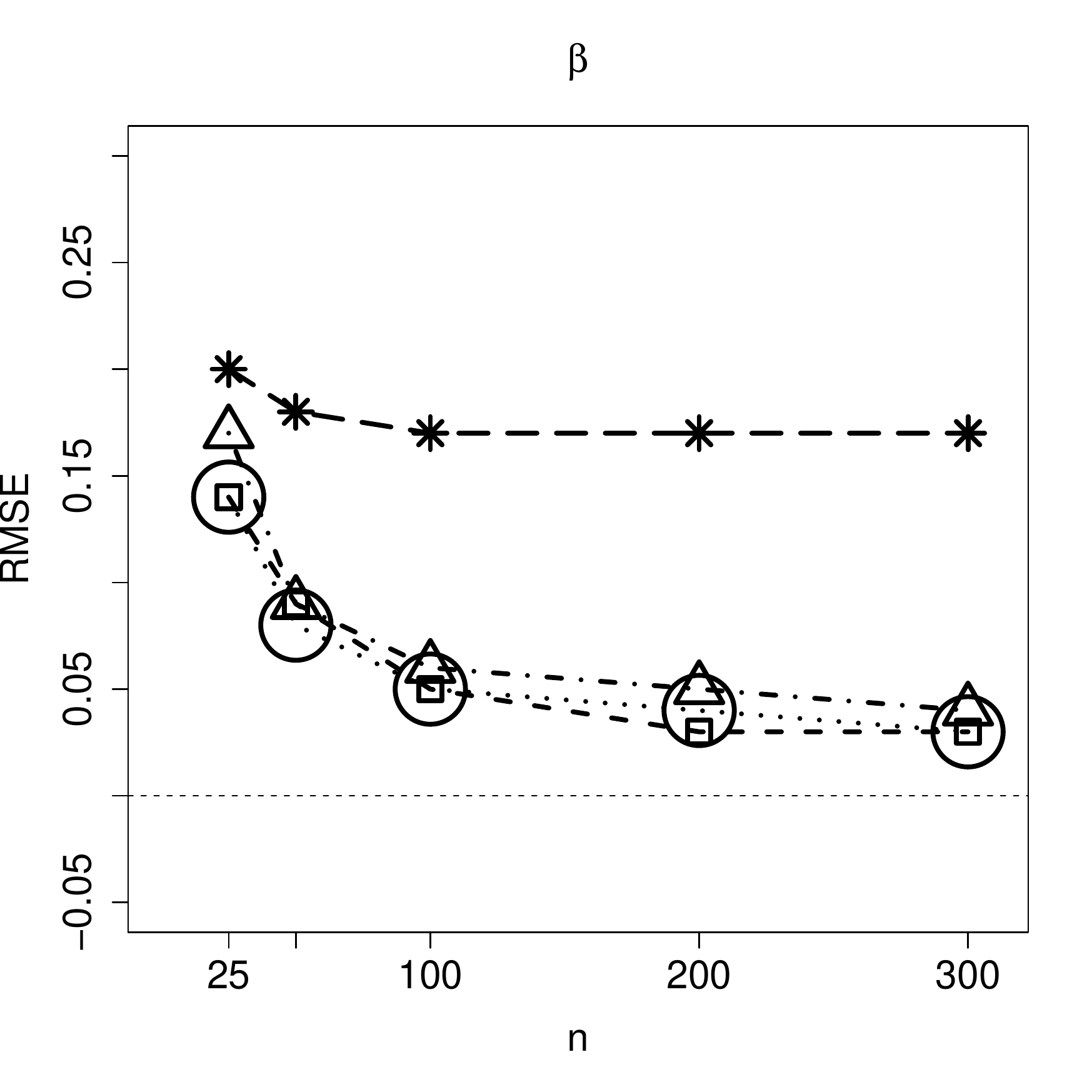}\\
\includegraphics[scale=0.25]{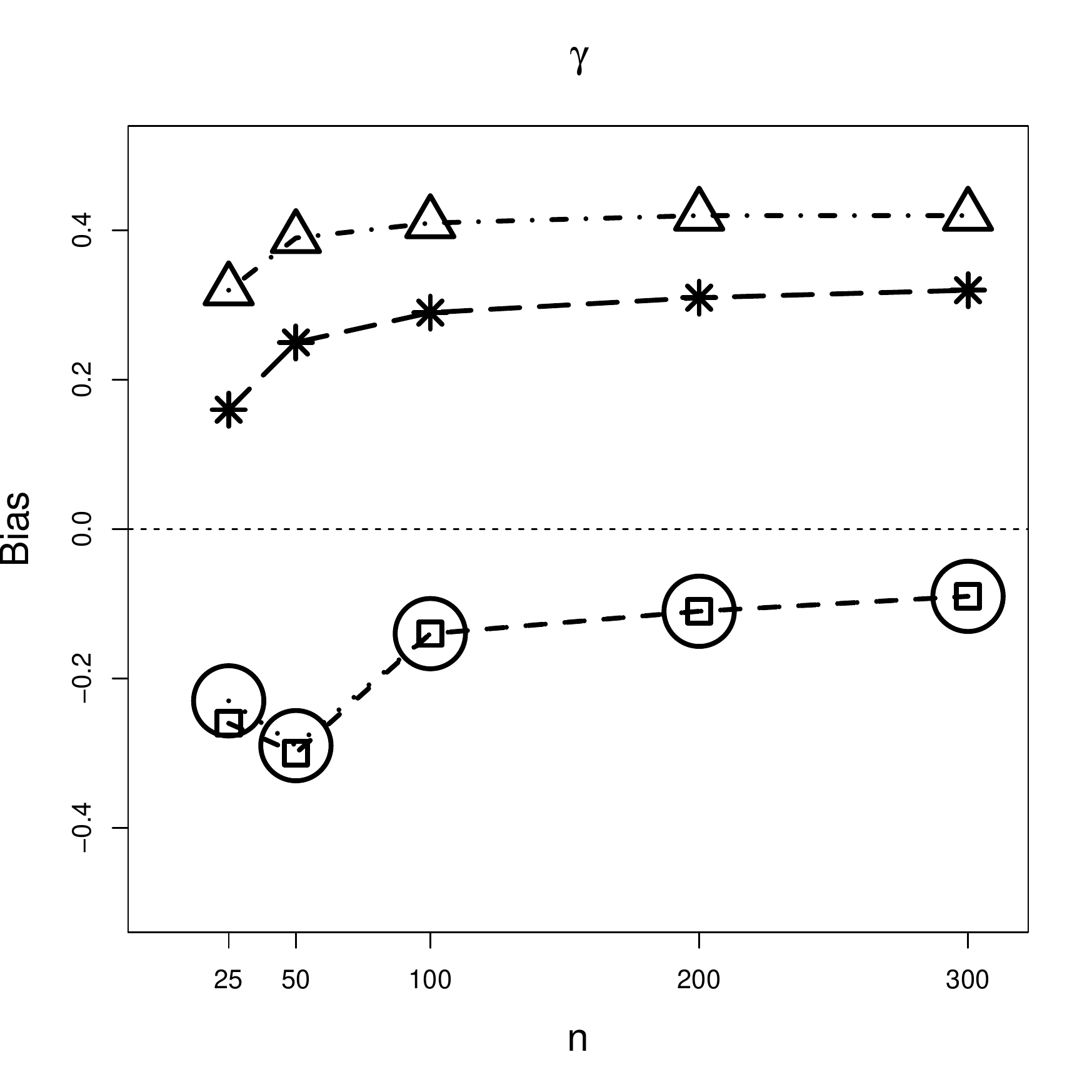}  \qquad \includegraphics[scale=0.25]{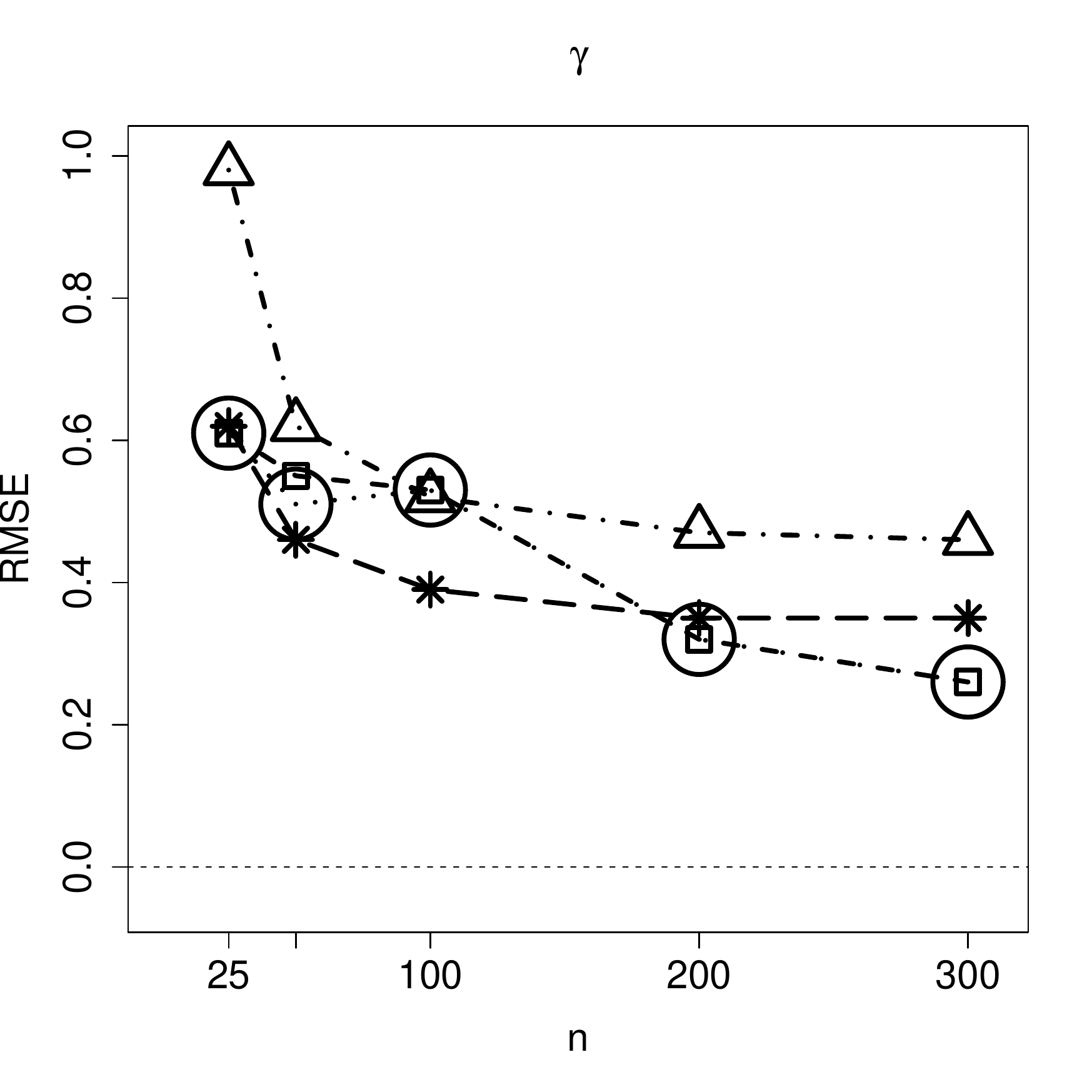}\\
\includegraphics[scale=0.25]{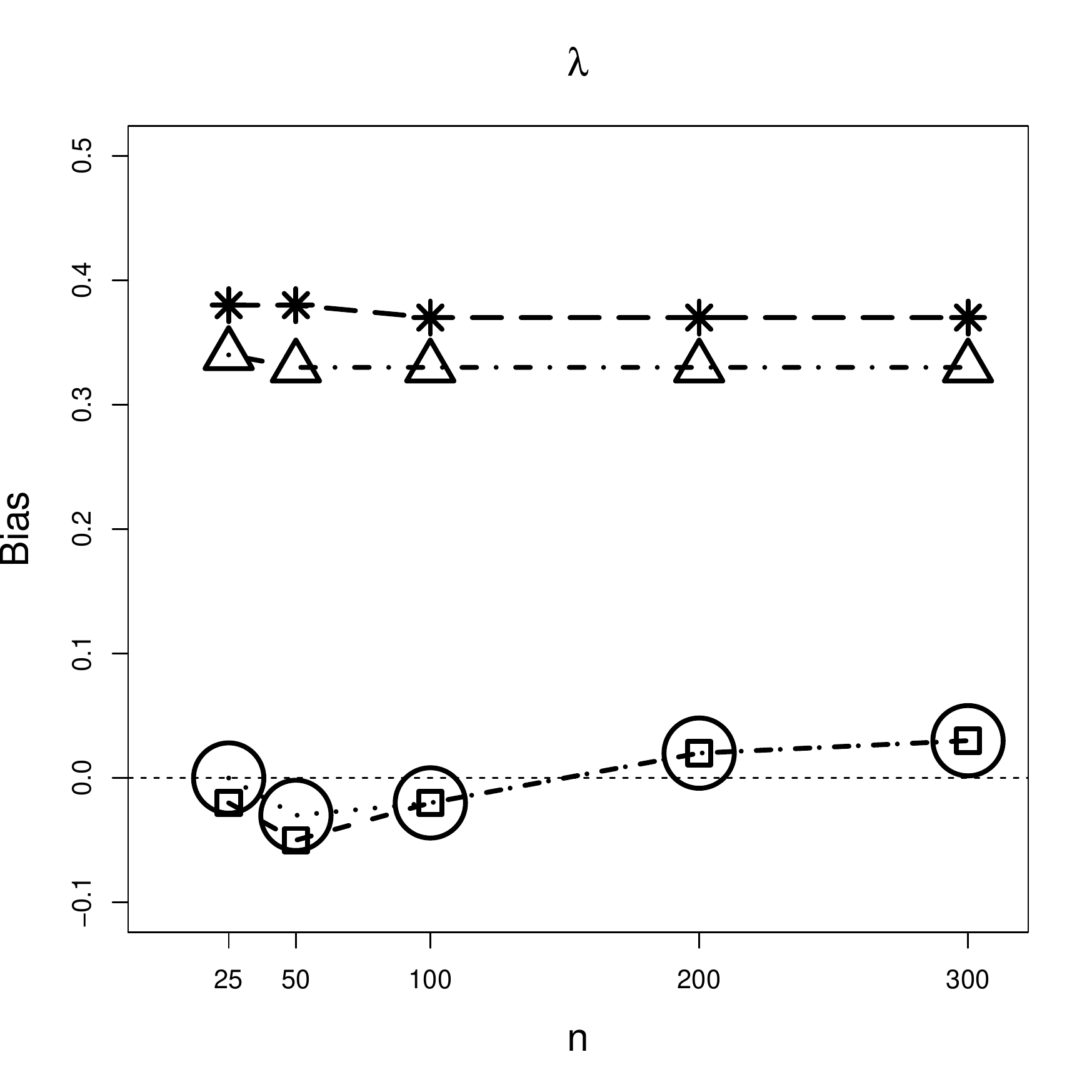} \qquad \includegraphics[scale=0.25]{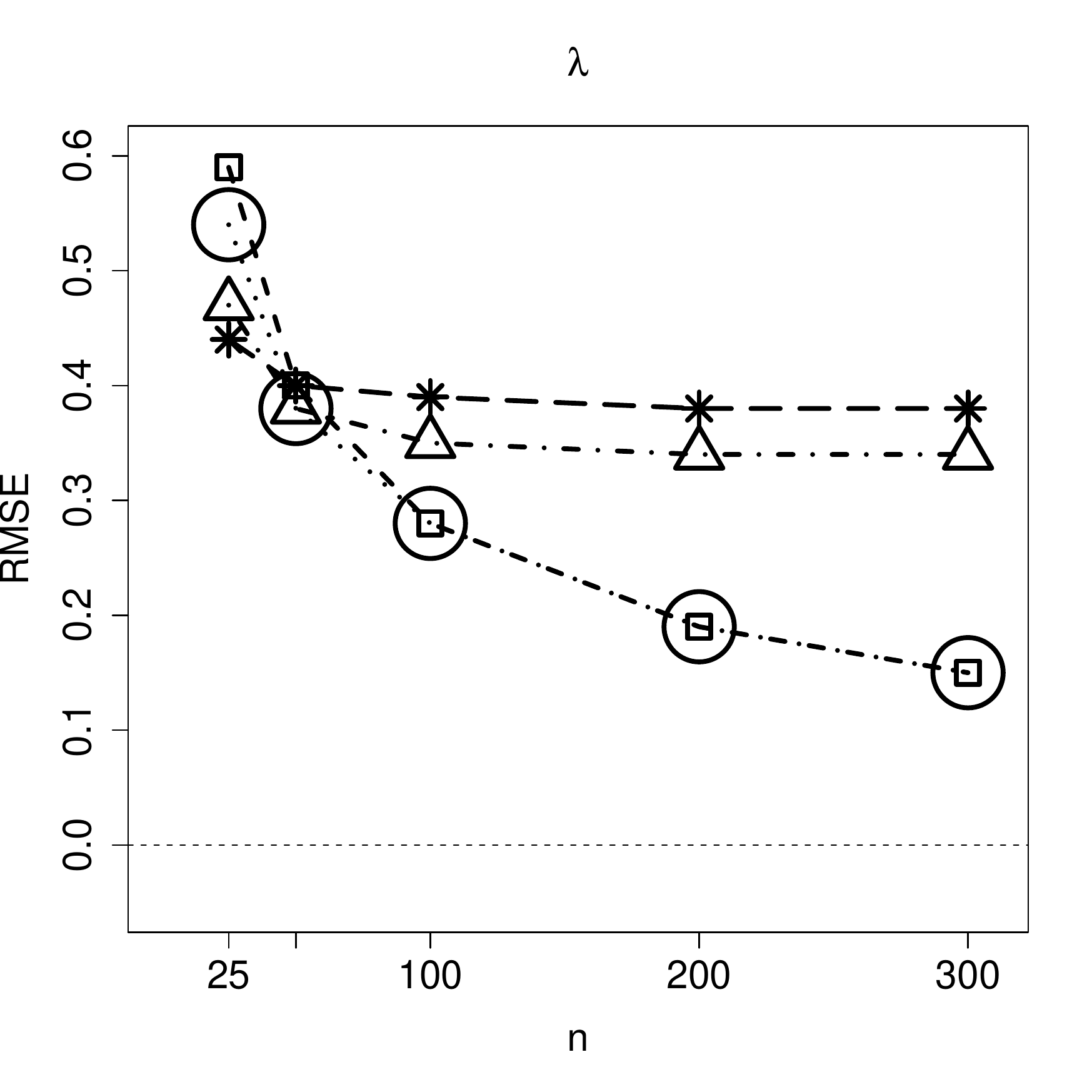}
\caption{Bias and RMSE for the estimators of $\alpha$, $\beta$, $\gamma$ and $\lambda$ for $k_{x}=0.75$, varying precision model; $\ell_{a}$ (square), $\ell_{p}$ (circle), $\ell_{rc}$ (triangle) and $\ell_{naive}$(star).}
\label{Bias075model2}
\end{figure}

\begin{figure}
\centering
\includegraphics[scale=0.25]{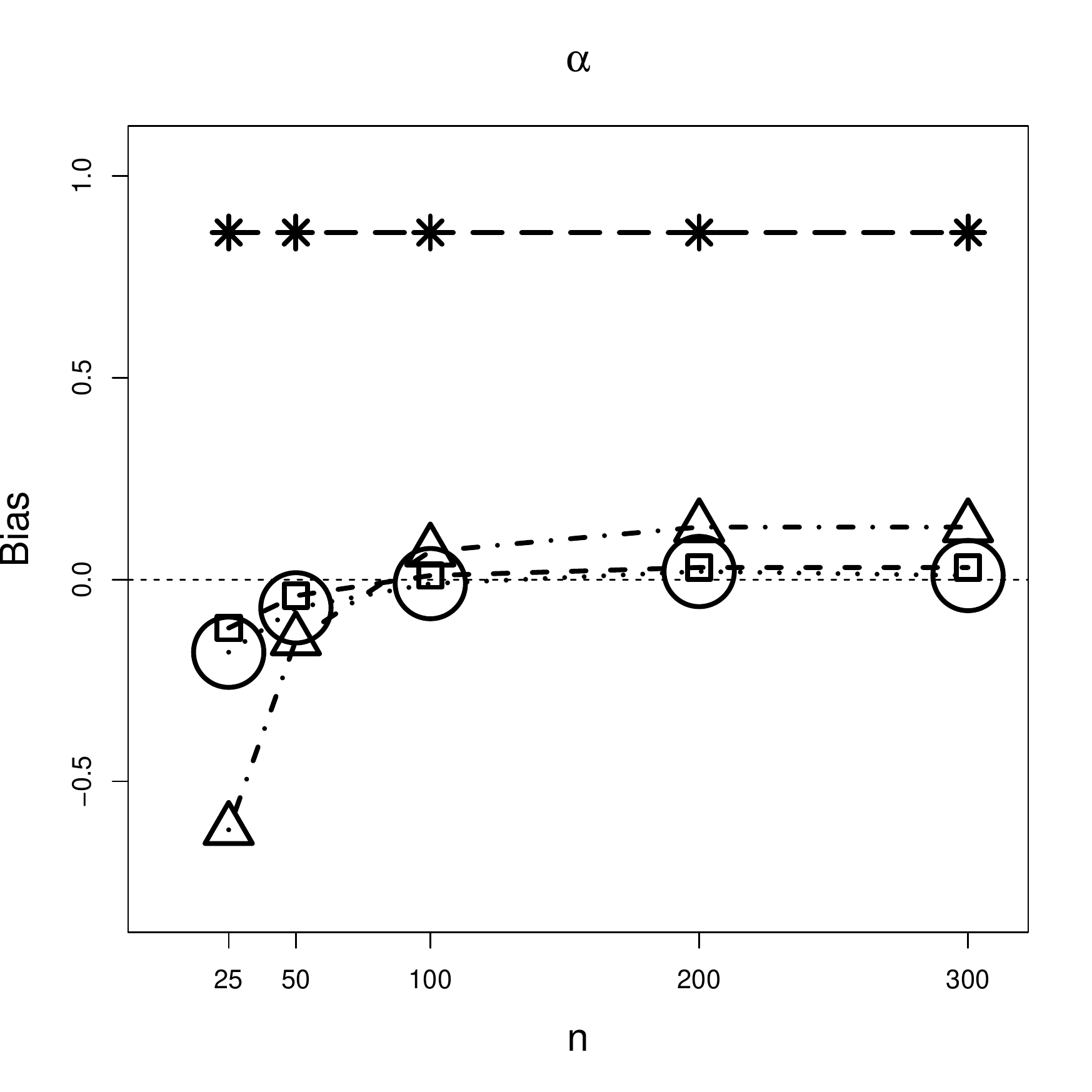} \qquad \includegraphics[scale=0.25]{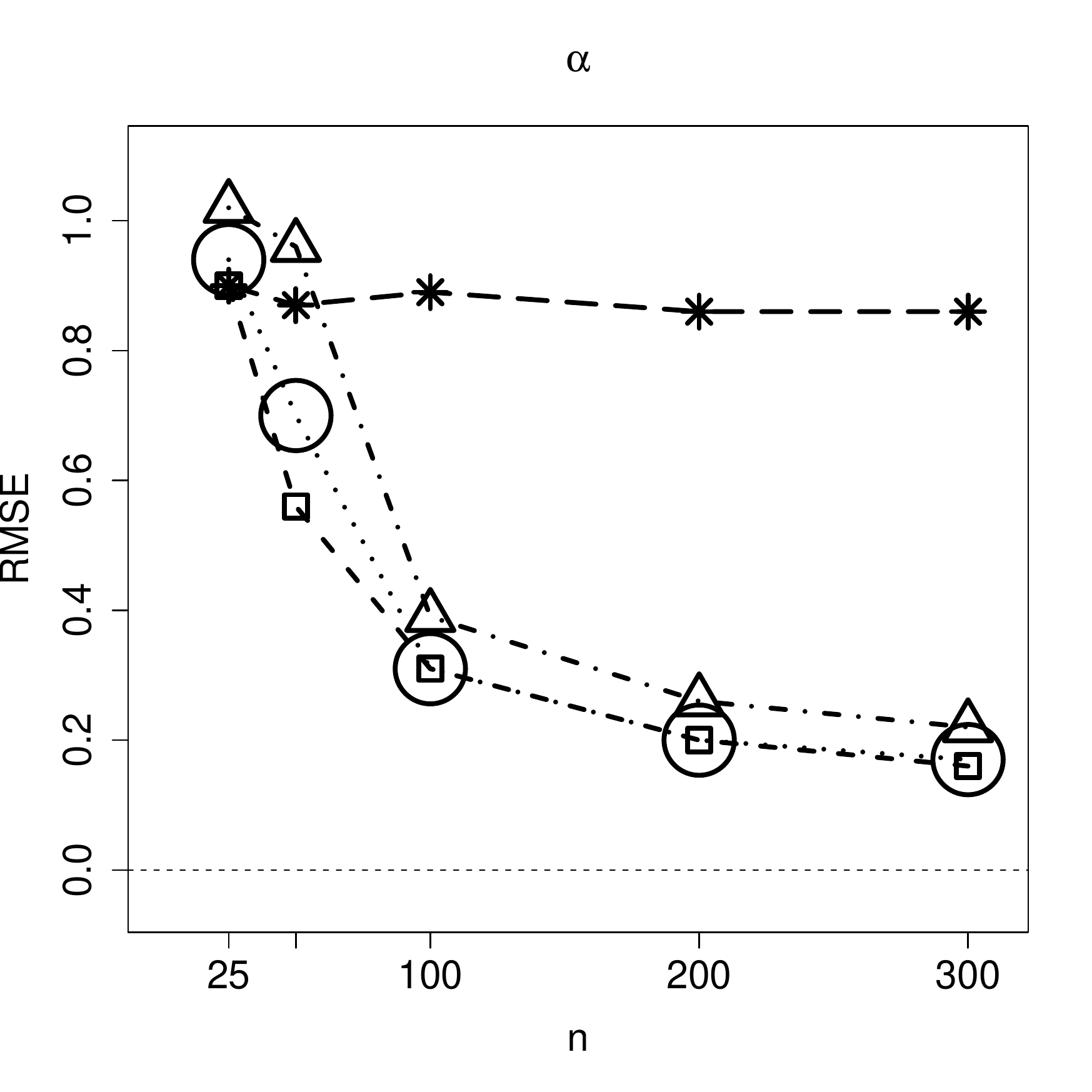}\\
\includegraphics[scale=0.25]{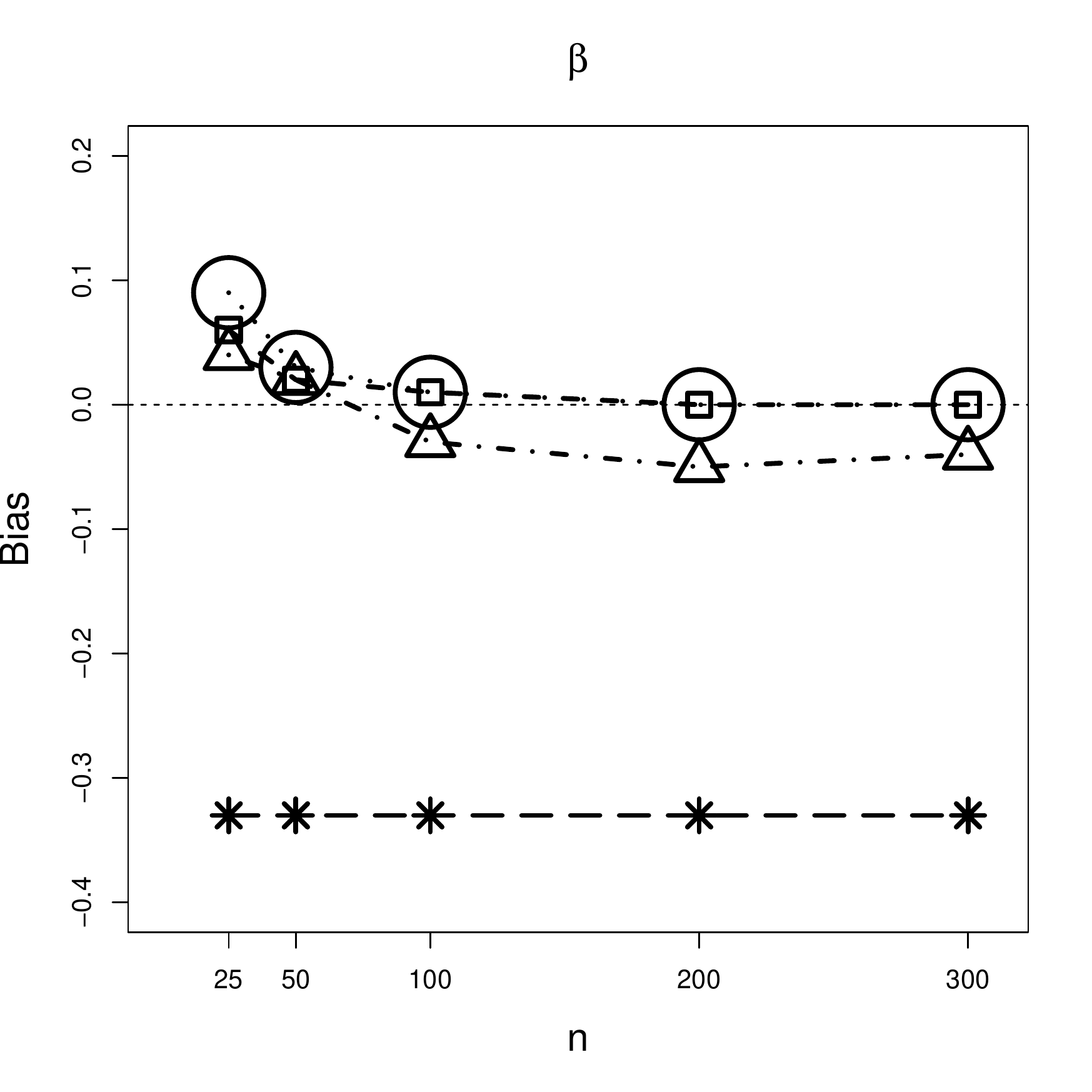}  \qquad \includegraphics[scale=0.25]{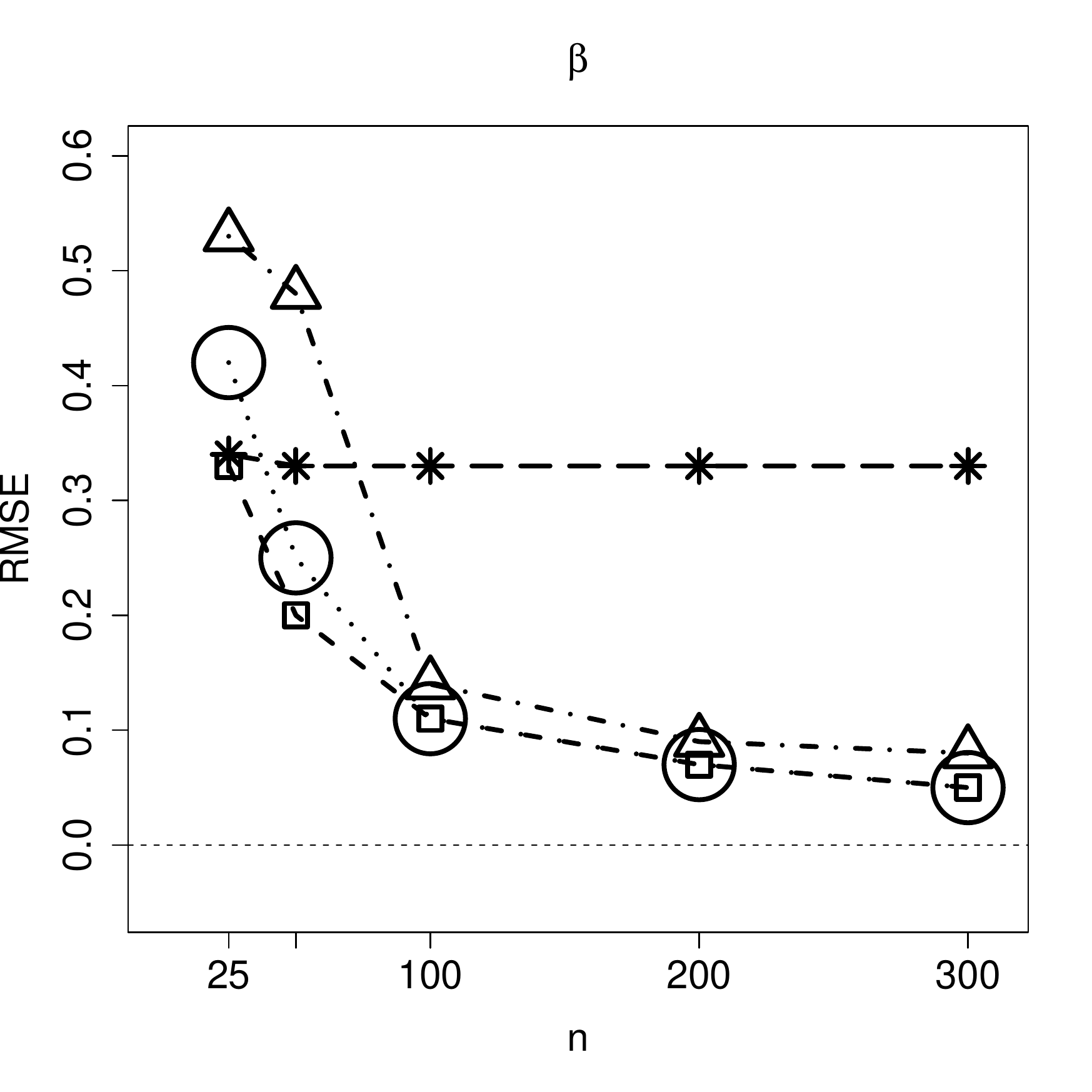}\\
\includegraphics[scale=0.25]{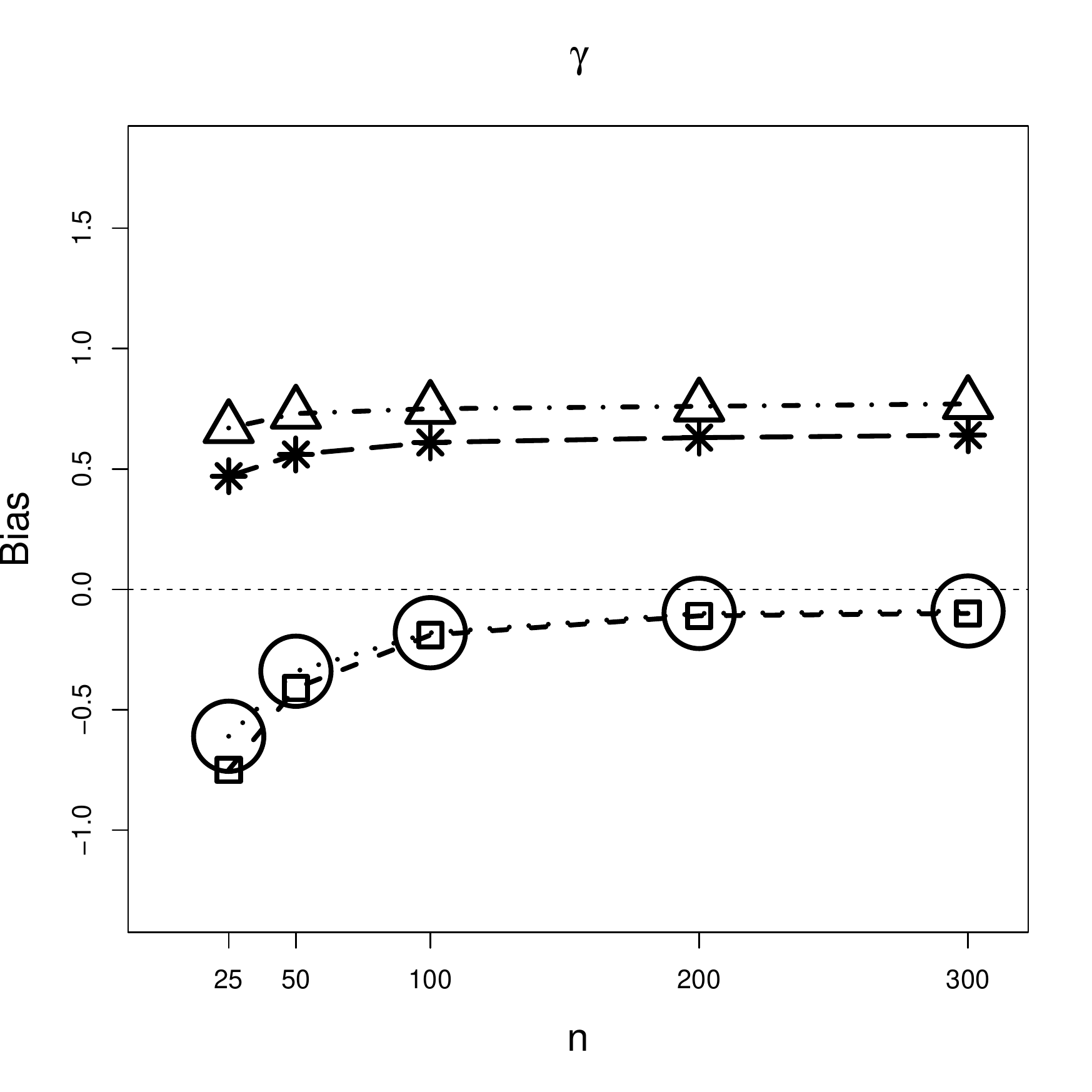}  \qquad \includegraphics[scale=0.25]{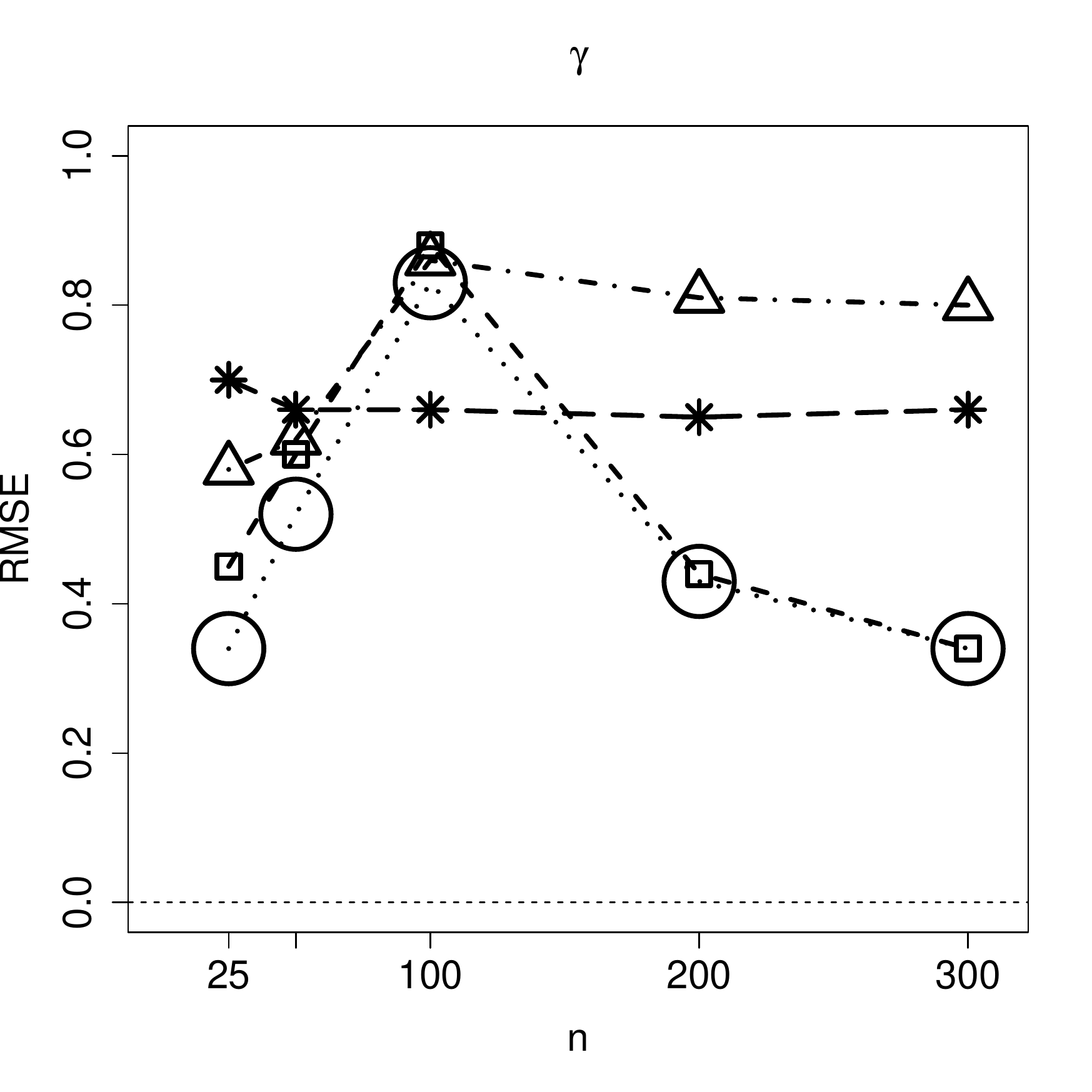}\\
\includegraphics[scale=0.25]{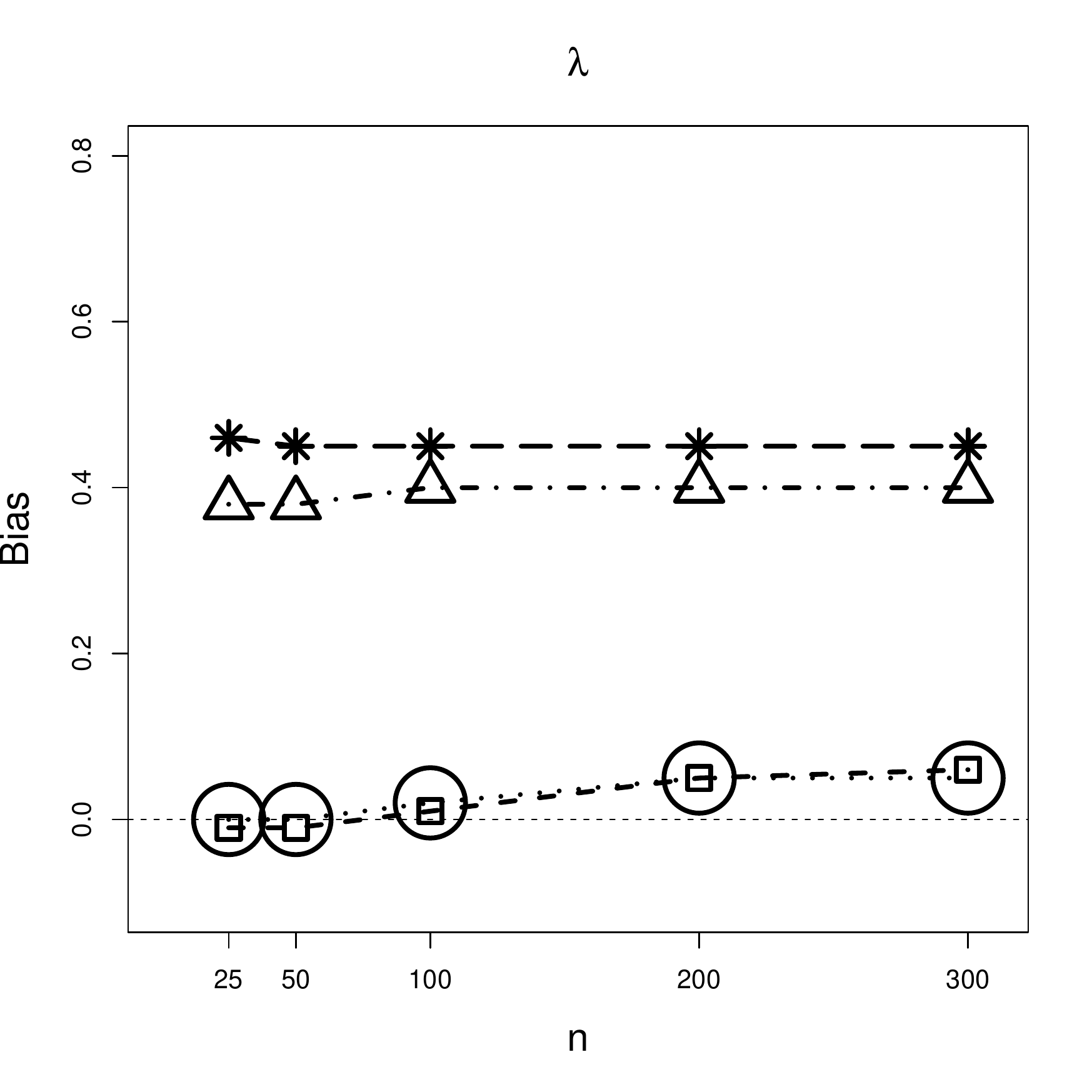} \qquad \includegraphics[scale=0.25]{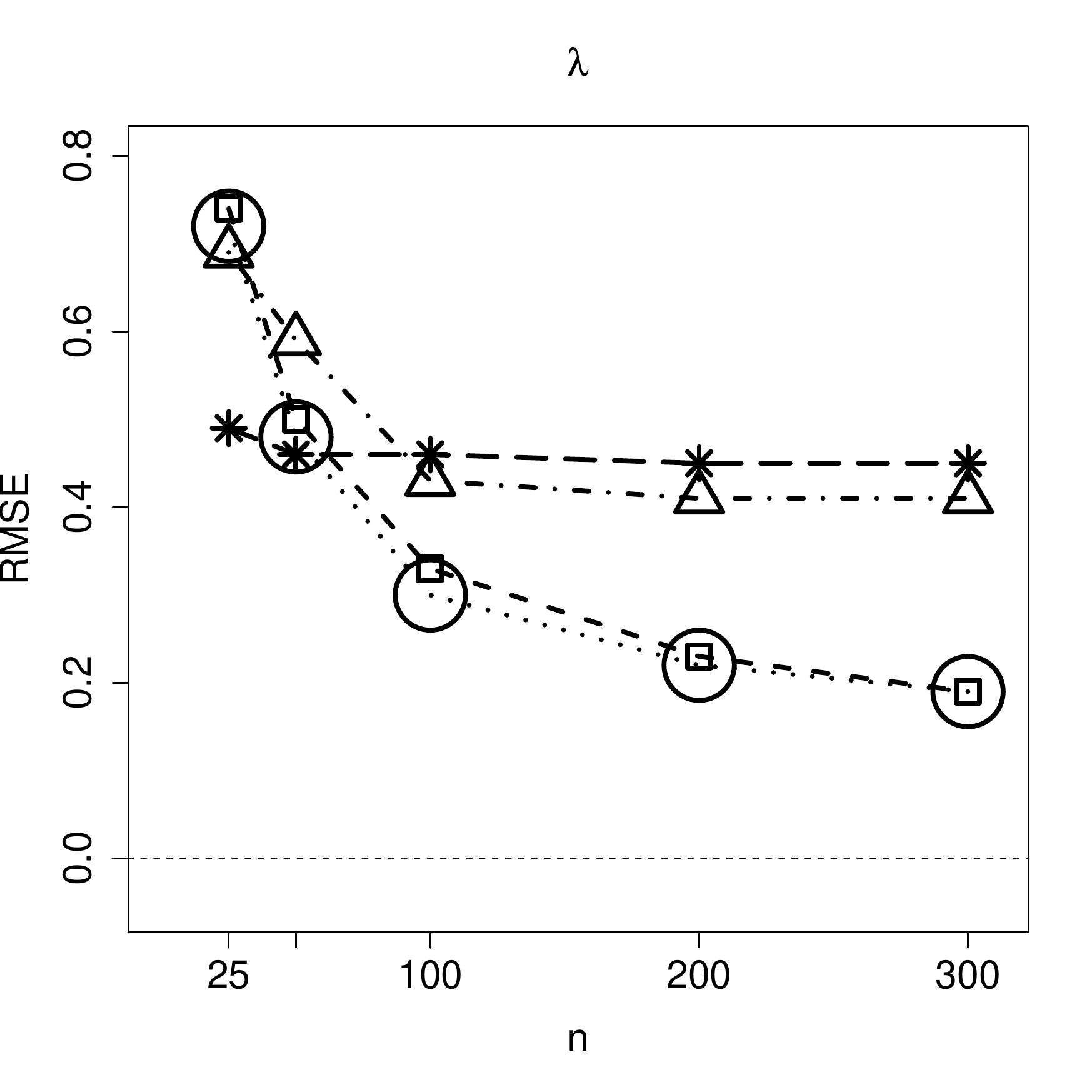}
\caption{Bias and RMSE for the estimators of $\alpha$, $\beta$, $\gamma$ and $\lambda$ for $k_{x}=0.50$, varying precision model; $\ell_{a}$ (square), $\ell_{p}$ (circle), $\ell_{rc}$ (triangle) and $\ell_{naive}$(star).}
\label{Bias050model2}
\end{figure}

\begin{figure}
\vspace{-0.7cm}
\centering
\includegraphics[scale=0.23]{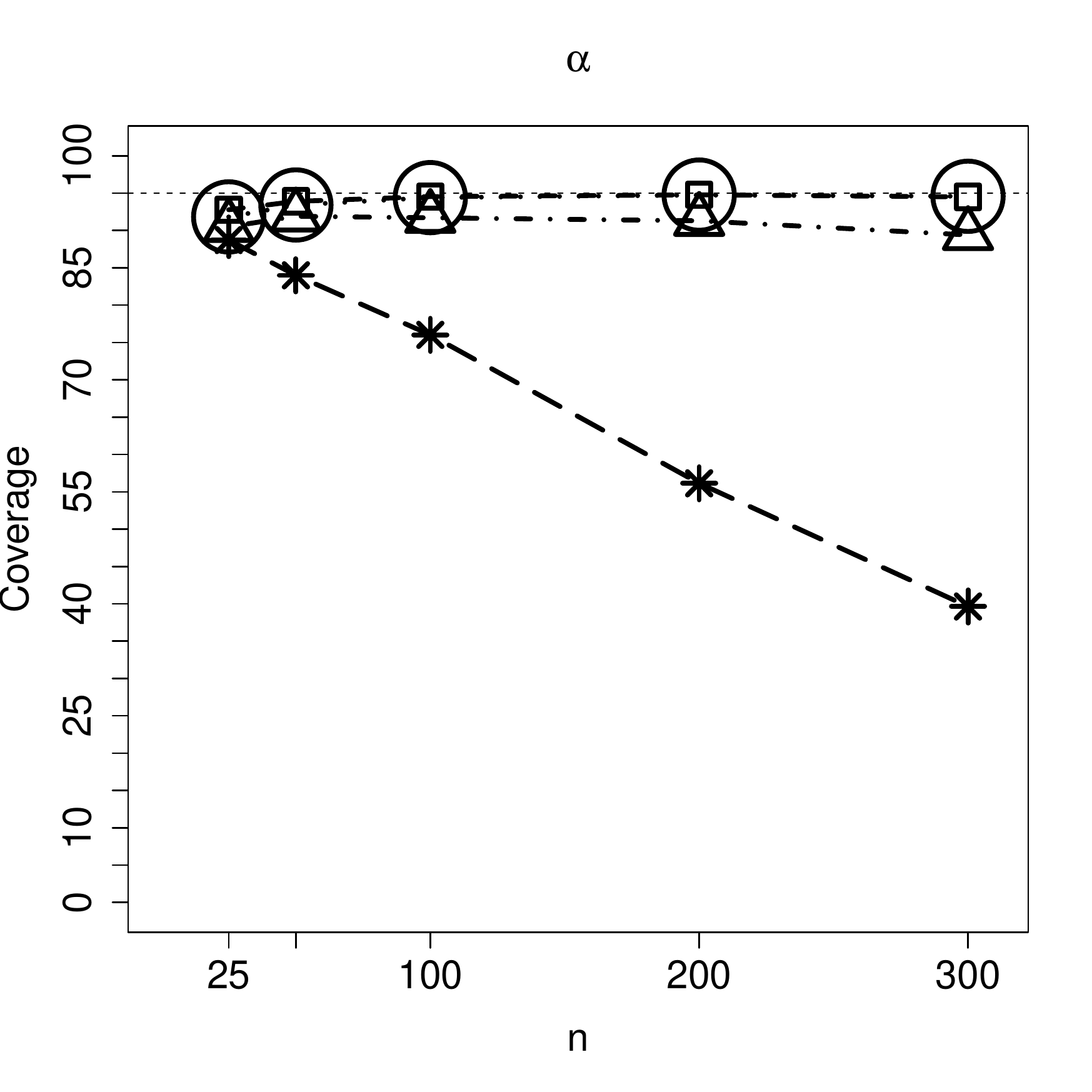} \quad \includegraphics[scale=0.23]{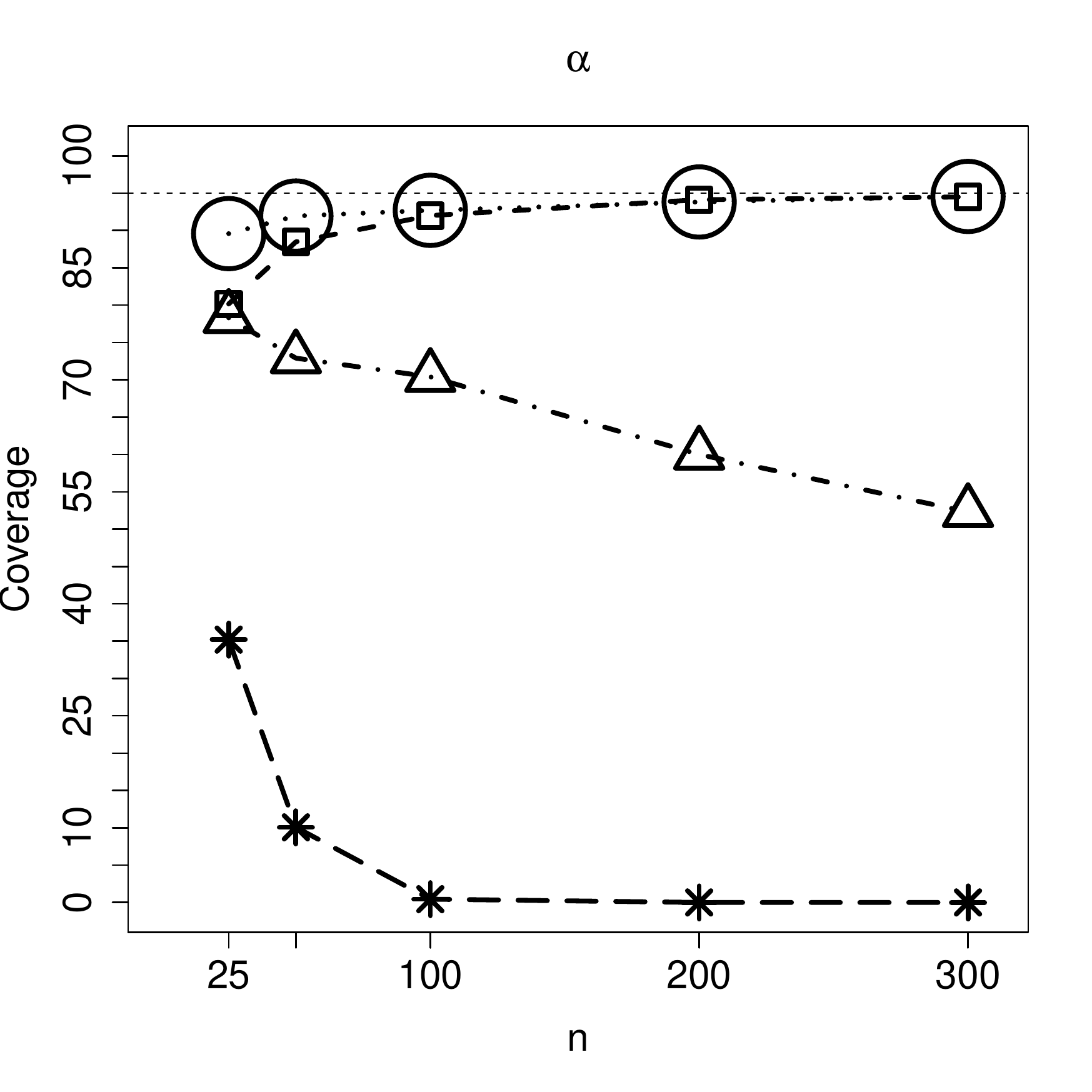} 
\quad \includegraphics[scale=0.23]{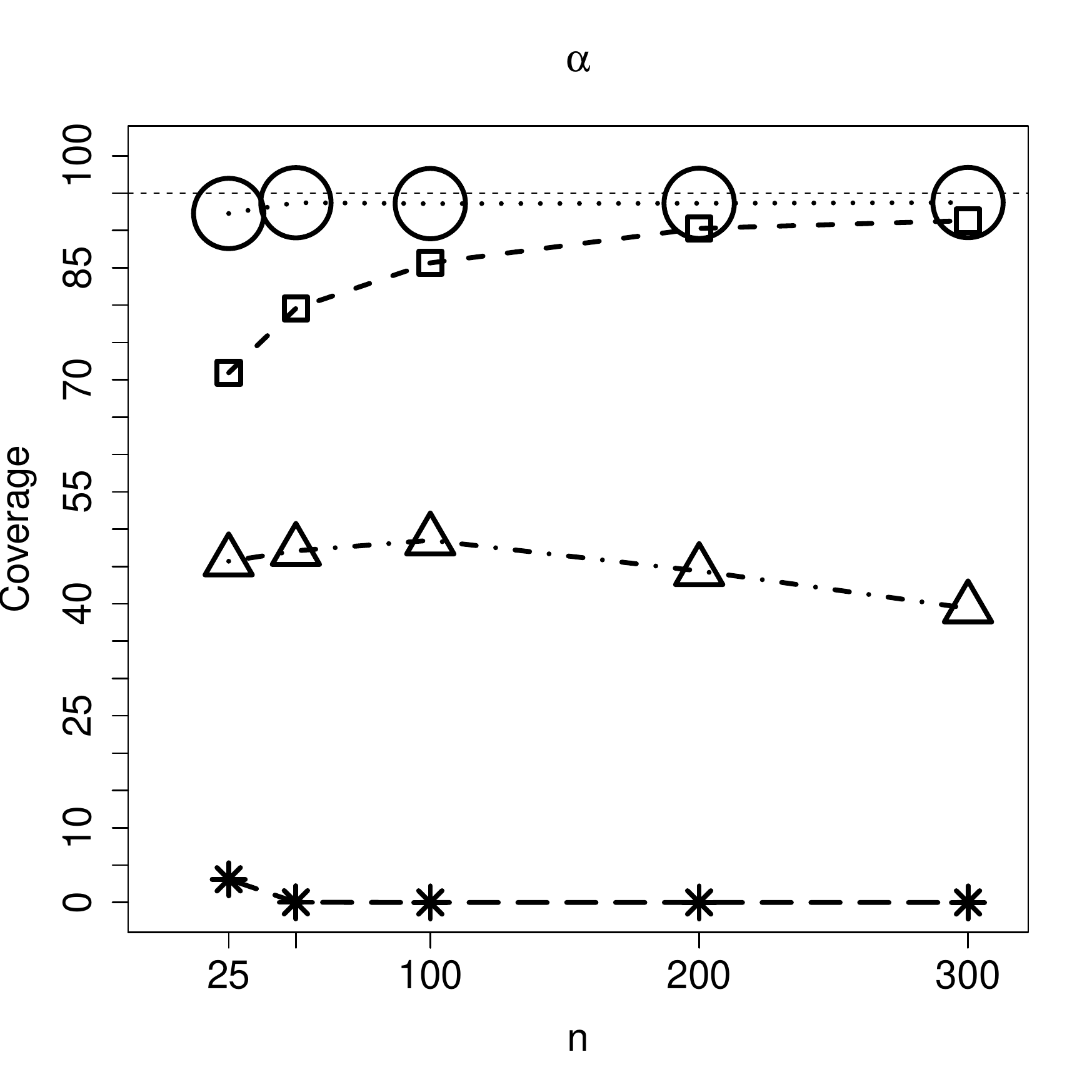}\\
\includegraphics[scale=0.23]{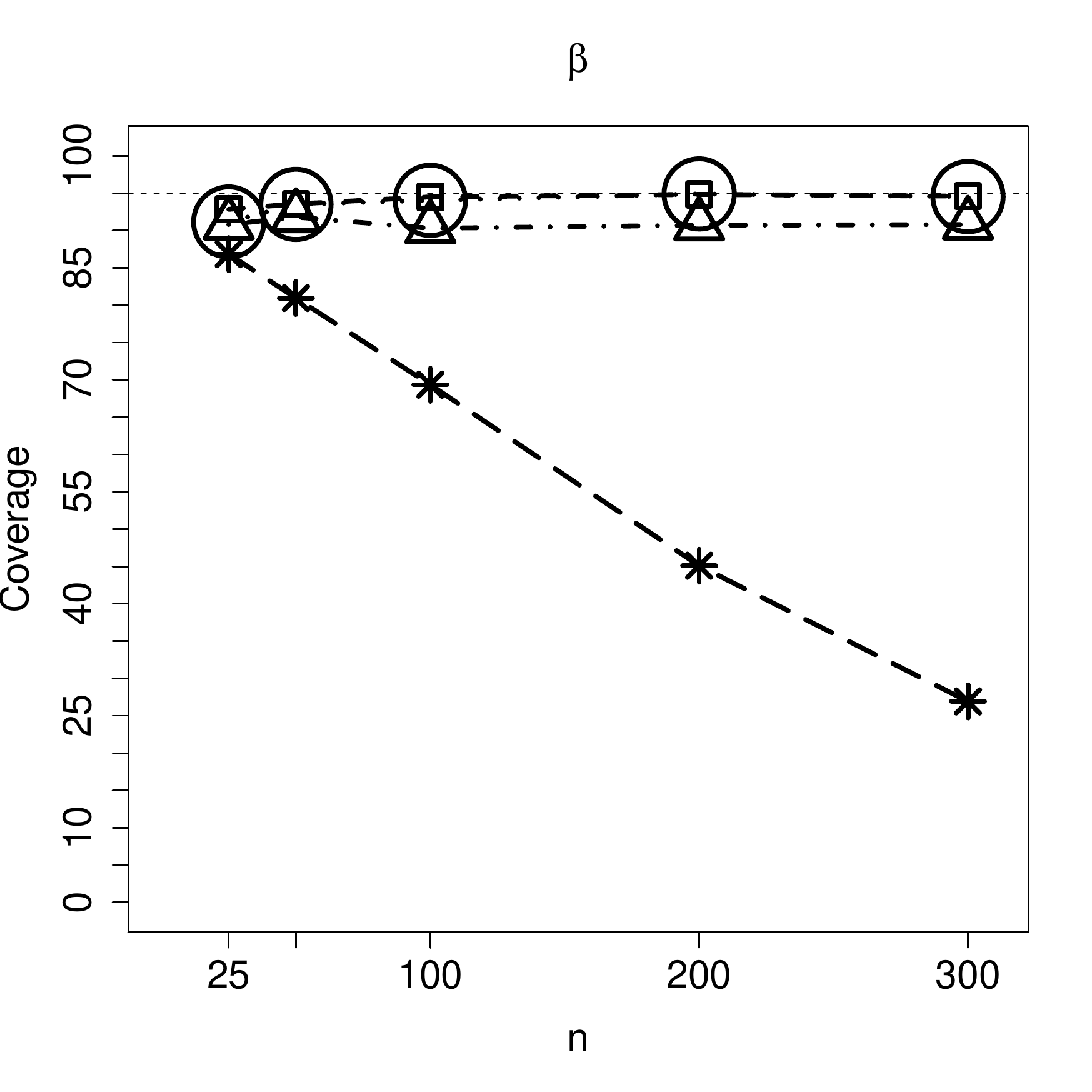} \quad \includegraphics[scale=0.23]{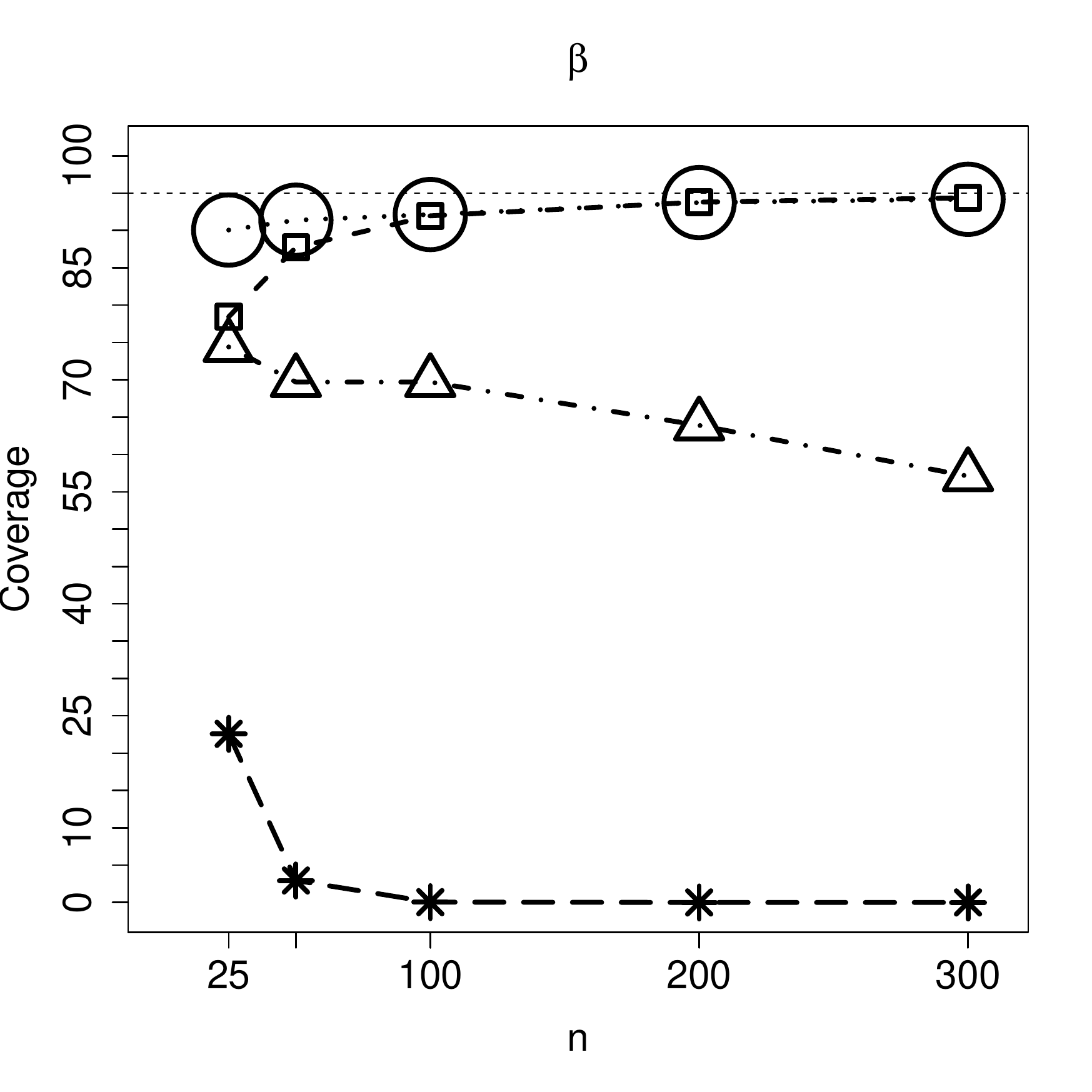} 
\quad \includegraphics[scale=0.23]{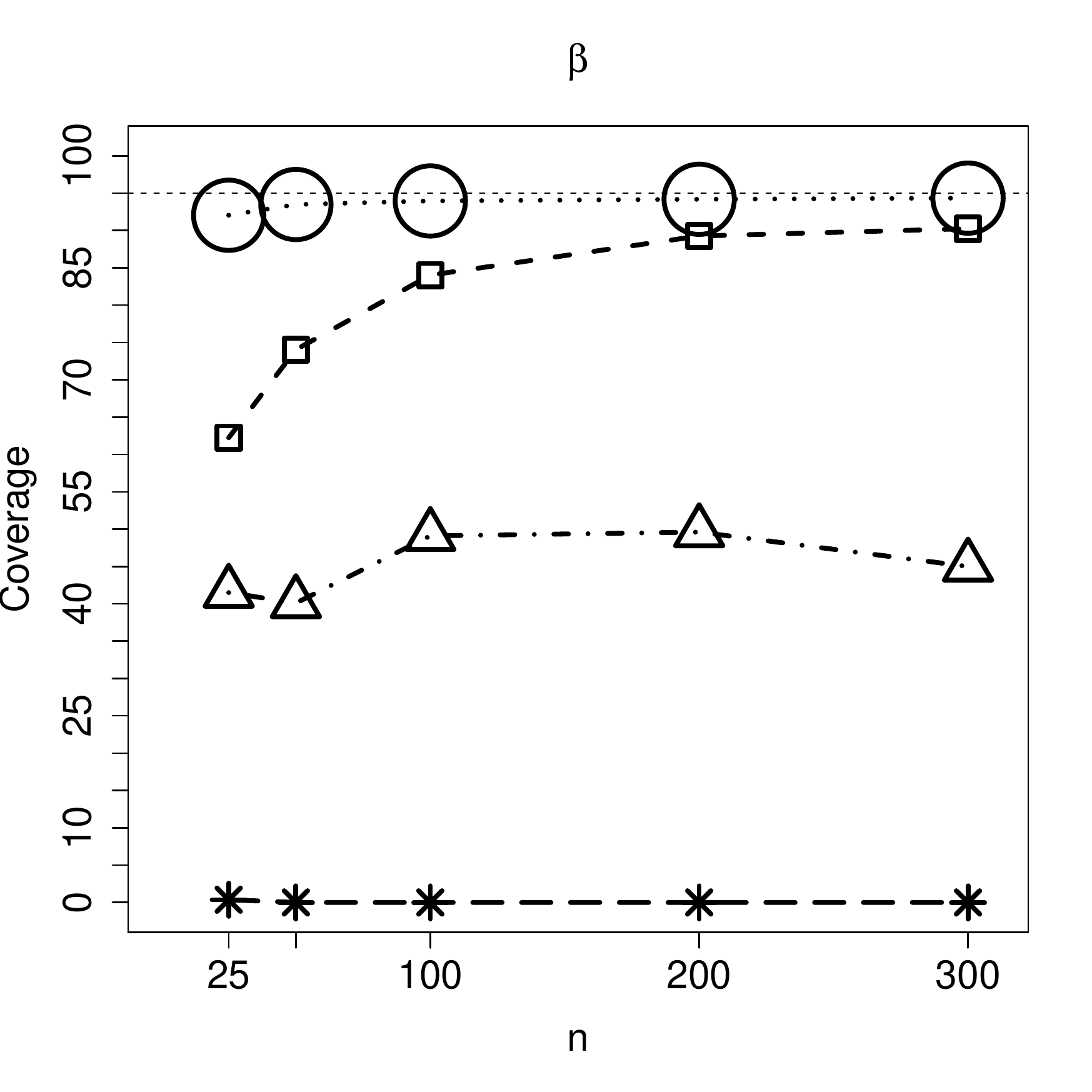}\\
\includegraphics[scale=0.23]{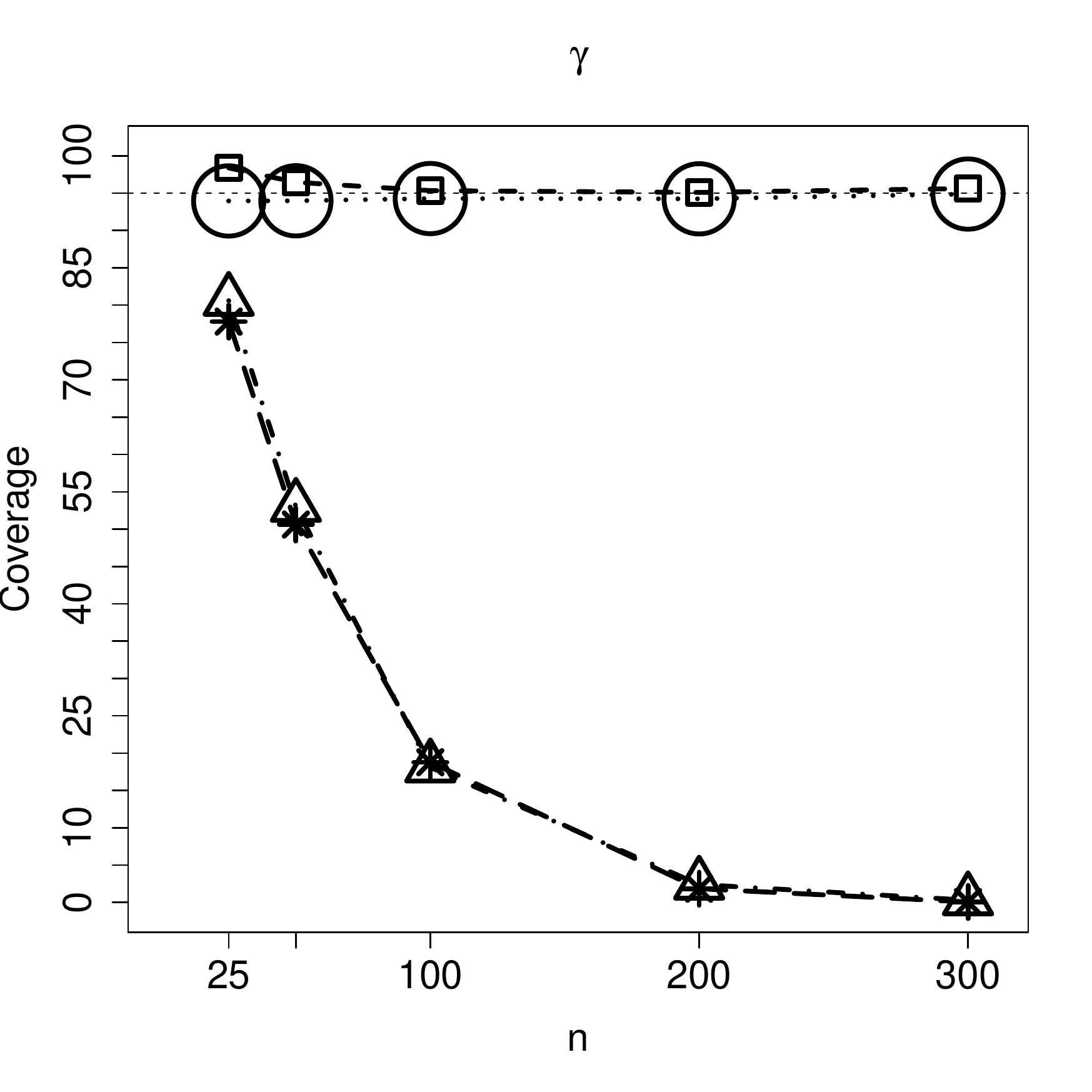} \quad \includegraphics[scale=0.23]{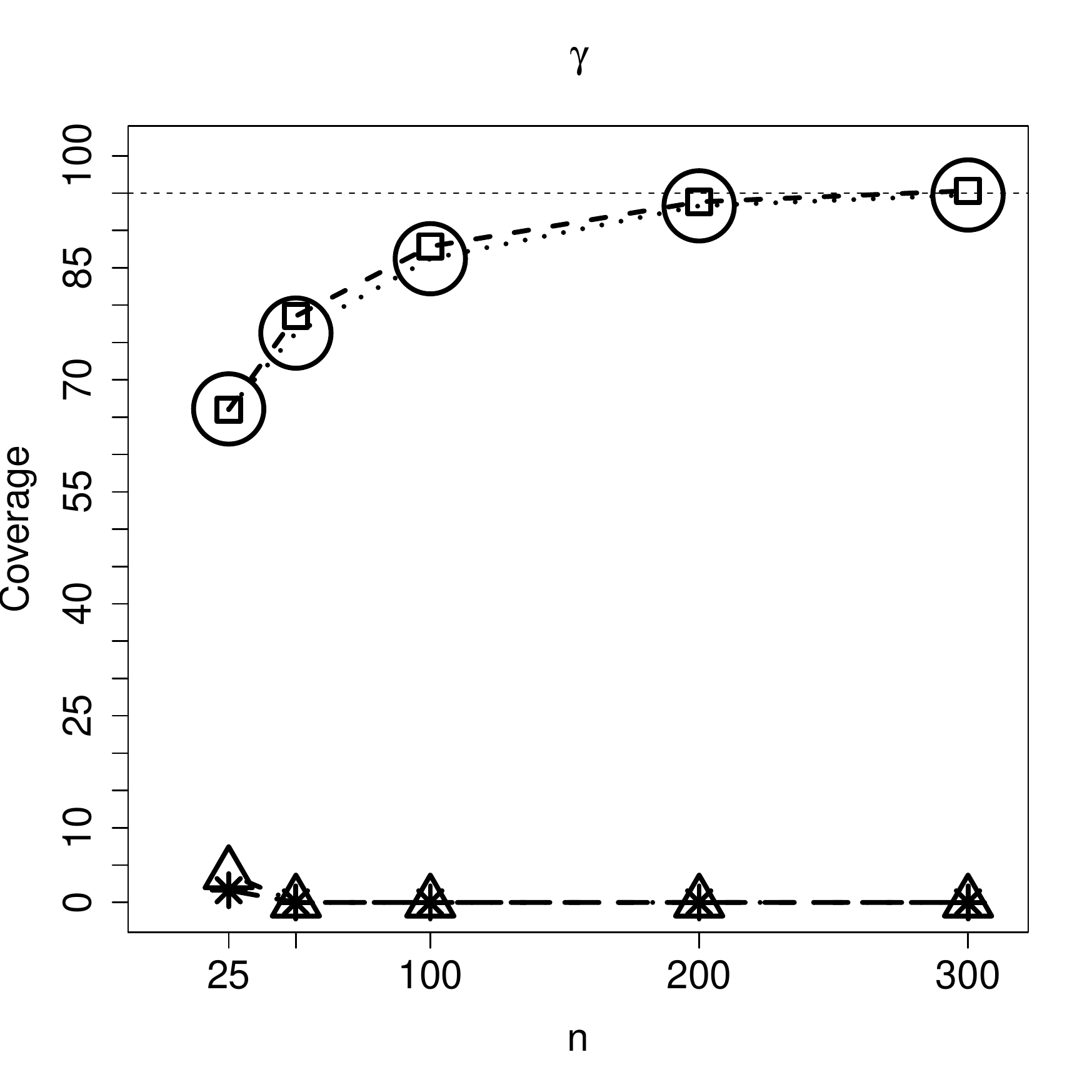}  
\quad \includegraphics[scale=0.23]{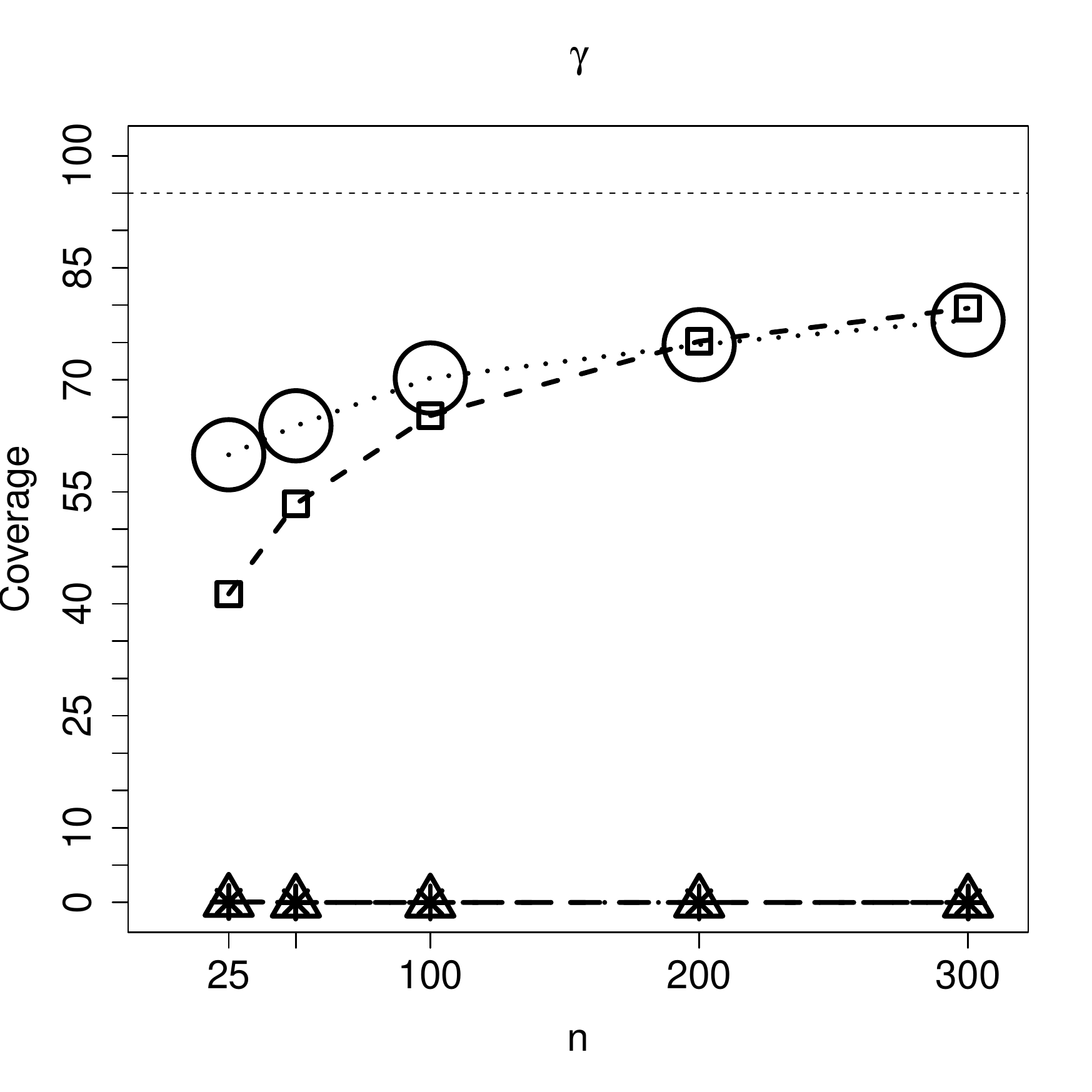}\\
  (a)            \hspace{4cm}       (b)  \hspace{3cm}   (c)            \\
\caption{Coverage of confidence intervals of $\alpha$, $\beta$, $\gamma$ and $\lambda$ for: column (a) $k_{x}=0.95$, constant precision model, column (b) $k_{x}=0.75$, constant precision model, and column (c) $k_{x}=0.50$, constant precision model; $\ell_{a}$ (square), $\ell_{p}$ (circle), $\ell_{rc}$ (triangle) and $\ell_{naive}$(star).}
\label{Coveragemodel1}
\end{figure}

\begin{figure}
\vspace{-0.7cm}
\centering
\includegraphics[scale=0.23]{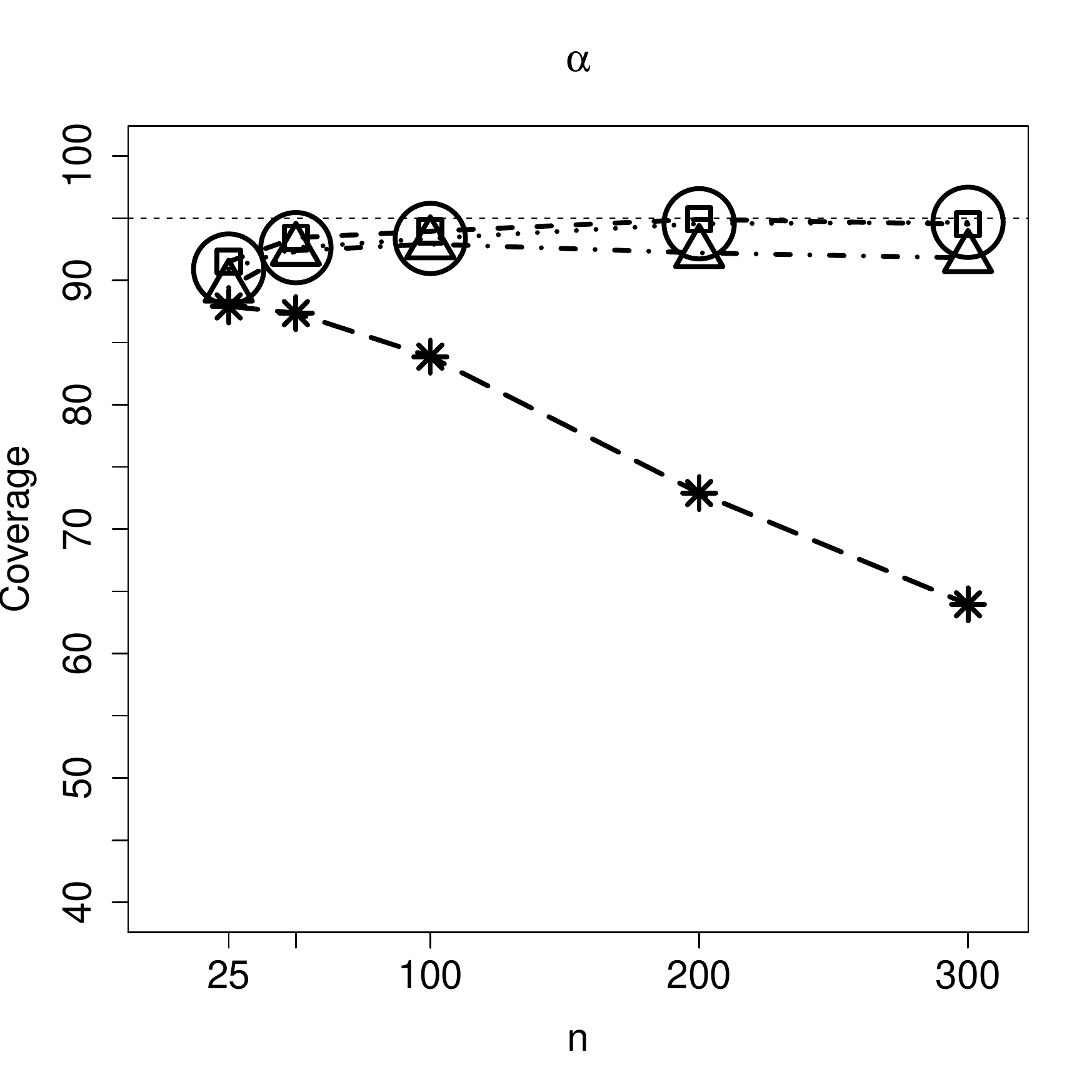} \quad \includegraphics[scale=0.23]{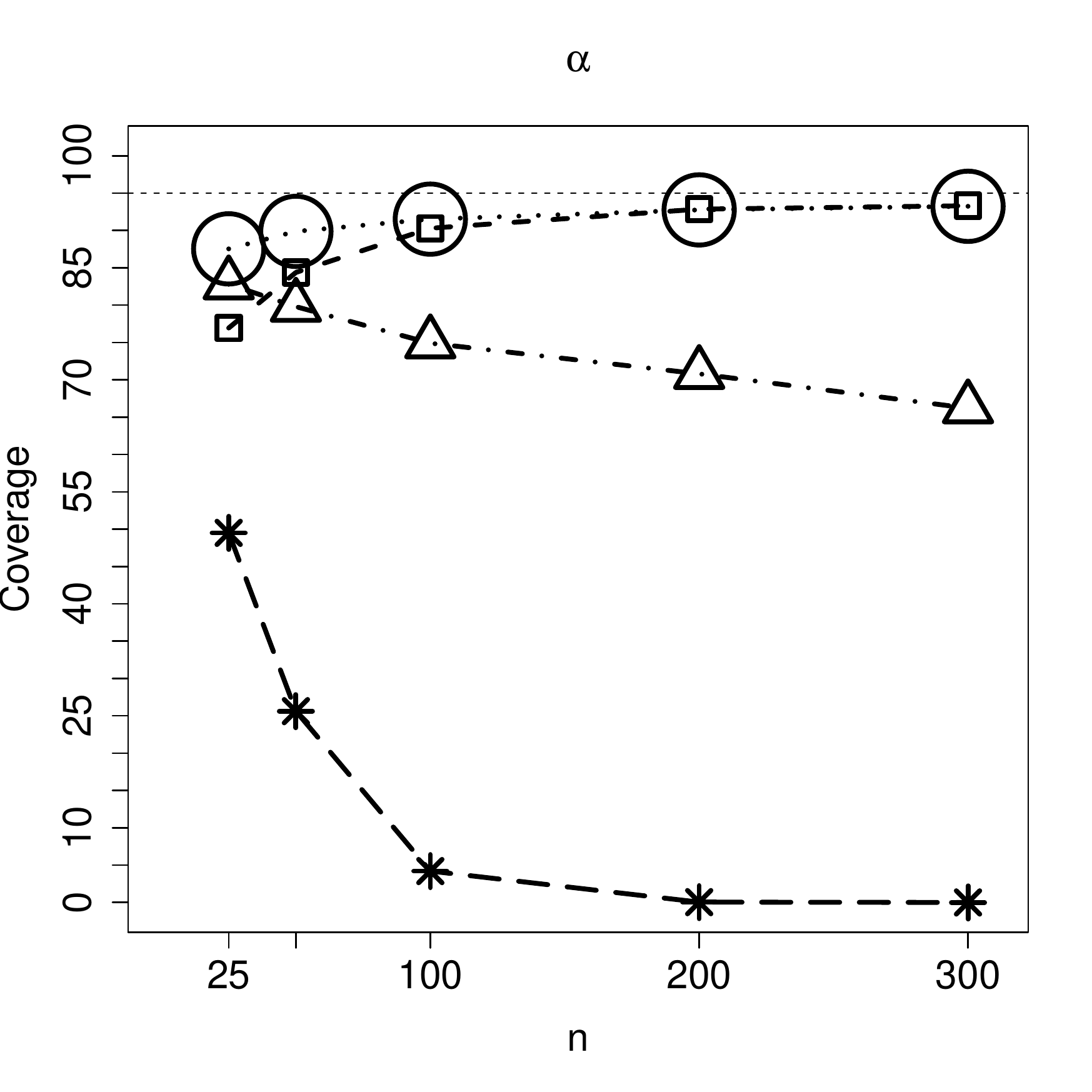} 
\quad \includegraphics[scale=0.23]{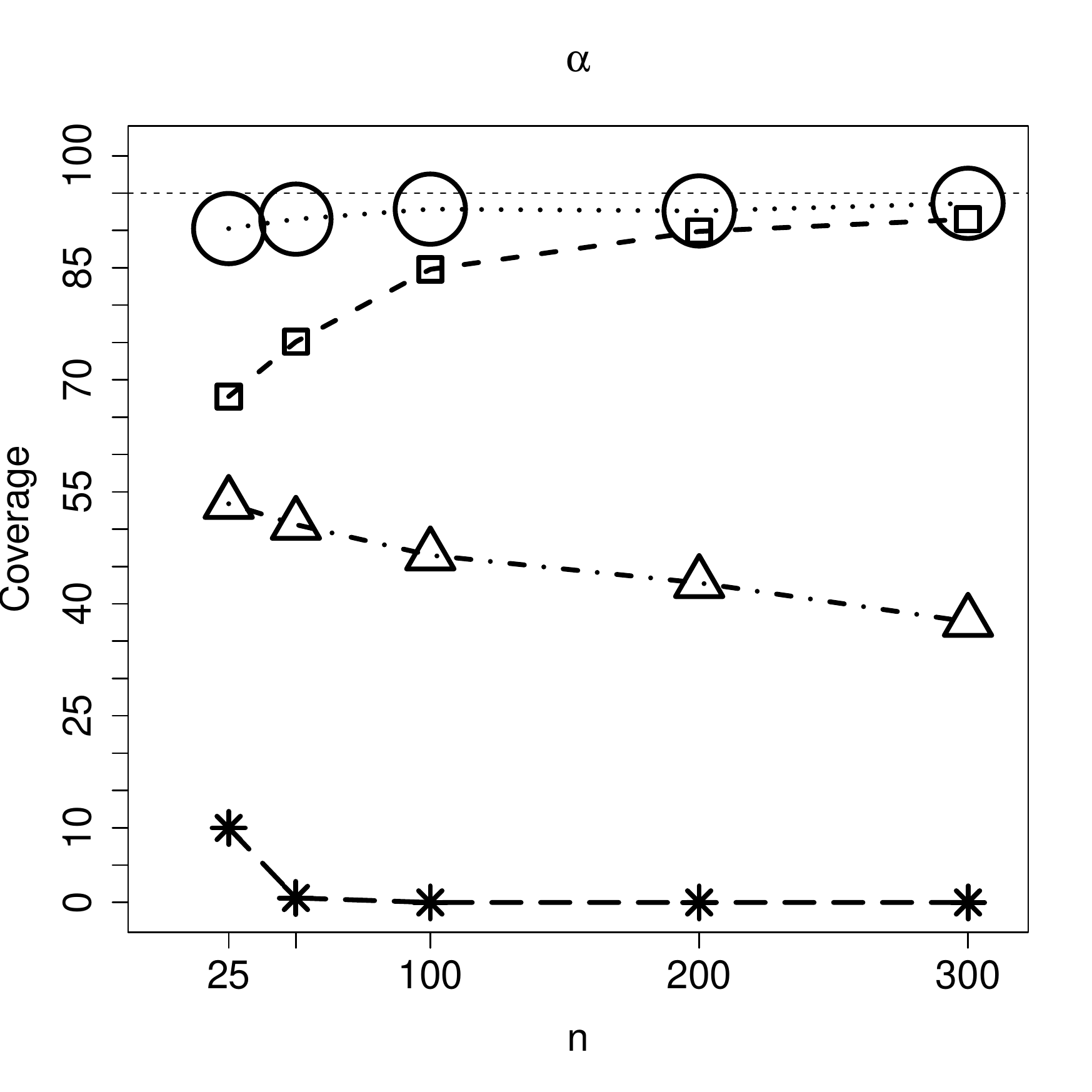}\\
\includegraphics[scale=0.23]{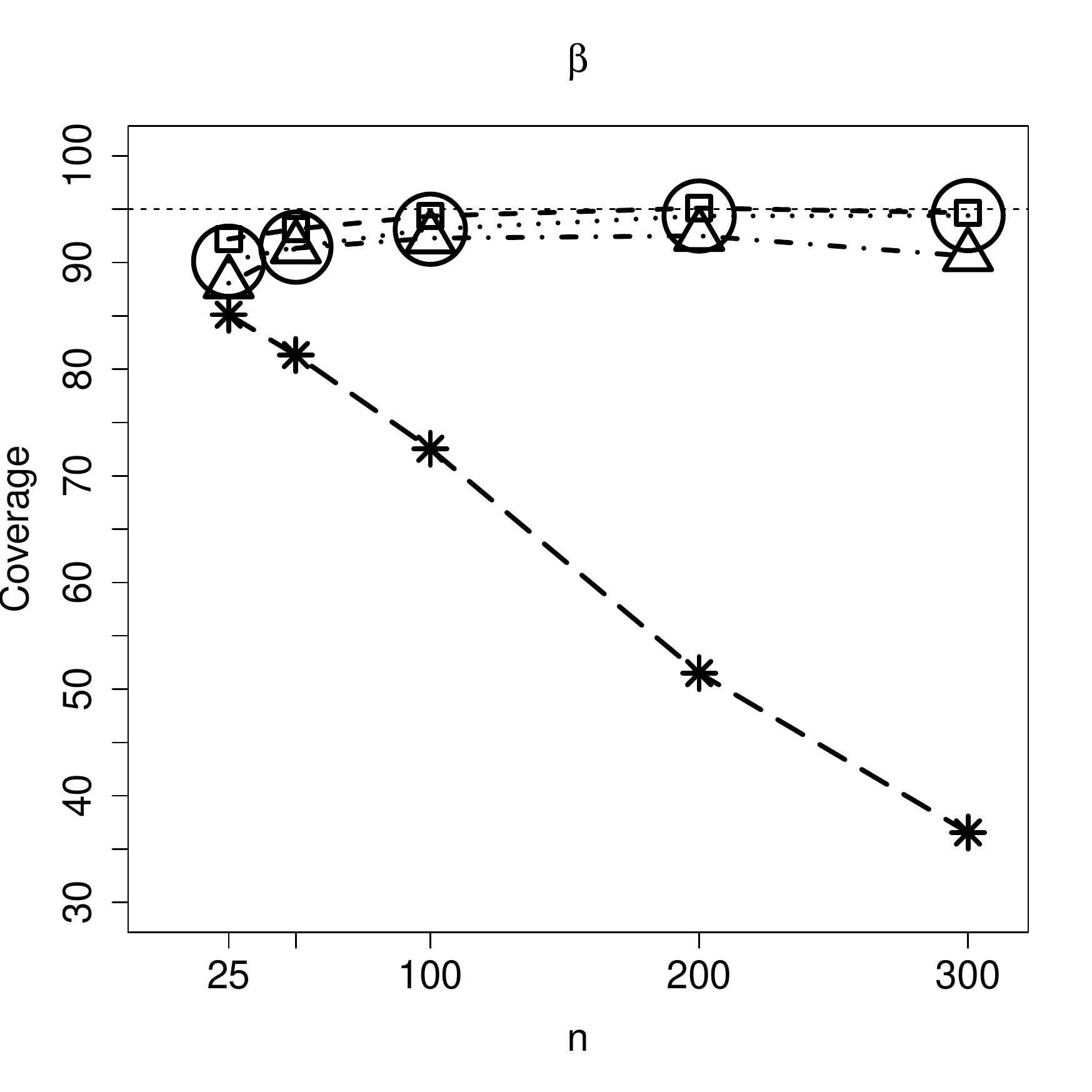} \quad \includegraphics[scale=0.23]{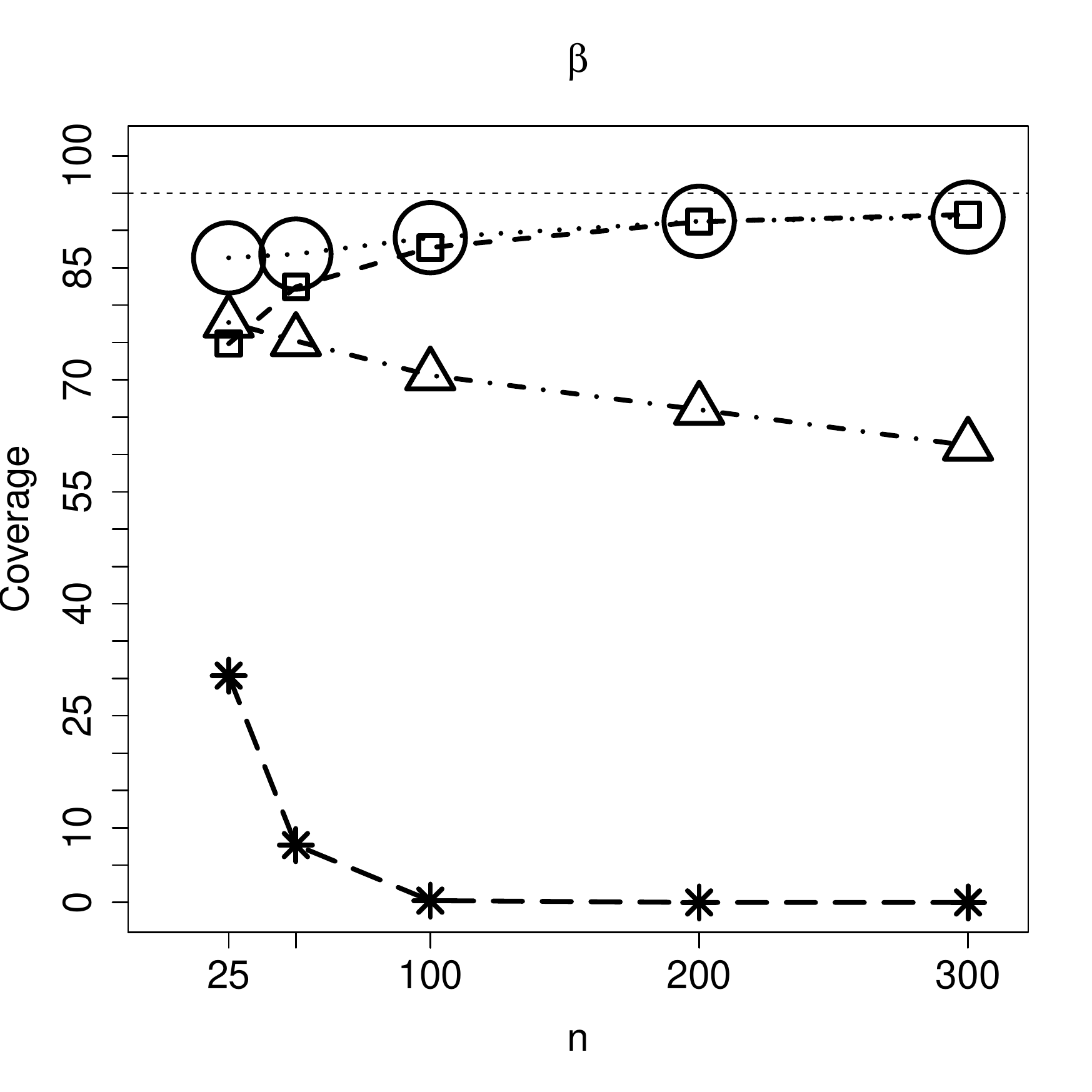} 
\quad \includegraphics[scale=0.23]{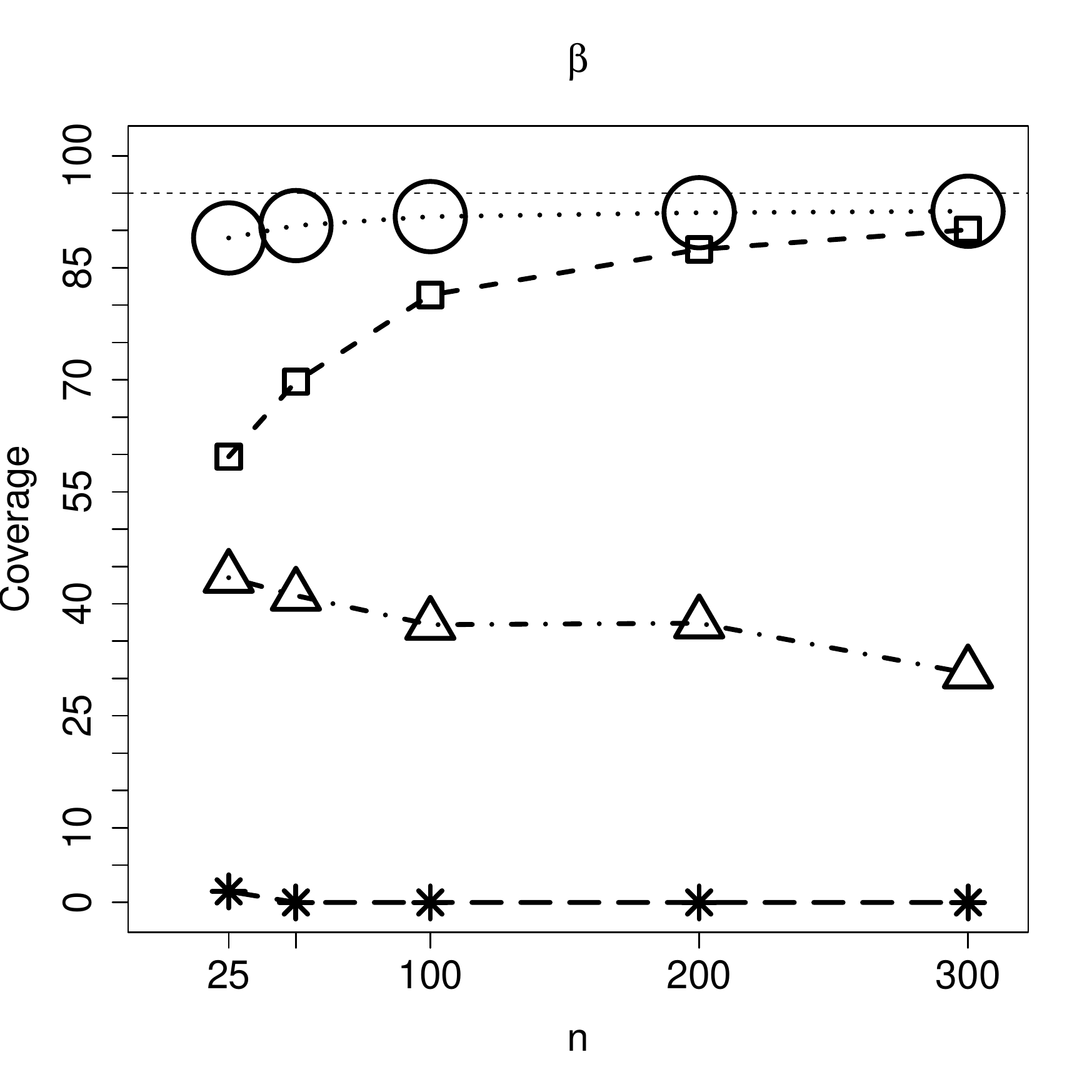}\\
\includegraphics[scale=0.23]{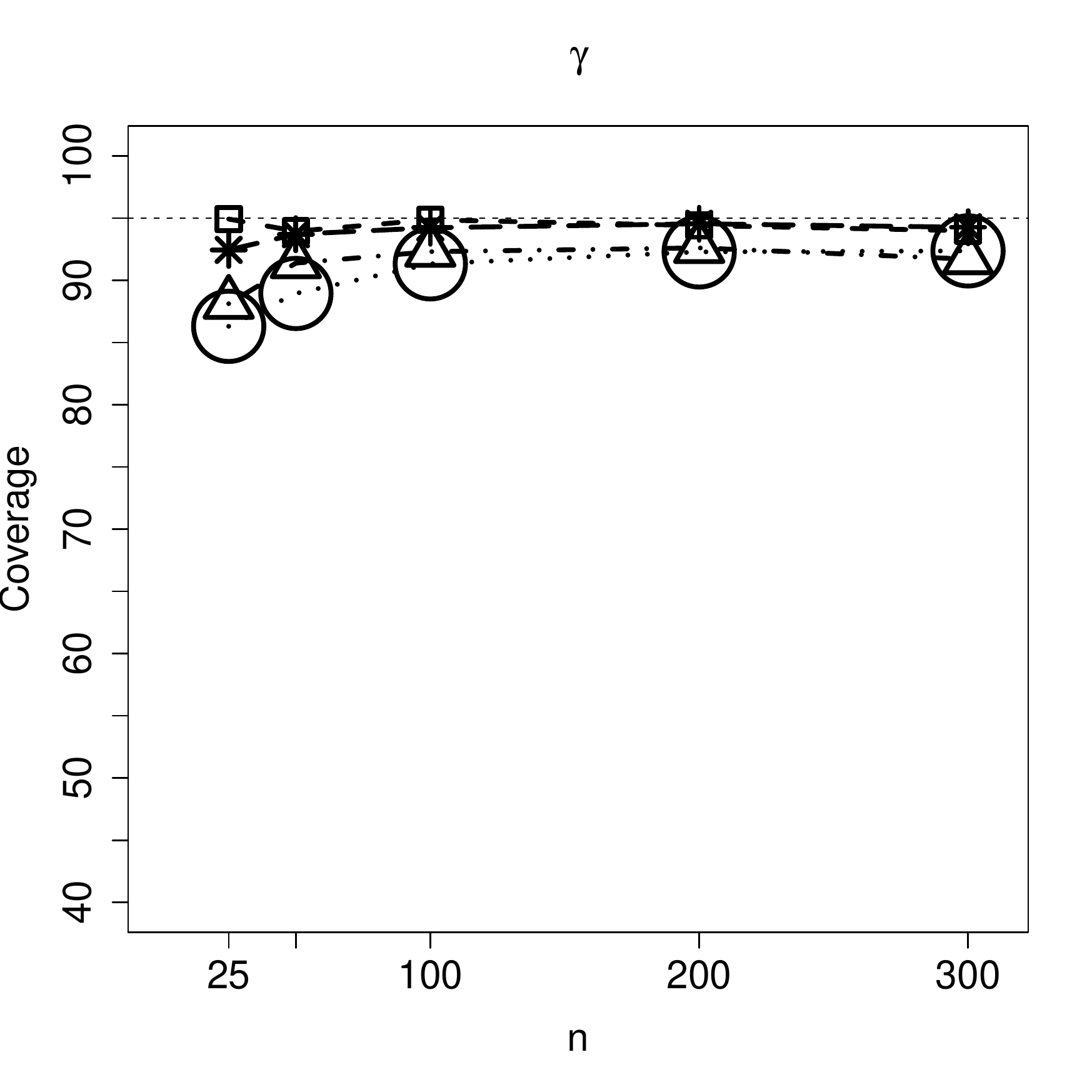} \quad \includegraphics[scale=0.23]{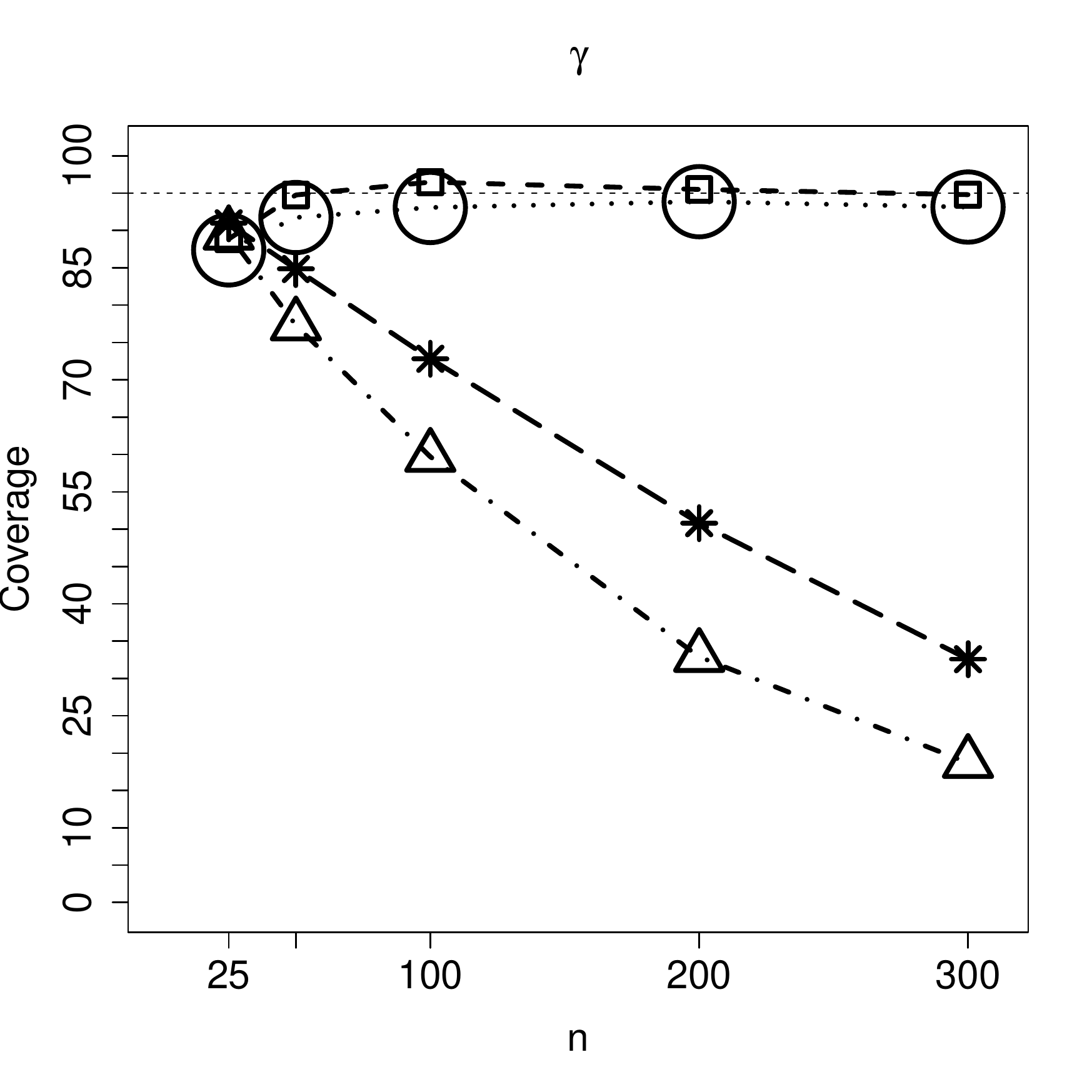}  
\quad \includegraphics[scale=0.23]{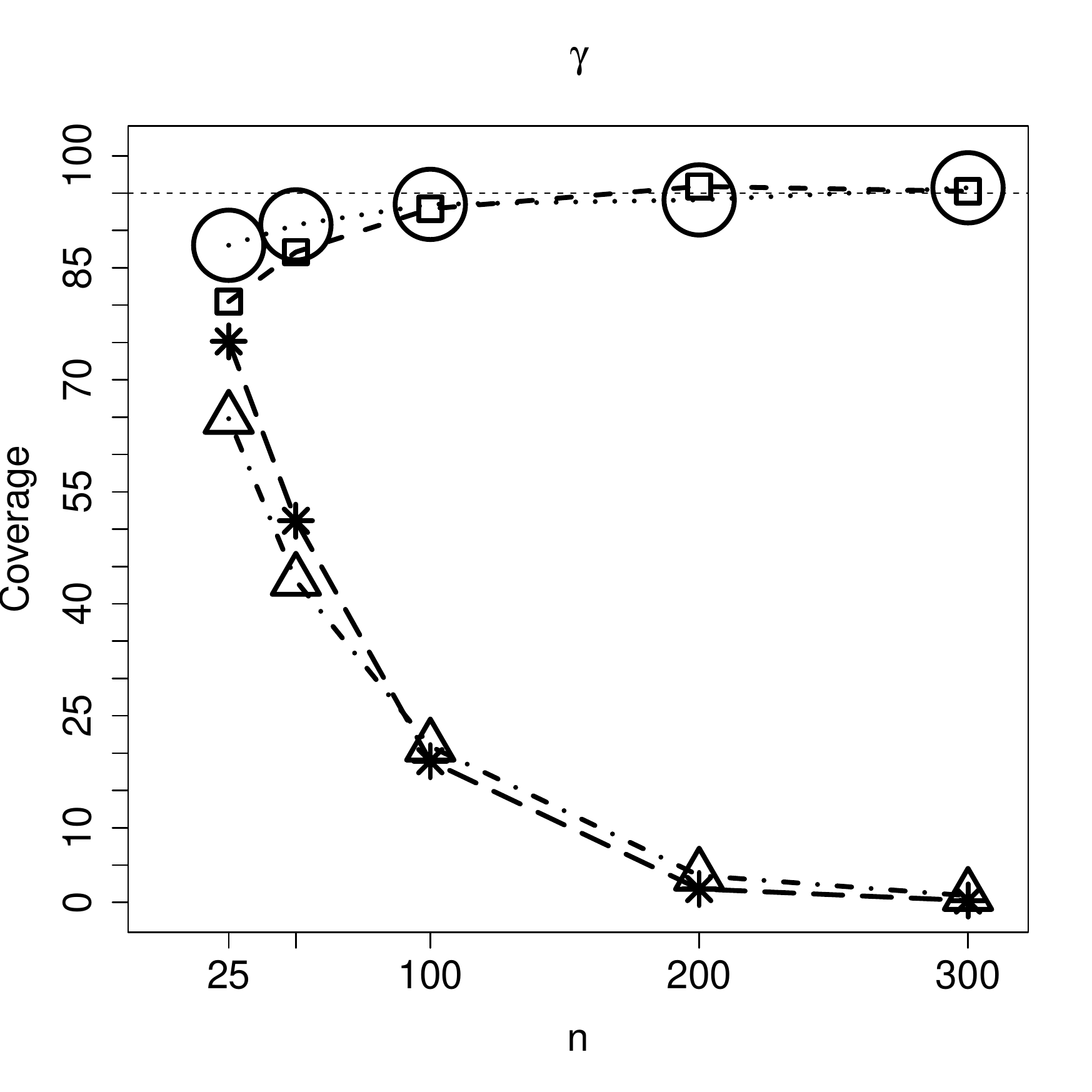}\\
\includegraphics[scale=0.23]{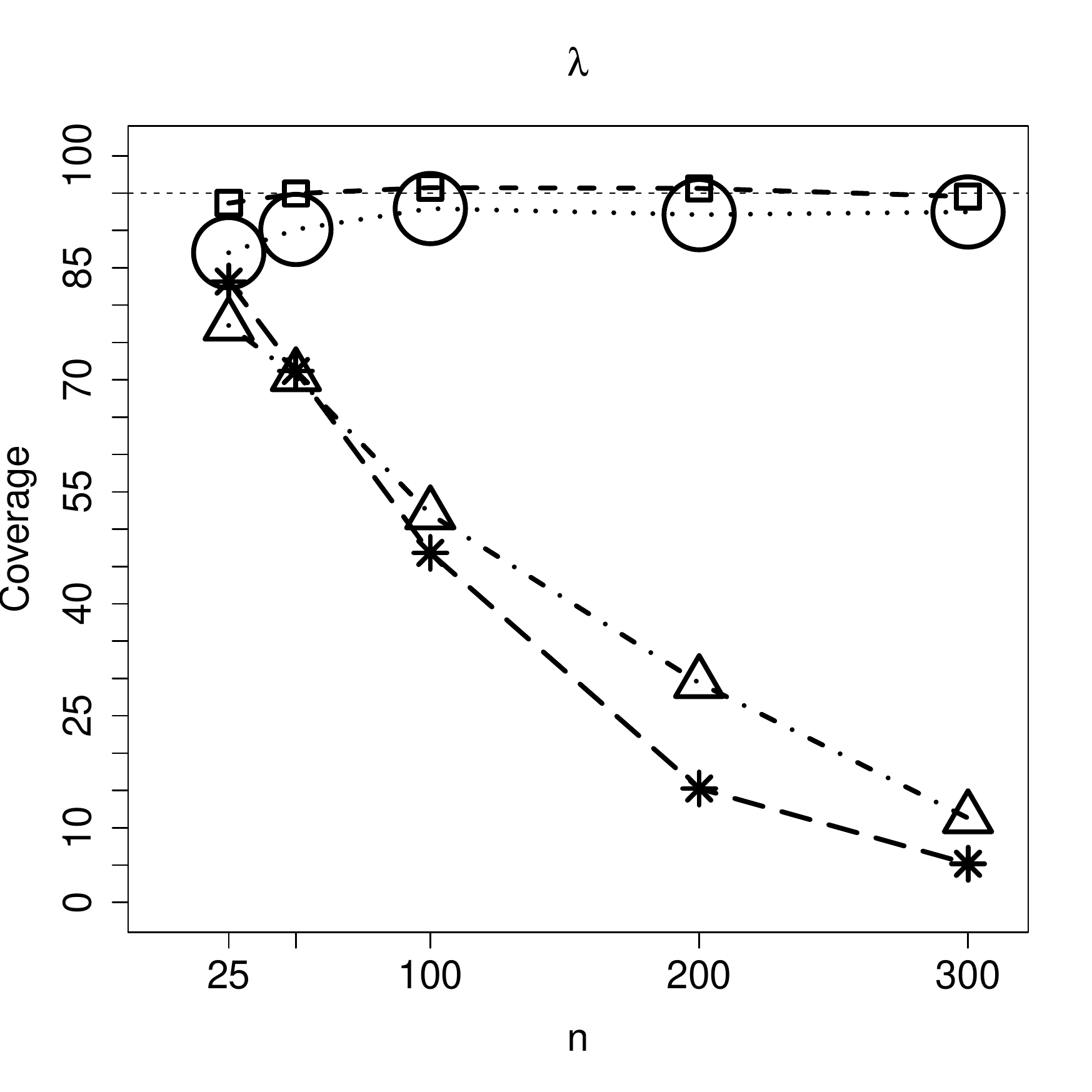} \quad \includegraphics[scale=0.23]{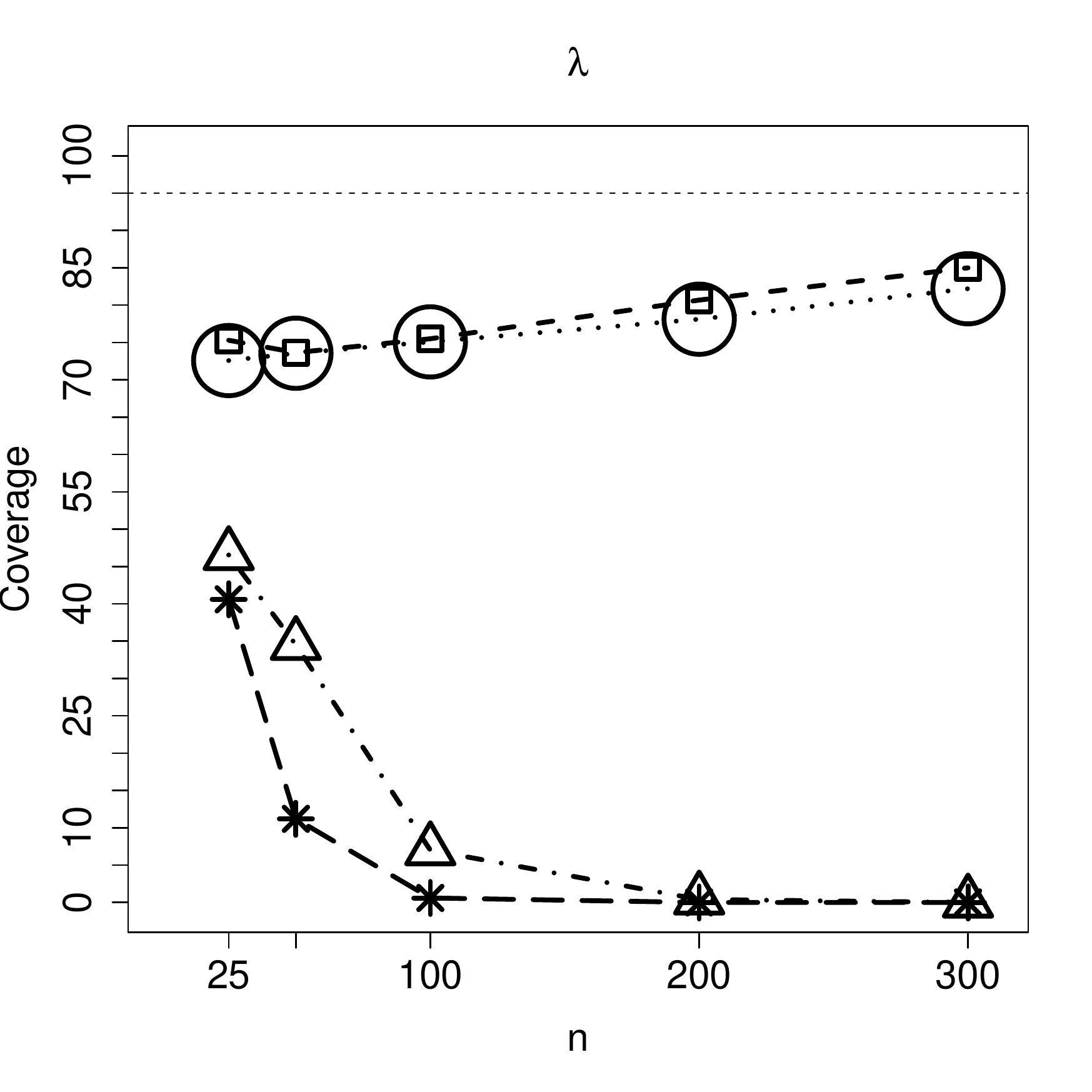} 
\quad \includegraphics[scale=0.23]{TCGama050.pdf}\\
  (a)            \hspace{4cm}       (b)  \hspace{3cm}   (c)            \\
\caption{Coverage of confidence intervals of $\alpha$, $\beta$, $\gamma$ and $\lambda$ for: column (a) $k_{x}=0.95$, varying precision model, column (b) $k_{x}=0.75$, varying precision model, and column (c) $k_{x}=0.50$, varying precision model; $\ell_{a}$ (square), $\ell_{p}$ (circle), $\ell_{rc}$ (triangle) and $\ell_{naive}$(star).}
\label{Coveragemodel2}
\end{figure}

\section{Residual analysis}
\label{sec:resid}

\citet{Espinheira02} and \citet{Ferrari+Espinheira+CribariNeto} proposed the use of standardized weighted residuals
as a diagnostic tool for beta regression models with constant and non-constant precision parameter, respectively.
Here, we modify the residuals defined by \citet{Ferrari+Espinheira+CribariNeto} to allow for measurement error in
covariates. We therefore define
\begin{eqnarray*}
r_{t}=\frac{y^{*}_{t}-\widehat{\mu}_{t}^{*}}{\sqrt{\widehat{\upsilon}_{t}(1-\widehat{h}^{*}_{tt})}},
\end{eqnarray*}
for $t=1,\ldots,n$, where 
%$y_{t}^{*}$ and $\mu_{t}^{*}$ are given in (\ref{y*}),
%\begin{eqnarray} \label{y*}
$y_t^{*}=\log({y_t}/{(1-y_t)})$,  $\mu^*=\psi(\mu_t\phi) - \psi[(1-\mu_t)\phi]$
%\end{eqnarray}
with $\psi(\cdot)$ being the digamma function, i.e., $\psi(z) =
{\rm d} \log \Gamma(z)/ {\rm d}z$ for $z > 0$,
$\upsilon_{t}=\psi^{'}(\mu_{t}\phi_{t})+\psi^{'}[(1-\mu_{t})\phi_{t}]$ and $h^{*}_{tt}$
is the $t$-th diagonal element of
\begin{eqnarray*}
\vm{H}^{*}=(\vm{M}\vg{\Phi})^{1/2}\vm{W}(\vm{W}^{\top}\vg{\Phi}\vm{M}\vm{W})^{-1}\vm{W}^{\top}(\vg{\Phi}\vm{M})^{1/2},
\end{eqnarray*}
in which $\vm{M}=\textrm{diag}\{m_{1},\ldots,m_{n}\}$ with $m_{t}=\phi_{t}\upsilon_{t}/[g^{'}(\mu_{t})]^{2}$, $\vg{\Phi}=\textrm{diag}\{\phi_{1},\ldots,\\ \phi_{n}\}$ and $\vm{W}$ is an $n \times (p_\alpha+p_\beta)$ matrix with the $t$-th row
given by $(\vm{z}_t^\top, \vm{w}_t^\top)$.
Here, hat indicates that the unknown parameters are replaced by estimates. We suggest the use of the
maximum pseudo-likelihood estimates since they performed well in our simulations and are computationally less
demanding than the approximate maximum likelihood estimates.

%Quando uma variável é considerada substituta ou apresenta erro de medição, é sugerido por \citet[Seção 2.2.2]{Fuller}, \citet{Carroll+Spiegelman} e \citet[Seção 4.7]{Buonaccorsi} avaliar o $t$-ésimo resíduo, baseado na estimativa corrigida (estimativa obtida quando consideramos a presença de erro de medição na variável) e nos valores observados. Utilizamos então, neste artigo, os resíduos padronizados padronizados proposta por \citet{Ferrari+Espinheira+CribariNeto}, sendo que as estimativas dos parâmetros são obtidos a partir do método de máxima pseudo-verossimilhança aproximada.

Plots of residuals versus observation indices are not always suitable for detecting lack of fit when measurement errors
are present; see, for instance, \citet[Section 2.2.2]{Fuller}, \citet{Carroll+Spiegelman} and \citet[Section 4.7]{Buonaccorsi}.
\citet[Section 2.2.2]{Fuller} suggests constructing plots of residuals versus consistent estimates of the expected
value of $x_{t}$ given $w_{t}$. \cite{Atkinson}, on the other hand, suggests the use of
simulated envelopes in normal probability plots to facilitate their interpretation.
%Os envelopes simulados são bandas de confiança obtidas por métodos de Monte Carlo a partir do modelo ajustado para avaliar a existência de afastamentos sérios da distribuição proposta.
The use of standardized weighted residuals plots proposed here will be illustrated in the next section.

\section{Real data application}
\label{apli:corazon}

We now illustrate our results in the dataset described in Section \ref{sec:intro}. Our aim is to model $RPP$ (relative pulse pressure) using $BMI$ (body mass index) and $LDL$ (low density lipoprotein cholesterol level) as covariates.
The total cholesterol level, $TC$, is easier to be accessed  than $LDL$, and provides a measure of $LDL$ plus unknown
quantities of other components as triglycerides and high density lipoprotein. Here, we shall consider $TC$ as a surrogate for $LDL$.

We shall assume that $y_{1}, y_{2}, \ldots, y_{n}$ are independent observations of the $PPR$ of the $n$ individuals in the sample,
and such that $y_{t}$ follows a beta distribution with mean $\mu_{t}$ and precision parameter $\phi_{t}$, with
\begin{eqnarray}
\label{mod1}
\log[\mu_{t}/(1-\mu_{t})]&=&\alpha_{0}+\alpha_{1}\textrm{BMI}_{t}+\beta \textrm{LDL}_{t}, \nonumber\\
\log(\phi_{t})&=&\gamma_{0}+\gamma_{1}\textrm{BMI}_{t}+\lambda \textrm{LDL}_{t},\nonumber\\
\textrm{TC}_{t}&=&\tau_{0}+\tau_{1}\textrm{LDL}_{t}+e_{t},\\
\textrm{LDL}_{t}& \stackrel{{\rm ind}}\sim& N(\mu_{x},\sigma_{x}^{2}), \ \ \ e_{t} \stackrel{{\rm ind}}\sim N(0,\sigma_{e}^{2}), \nonumber
\end{eqnarray}
for $t=1,\ldots,182$. Also, $\textrm{LDL}_{t}$ and $e_{t'}$, for $t,t'=1,\dots,n$ are assumed to be independent.
Here, the interest parameter vector is
$\vg{\theta}=(\alpha_{0},\alpha_{1},\beta,\gamma_{0},\gamma_{1},\lambda)^{\top}$.

For this particular dataset, the values of $LDL$ and $TC$ are available for all the individuals in the sample. We then use these data
to mimic a situation where both the true and surrogate covariates are observed for some but not all individuals in the sample. We
randomly selected a subsample of $21$ individuals for which we consider the corresponding observed values of $LDL$ and $TC$; for the
remaining individuals, only the observations on $TC$ are considered in the analysis.
A scatter plot of $TC$ versus $LDL$ for the selected individuals (not shown) suggests a clear approximate linear tendency.

It is possible to estimate $\tau_{0}$, $\tau_{1}$ and $\sigma_{e}^{2}$ from (\ref{mod1}), since we have observations on both $TC$ and $LDL$ for some individuals. We obtained the following estimates: $\widehat{\tau}_{0}=0.7351$, $\widehat{\tau}_{1}=1.062$ and $\widehat{\sigma}_{e}^{2}=0.030$.  To estimate the parameters of interest we used the approximate maximum likelihood, maximum pseudo-likelihood, regression calibration and na\"{i}ve methods. Also, as a gold standard to compare with these methods, we fitted a beta regression model in which $LDL$ is used as the true covariate, measured without error, for all individuals.

Table \ref{estimativasM11} shows the estimates, standard errors and $p$-values for the parameters of interest.
All the approaches produce similar inferences on the parameters of the mean submodel ($\alpha_0$, $\alpha_1$ and $\beta$)
and on the intercept ($\gamma_0$) and the coefficient $\gamma_1$ of BMI, which is the covariate measured without error,  in the
precision submodel. Inference on $\lambda$, the coefficient of $LDL$ (the covariate measured with error) in the
precision submodel, varies depending on the approach being used.
The gold standard and the approximate maximum likelihood and pseudo-likelihood approaches indicate that the null hypothesis
$\mathcal{H}_{0}: \lambda=0$ should be rejected at the $5\%$ nominal level ($p$-value$= 0.013, \ 0.025, \ 0.022$,
respectively), while $\mathcal{H}_{0}$ is not rejected when the regression calibration and the na\"{i}ve methods are
employed ($p$-value$=0.064, \ 0.064$). In other words, at the 5\% nominal level, the approximate maximum likelihood and maximum pseudo-likelihood
approaches agree with the gold standard in that they indicate that the precision varies with $LDL$,
unlike the regression calibration and na\"{i}ve methods. Finally, it can be noticed that the results for the
approximate maximum likelihood and maximum pseudo-likelihood methods are very close.

\begin{table}[ht!]
\centering {\caption{Estimates, standard errors and $p$-values}
\vspace*{0.3cm}
\label{estimativasM11}
\setlength{\tabcolsep}{9pt}
\renewcommand{\arraystretch}{0.97}
\begin{small}
\begin{tabular}{ccrcrc}
\hline
Method & Parameter & Estimate & Standard error & \textit{z} stat &$p$-value  \\
\hline
\multicolumn{1}{c}{\multirow{6}{*}{Gold standard}} &  $\alpha_{0}$   &$-$0.354 & 0.124 & $-$2.855 & 0.004\\
\multicolumn{1}{c}{}                               &  $\alpha_{1}$   &$-$0.009 & 0.004 & $-$2.250 & 0.023\\
\multicolumn{1}{c}{}                               &  $\beta$        &   0.107 & 0.050 &  2.140 & 0.034\\
\cline{2-6}
\multicolumn{1}{c}{}                               &  $\gamma_{0}$   &   5.905 & 0.764 &    7.729 & 0.000\\
\multicolumn{1}{c}{}                               &  $\gamma_{1}$   &$-$0.022 & 0.024 & $-$0.916 & 0.356\\
\multicolumn{1}{c}{}                               &  $\lambda$      &$-$0.729 & 0.292 & $-$2.497 & 0.013  \\
\hline\hline
\multicolumn{1}{c}{\multirow{6}{*}{$\ell_{a}$}} &  $\alpha_{0}$   &$-$0.366  & 0.113 & $-$3.239 & 0.006\\
\multicolumn{1}{c}{}                            &  $\alpha_{1}$   &$-$0.009  & 0.004 & $-$2.250 & 0.024\\
\multicolumn{1}{c}{}                            &  $\beta$        &   0.118  & 0.062 &    1.903 & 0.058\\
\cline{2-6}
\multicolumn{1}{c}{}                            &  $\gamma_{0}$   &   6.109  & 0.908 &    6.728 & 0.000\\
\multicolumn{1}{c}{}                            &  $\gamma_{1}$   &$-$0.028  & 0.028 & $-$1.000 & 0.310\\
\multicolumn{1}{c}{}                            &  $\lambda$      &$-$0.751  & 0.336 & $-$2.235 & 0.025\\
\hline\hline
\multicolumn{1}{c}{\multirow{6}{*}{$\ell_{p}$}} &  $\alpha_{0}$   &$-$0.366  & 0.133 & $-$2.759 & 0.006\\
\multicolumn{1}{c}{}                            &  $\alpha_{1}$   &$-$0.009  & 0.004 & $-$2.250 & 0.013\\
\multicolumn{1}{c}{}                            &  $\beta$        &   0.118  & 0.068 &    1.735 & 0.082\\
\cline{2-6}
\multicolumn{1}{c}{}                            &  $\gamma_{0}$   &  6.109   & 0.994 &    6.146 & 0.000\\
\multicolumn{1}{c}{}                            &  $\gamma_{1}$   &$-$0.028  & 0.036 & $-$0.778 & 0.434\\
\multicolumn{1}{c}{}                            &  $\lambda$      &$-$0.751  & 0.329 & $-$2.283 & 0.022\\
\hline\hline
\multicolumn{1}{c}{\multirow{6}{*}{$\ell_{rc}$}} &  $\alpha_{0}$   &$-$0.356 & 0.134 & $-$2.657 & 0.008\\
\multicolumn{1}{c}{}                             &  $\alpha_{1}$   &$-$0.009 & 0.004 & $-$2.250 & 0.023\\
\multicolumn{1}{c}{}                             &  $\beta$        &   0.112 & 0.062 &    1.806 & 0.068\\
\cline{2-6}
\multicolumn{1}{c}{}                             &  $\gamma_{0}$  &   5.945  & 0.804 &    7.394 & 0.000\\
\multicolumn{1}{c}{}                             &  $\gamma_{1}$  &$-$0.028  & 0.024 & $-$1.167 & 0.239\\
\multicolumn{1}{c}{}                             &  $\lambda$     &$-$0.649  & 0.351 & $-$1.849 & 0.064\\
\hline\hline
\multicolumn{1}{c}{\multirow{6}{*}{$\ell_{naive}$}} &  $\alpha_{0}$ &$-$0.389 & 0.144 & $-$2.701 & 0.007\\
\multicolumn{1}{c}{}                                &  $\alpha_{1}$ &$-$0.009 & 0.004 & $-$2.250 & 0.023\\
\multicolumn{1}{c}{}                                &  $\beta$      &   0.085 & 0.047 &    1.809 & 0.068\\
\cline{2-6}
\multicolumn{1}{c}{}                                &$\gamma_{0}$   &   6.136 & 0.864 &  7.102 & 0.000\\
\multicolumn{1}{c}{}                                &  $\gamma_{1}$ &$-$0.028 & 0.024 & $-$1.167 & 0.239\\
\multicolumn{1}{c}{}                                &  $\lambda$    &$-$0.490 & 0.265 & $-$1.849 & 0.064\\
\hline
\end{tabular}
\end{small}
}
\end{table}

\begin{figure}%[ht!]
\begin{minipage}[b]{0.5\linewidth}
\includegraphics[width=\linewidth,height=7cm]{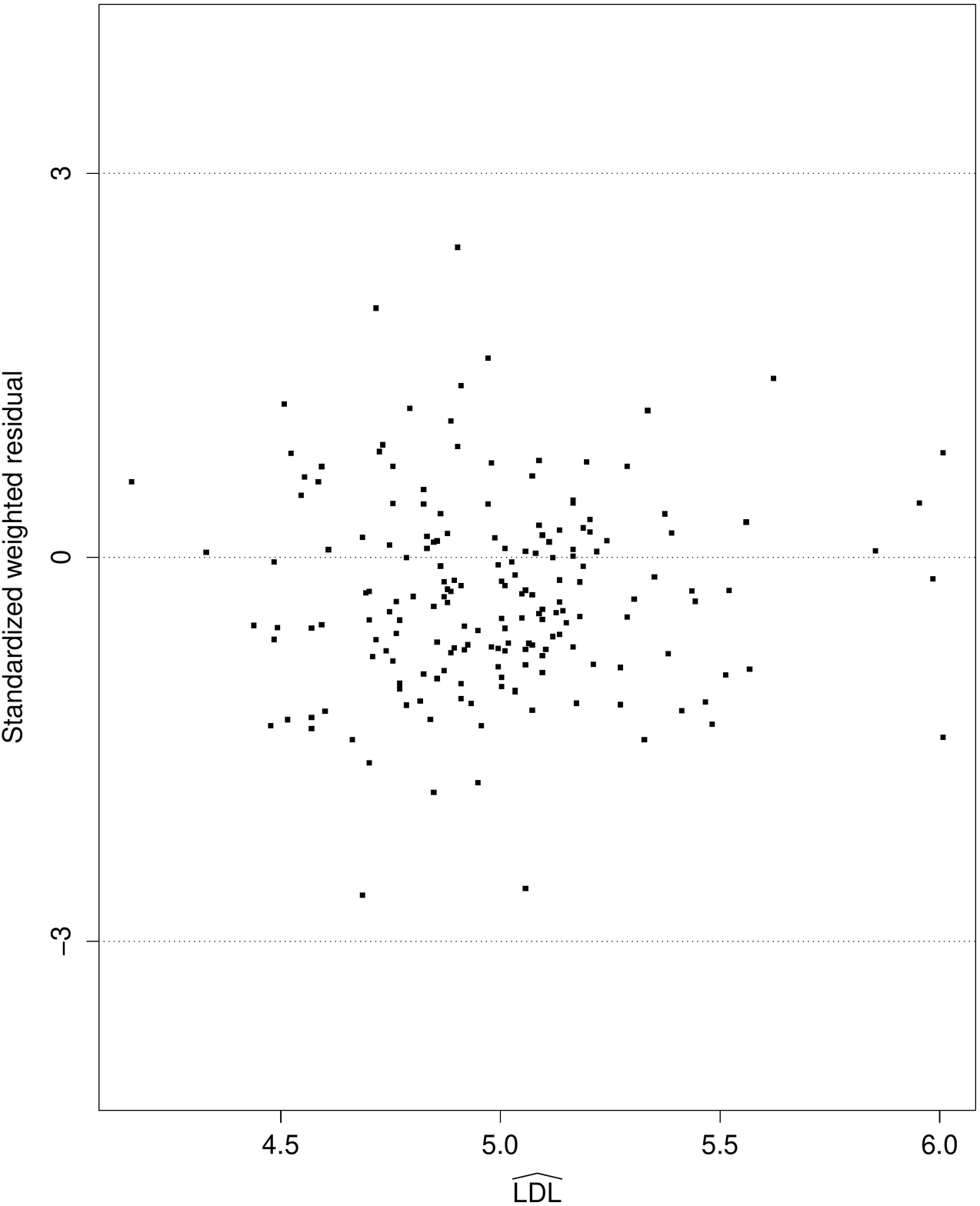}
\begin{center}
(a)
\end{center}
\end{minipage}
\hfill
\begin{minipage}[b]{0.5\linewidth}
\includegraphics[width=\linewidth,height=7cm]{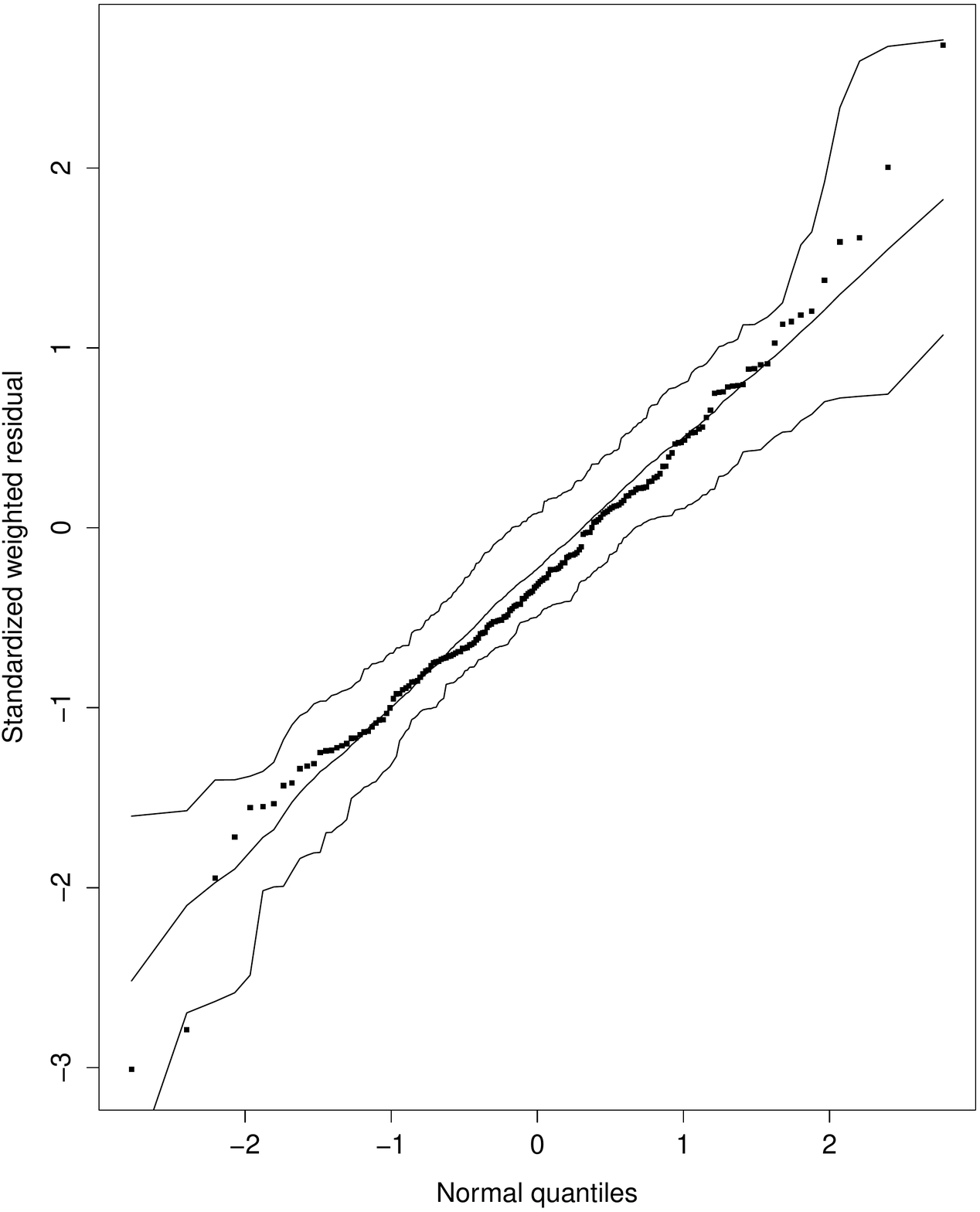}
\begin{center}
(b)
\end{center}
\end{minipage}
\caption{Plot of standardized weighted residuals  versus predicted values of $LDL$ (a), and normal probability plot of standardized weighted residuals (b).}
\label{residuo_aplica2}
\end{figure}

We now use the standardized weighted residual presented in Section \ref{sec:resid} to investigate the presence of outliers or any indication of lack of fit. Figure \ref{residuo_aplica2} shows residual plots for the model fitted using the maximum
pseudo-likelihood approach.
Figure \ref{residuo_aplica2}(a) shows the plot of the residuals against predicted values of $LDL$, $\widehat{LDL}$, and
Figure \ref{residuo_aplica2}(b) shows a normal probability plot with simulated envelope.
There is no indication of outliers or any apparent pattern. This indicates that the errors-in-variables model
considered here fits the data well.

\section{Concluding remarks}
\label{conclu}

In this paper we proposed and studied errors-in-variables beta regression models. We proposed three different estimation methods,
namely, the approximate maximum likelihood, maximum \ pseudo-likelihood and regression \ calibration methods. 
We performed a Monte Carlo simulation study to compare the performance of the estimators in terms of bias, root-mean-square errors and
coverage of confidence intervals. Overall, we reached the following conclusions. First, ignoring the measurement error
may lead to severely biased inference. Second, the regression calibration approach is very simple and seems to be reliable for
estimating the parameters of the mean submodel when the measurement error variance is small. However, there is clear indication 
that  it is not consistent for estimating the parameters that model the precision of the data.Third, the approximate maximum
likelihood and maximum pseudo-likelihood approaches perform well, the later being less computationally demanding than the
former. We, therefore, recommend the maximum pseudo-likelihood estimation for practical applications.  
We emphasize that the maximum pseudo-likelihood estimator coincides with the improved regression calibration estimator proposed by 
\citet{Skrondal+Kuha}. Its consistency and asymptotic normality are justified by these authors.
We also proposed a standardized weighted residual for diagnostic purposes. All our results were illustrated in the analysis of a real data set.

An alternative estimation method that could be applied in errors-in-variables beta regression models was recently proposed by 
\citet{Kukush+Malenko+Schneeweiss}; see Section 4.2 in their paper. It is a quasi-score estimator, which is optimal within a class of estimators 
based on unbiased estimating functions that are linear in the response variable. Although the method is simple, for measurement error 
beta regression models it involves moments of nonlinear functions of the latent variable, which cannot be obtained analytically. 
We leave this interesting topic for future research.

\section*{Acknowledgements}
We gratefully acknowledge the financial support from CAPES-Brazil, CNPq-Brazil, FAPESP-Brazil and FONDECYT(1120121)-Chile. We also thank two anonymous referees for constructive comments and suggestions.

%\section*{Apêndice}

%Ver arquivo complemento.pdf

% ---------------------------------------------------------------------------- %
% Bibliografia
\singlespacing           % espaçamento simple
%\bibliographystyle{natbib} % citação bibliográfica textual

% ---------------------------------------------------------------------------- %

\newpage

\setcounter{page}{1}

\begin{frontmatter}

%\title{Beta regression models with measurement errors}
\title{Supplementary material to ``Errors-in-variables beta regression models''}
%\author{Jalmar M. F. Carrasco}
%\address{Departamento de Estatística, Universidade Federal da Bahia, Brazil}
%\author{Silvia L. P. Ferrari
%\footnote
%{Corresponding author: Departamento de Estatística, Universidade de São Paulo,
%Rua do Matão, 1010, 05508-090,  São Paulo, SP, Brazil.
%e-mail:silviaferrari.usp@gmail.com}
%}
%\address{Departamento de Estatística, Universidade de São Paulo, Brazil}
%\author{Reinaldo B. Arellano-Valle}
%\address{Departamento de Estadística, Pontifícia Universidad Católica de Chile, Chile}
\date{}

\begin{abstract}
We present numerical tables used to produce Figures 1-8. 

\end{abstract}

\end{frontmatter}

%..................................................................................................

\addtocounter{table}{-1}

\begin{table}
\centering {\caption{Bias and root-mean-square error; $k_{x}$=0.95, constant precision model}
\label{simu21}
\setlength{\tabcolsep}{12pt}
\renewcommand{\arraystretch}{1.50}
\begin{tiny}
\begin{tabular}{cccrrr}
\hline
$n$   &  Method                      &         &  \multicolumn{1}{c}{$\alpha$}  & \multicolumn{1}{c}{$\beta$} & \multicolumn{1}{c}{$\phi$}  \\
\hline
      & \multicolumn{1}{c}{\multirow{2}{*}{$\ell_{a}$}}    &  Bias  & -0.02  &  0.01  &  -0.26   \\
      & \multicolumn{1}{c}{}                               &  RMSE  &  0.17  &  0.06  &   0.64  \\
	    \cline{2-6}
      & \multicolumn{1}{c}{\multirow{2}{*}{$\ell_{p}$}}    &  Bias  & -0.02  & 0.01  &   -0.26  \\
25    & \multicolumn{1}{c}{}                               &  RMSE  &  0.17  & 0.05  &    0.64 \\
      \cline{2-6}
      & \multicolumn{1}{c}{\multirow{2}{*}{$\ell_{rc}$}}   &  Bias  &  0.01 &  -0.01 &  0.29   \\
      & \multicolumn{1}{c}{}                               &  RMSE  &  0.16 &   0.05 &  0.41   \\
      \cline{2-6}
      & \multicolumn{1}{c}{\multirow{2}{*}{$\ell_{naive}$}}& Bias   & 0.09  & -0.04  &  0.31  \\
\hline
\end{tabular}
\end{tiny}
}
\end{table}

\begin{table}
\centering {\caption{Bias and root-mean-square error; $k_{x}$=0.75, constant precision model}
\label{simu22}
\setlength{\tabcolsep}{12pt}
\renewcommand{\arraystretch}{1.50}
\begin{tiny}
\begin{tabular}{cccrrr}
\hline
$n$   &  Method                      &         &  \multicolumn{1}{c}{$\alpha$}  & \multicolumn{1}{c}{$\beta$} & \multicolumn{1}{c}{$\phi$}  \\
\hline
      & \multicolumn{1}{c}{\multirow{2}{*}{$\ell_{a}$}}    &  Bias  & -0.06  & 0.02  &  -1.02   \\
      & \multicolumn{1}{c}{}                               &  RMSE  &  0.37  & 0.13  &   1.92  \\
      \cline{2-6}
      & \multicolumn{1}{c}{\multirow{2}{*}{$\ell_{p}$}}    &  Bias  & -0.06 & 0.03  &  -0.91   \\
25    & \multicolumn{1}{c}{}                               &  RMSE  &  0.39 & 0.13  &   1.63  \\
      \cline{2-6}
      & \multicolumn{1}{c}{\multirow{2}{*}{$\ell_{rc}$}}   &  Bias  &  0.02 & 0.02  &  1.18  \\
      & \multicolumn{1}{c}{}                               &  RMSE  &  0.35 & 0.12  &  1.22   \\
      \cline{2-6}
      & \multicolumn{1}{c}{\multirow{2}{*}{$\ell_{naive}$}}& Bias   & 0.47  & -0.17  &  1.21   \\
\hline
\end{tabular}
\end{tiny}
}
\end{table}

\begin{table}
\centering {\caption{Bias and root-mean-square error; $k_{x}$=0.50, constant precision model}
\label{simu23}
\setlength{\tabcolsep}{12pt}
\renewcommand{\arraystretch}{1.50}
\begin{tiny}
\begin{tabular}{cccrrr}
\hline
$n$   &  Method                      &         &  \multicolumn{1}{c}{$\alpha$}  & \multicolumn{1}{c}{$\beta$} & \multicolumn{1}{c}{$\phi$}  \\
\hline
      & \multicolumn{1}{c}{\multirow{2}{*}{$\ell_{a}$}}    &  Bias  & -0.16  & 0.06  &  -1.07   \\
      & \multicolumn{1}{c}{}                               &  RMSE  &  0.99  & 0.37  &   2.01  \\			
      \cline{2-6}
      & \multicolumn{1}{c}{\multirow{2}{*}{$\ell_{p}$}}    &  Bias  &  -0.23 &  0.09 &   -0.63  \\
25    & \multicolumn{1}{c}{}                               &  RMSE  &   1.17 &  0.45 &    1.50 \\
      \cline{2-6}
      & \multicolumn{1}{c}{\multirow{2}{*}{$\ell_{rc}$}}   &  Bias  &  0.04  &  0.04 &  1.74   \\
      & \multicolumn{1}{c}{}                               &  RMSE  &  0.60 &   0.24 &  1.77   \\		
      \cline{2-6}
      & \multicolumn{1}{c}{\multirow{2}{*}{$\ell_{naive}$}}& Bias   & 0.88  & -0.33  &  1.75   \\
\hline
\end{tabular}
\end{tiny}
}
\end{table}

%..................................................................................................

\begin{table}
\centering {\caption{Bias and root-mean-square error; $k_{x}$=0.95, varying precision model}
\label{simu11}
\setlength{\tabcolsep}{12pt}
\renewcommand{\arraystretch}{1.50}
\begin{tiny}
\begin{tabular}{cccrrrr}
\hline
$n$   &  Method                      &         &  \multicolumn{1}{c}{$\alpha$}  & \multicolumn{1}{c}{$\beta$} & \multicolumn{1}{c}{$\gamma$} & \multicolumn{1}{c}{$\lambda$}  \\
\hline
      & \multicolumn{1}{c}{\multirow{2}{*}{$\ell_{a}$}}    &  Bias  & 0.00  &  0.00  & -0.23  & -0.01  \\
      & \multicolumn{1}{c}{}                               &  RMSE  & 0.22  &  0.06  &  0.16  &  0.12 \\
      \cline{2-7}
      & \multicolumn{1}{c}{\multirow{2}{*}{$\ell_{p}$}}    &  Bias  & 0.00  &  0.00  & -0.23  & 0.00  \\
25    & \multicolumn{1}{c}{}                               &  RMSE  & 0.22  &  0.06  &  0.16 &  0.12 \\
      \cline{2-7}
      & \multicolumn{1}{c}{\multirow{2}{*}{$\ell_{rc}$}}   &  Bias  & 0.02  &  -0.01  &  -0.06 &  0.15 \\
      & \multicolumn{1}{c}{}                               &  RMSE  & 0.22  &   0.06  &   0.66 &  0.28  \\
      \cline{2-7}
      & \multicolumn{1}{c}{\multirow{2}{*}{$\ell_{naive}$}}& Bias   &  0.10 &  -0.04  & -0.11  & 0.17   \\
\hline
\end{tabular}
\end{tiny}
}
\end{table}

\begin{table}
\centering {\caption{Bias and root-mean-square error; $k_{x}$=0.75, varying precision model}
\label{simu12}
\setlength{\tabcolsep}{12pt}
\renewcommand{\arraystretch}{1.50}
\begin{tiny}
\begin{tabular}{cccrrrr}
\hline
$n$   &  Method                      &         &  \multicolumn{1}{c}{$\alpha$}  & \multicolumn{1}{c}{$\beta$} & \multicolumn{1}{c}{$\gamma$} & \multicolumn{1}{c}{$\lambda$}  \\
\hline
      & \multicolumn{1}{c}{\multirow{2}{*}{$\ell_{a}$}}    &  Bias  &  -0.03 & 0.01   & -0.26  &  -0.02 \\
      & \multicolumn{1}{c}{}                               &  RMSE  &   0.44 & 0.14   &  0.61  &   0.59\\	
      \cline{2-7}
      & \multicolumn{1}{c}{\multirow{2}{*}{$\ell_{p}$}}    &  Bias  & -0.03  & 0.01   & -0.23  &  0.00 \\
25    & \multicolumn{1}{c}{}                               &  RMSE  &  0.43  & 0.14   &  0.61  &  0.54 \\
      \cline{2-7}
      & \multicolumn{1}{c}{\multirow{2}{*}{$\ell_{rc}$}}   &  Bias  & 0.01  &  0.00  & 0.32  &  0.34 \\
      & \multicolumn{1}{c}{}                               &  RMSE  & 0.48  &  0.17  & 0.98  &  0.47  \\
      \cline{2-7}
      & \multicolumn{1}{c}{\multirow{2}{*}{$\ell_{naive}$}}& Bias   &  0.46 &  -0.17  & 0.16  &  0.38  \\
\hline
\end{tabular}
\end{tiny}
}
\end{table}

\begin{table}
\centering {\caption{Bias and root-mean-square error; $k_{x}$=0.50, varying precision model}
\label{simu13}
\setlength{\tabcolsep}{12pt}
\renewcommand{\arraystretch}{1.50}
\begin{tiny}
\begin{tabular}{cccrrrr}
\hline
$n$   &  Method                      &         &  \multicolumn{1}{c}{$\alpha$}  & \multicolumn{1}{c}{$\beta$} & \multicolumn{1}{c}{$\gamma$} & \multicolumn{1}{c}{$\lambda$}  \\
\hline
      & \multicolumn{1}{c}{\multirow{2}{*}{$\ell_{a}$}}    &  Bias  & -0.12  & 0.04  & -0.75  & -0.01  \\
      & \multicolumn{1}{c}{}                               &  RMSE  &  0.90  & 0.33  &  0.45  &  0.74 \\		
      \cline{2-7}
      & \multicolumn{1}{c}{\multirow{2}{*}{$\ell_{p}$}}    &  Bias  &  -0.18 &  0.06  & -0.61  & 0.00  \\
25    & \multicolumn{1}{c}{}                               &  RMSE  &   0.14 &  0.42  &  0.34  & 0.72  \\
      \cline{2-7}
      & \multicolumn{1}{c}{\multirow{2}{*}{$\ell_{rc}$}}   &  Bias  & -0.62  &  0.25  & 0.67  & 0.38  \\
      & \multicolumn{1}{c}{}                               &  RMSE  &  1.02  &  0.53  & 0.58  & 0.69   \\
      \cline{2-7}
      & \multicolumn{1}{c}{\multirow{2}{*}{$\ell_{naive}$}}& Bias   &  0.86 &  -0.33  & 0.47  &  0.46  \\
\hline
\end{tabular}
\end{tiny}
}
\end{table}

%..................................................................................................

%..................................................................................................

\begin{table}
\centering {\caption{Coverage of confidence intervals ($\%$)}
\vspace{0.3cm}
\label{Cover11}
\setlength{\tabcolsep}{12pt}
\renewcommand{\arraystretch}{1.50}
\begin{tiny}
\begin{tabular}{ccccccc}
\hline
  $k_{x}$  &  $n$  &  Method  &  $\alpha$  &  $\beta$  &  $\gamma$  & $\lambda$  \\
\hline
\multicolumn{1}{c}{\multirow{24}{*}{0.95}}&\multicolumn{1}{c}{\multirow{4}{*}{25}}&$\ell_{a}$
& 91.50 & 92.20 & 94.92 & 93.66  \\
\multicolumn{1}{c}{}                      &           \multicolumn{1}{c}{}        &$\ell_{p}$
& 90.90 & 90.12 & 86.30 & 87.02  \\
\multicolumn{1}{c}{}                      &           \multicolumn{1}{c}{}       &$\ell_{rc}$
& 89.42 & 88.06 & 88.12 & 77.28  \\
\multicolumn{1}{c}{}                      &           \multicolumn{1}{c}{}      &$\ell_{naive}$
& 87.90 & 85.10 & 92.44 & 83.16  \\
\cline{2-7}
    &  \multicolumn{1}{c}{\multirow{4}{*}{50}}       &  $\ell_{a}$
& 93.42 & 93.14 & 93.94 & 94.94  \\
    &           \multicolumn{1}{c}{}                 &  $\ell_{p}$
& 92.62 & 91.44 & 88.94 & 90.14  \\
    &           \multicolumn{1}{c}{}                 &  $\ell_{rc}$
& 92.38 & 91.34 & 91.38 & 70.50  \\
\hline
\end{tabular}
\end{tiny}
}
\end{table}

\begin{table}
\centering {\caption{Coverage of confidence intervals ($\%$) (cont.)}
\vspace{0.3cm}
\label{Cover12}
\setlength{\tabcolsep}{12pt}
\renewcommand{\arraystretch}{1.5}
\begin{tiny}
\begin{tabular}{ccccccc}
\hline
  $k_{x}$  &  $n$  &  Method  &  $\alpha$  &  $\beta$  &  $\gamma$  & $\lambda$  \\
\hline
\multicolumn{1}{c}{\multirow{24}{*}{0.50}}&\multicolumn{1}{c}{\multirow{4}{*}{25}}&$\ell_{a}$
& 67.76 & 59.70 & 80.48 & 70.62  \\
\multicolumn{1}{c}{}                      &           \multicolumn{1}{c}{}        &$\ell_{p}$
& 90.26 & 89.00 & 88.02 & 73.28  \\			
\multicolumn{1}{c}{}                      &           \multicolumn{1}{c}{}       &$\ell_{rc}$
& 53.40 & 43.50 & 64.80 & 42.00  \\
\multicolumn{1}{c}{}                      &           \multicolumn{1}{c}{}      &$\ell_{naive}$
& 9.98 & 1.48 & 75.14 & 14.26  \\
\cline{2-7}
    &  \multicolumn{1}{c}{\multirow{4}{*}{50}}       &  $\ell_{a}$
& 75.10 & 69.74 & 87.12 & 68.92  \\
    &           \multicolumn{1}{c}{}                 &  $\ell_{p}$
& 91.52 & 90.66 & 90.76 & 72.16  \\
    &           \multicolumn{1}{c}{}                 &  $\ell_{rc}$
& 50.60 & 41.10 & 43.10 & 21.50  \\
\hline
\end{tabular}
\end{tiny}
}
\end{table}

%........................................................................................
%........................................................................................

\begin{table}
\centering {\caption{Coverage of confidence intervals ($\%$)}
\vspace{0.3cm}
\label{Cover21}
\setlength{\tabcolsep}{12pt}
\renewcommand{\arraystretch}{1.50}
\begin{tiny}
\begin{tabular}{cccccc}
\hline
  $k_{x}$  &  $n$  &  Method  &  $\alpha$  &  $\beta$  &  $\phi$    \\
\hline
\multicolumn{1}{c}{\multirow{24}{*}{0.95}}&\multicolumn{1}{c}{\multirow{4}{*}{25}}&$\ell_{a}$
& 92.78 & 92.84 & 98.38   \\
\multicolumn{1}{c}{}                      &           \multicolumn{1}{c}{}        &$\ell_{p}$
& 91.82 & 91.14 & 93.96   \\
\multicolumn{1}{c}{}                      &           \multicolumn{1}{c}{}       &$\ell_{rc}$
& 90.50 & 90.80 & 80.60   \\
\multicolumn{1}{c}{}                      &           \multicolumn{1}{c}{}      &$\ell_{naive}$
& 88.72 & 86.84 & 77.80   \\
\cline{2-6}
    &  \multicolumn{1}{c}{\multirow{4}{*}{50}}       &  $\ell_{a}$
& 93.90 & 93.56 & 96.46   \\
    &           \multicolumn{1}{c}{}                 &  $\ell_{p}$
& 93.42 & 93.50 & 93.98  \\
    &           \multicolumn{1}{c}{}                 &  $\ell_{rc}$
& 91.90 & 91.80 & 53.00   \\
\hline
\end{tabular}
\end{tiny}
}
\end{table}

\begin{table}
\centering {\caption{Coverage of confidence intervals ($\%$) (cont.)}
\vspace{0.3cm}
\label{Cover22}
\setlength{\tabcolsep}{12pt}
\renewcommand{\arraystretch}{1.50}
\begin{tiny}
\begin{tabular}{cccccc}
\hline
  $k_{x}$  &  $n$  &  Method  &  $\alpha$  &  $\beta$  &  $\phi$    \\
\hline
\multicolumn{1}{c}{\multirow{24}{*}{0.50}}&\multicolumn{1}{c}{\multirow{4}{*}{25}}&$\ell_{a}$
& 70.94 & 62.24 & 41.32   \\	
\multicolumn{1}{c}{}                      &           \multicolumn{1}{c}{}        &$\ell_{p}$
& 92.28 & 92.04 & 59.96   \\
\multicolumn{1}{c}{}                      &           \multicolumn{1}{c}{}       &$\ell_{rc}$
& 45.70 & 41.50 & 0.10   \\
\multicolumn{1}{c}{}                      &           \multicolumn{1}{c}{}      &$\ell_{naive}$
& 3.10 & 0.36 & 0.06   \\
\cline{2-6}
    &  \multicolumn{1}{c}{\multirow{4}{*}{50}}       &  $\ell_{a}$
& 79.56 & 74.04 & 53.38   \\
    &           \multicolumn{1}{c}{}                 &  $\ell_{p}$
& 93.72 & 93.48 & 63.84   \\
    &           \multicolumn{1}{c}{}                 &  $\ell_{rc}$
& 47.10 & 40.10 & 0.00   \\
\hline
\end{tabular}
\end{tiny}
}
\end{table}

\end{document}